\documentclass[preprint,review,12pt]{elsarticle}

\usepackage{amssymb}
\usepackage{amsmath,bm}
\usepackage{colortbl,booktabs}
\usepackage{tabularx}
\usepackage{threeparttable}
\usepackage{multicol,multirow}
\usepackage{subfigure}
\usepackage [font=footnotesize,labelfont=bf]{caption}
\usepackage{rotating}
\usepackage{tikz}
\usepackage{hyperref}
\usepackage{makecell}

\usepackage{color}

\journal{Elsevier}

\begin{document}

\begin{frontmatter}

	\title{A multi-scale approach for wall-bounded WCSPH--RANS simulations}

	\author{Feng Wang}
	\ead{feng.wang.aer@tum.de}

	\author{Nikolaus Adams}
	\ead{nikolaus.adams@tum.de}

	\author{Xiangyu Hu\texorpdfstring{\corref{mycorrespondingauthor}}{}}
	\cortext[mycorrespondingauthor]{Corresponding author.}
	\ead{xiangyu.hu@tum.de}

	\address{Department of Mechanical Engineering, Technical University of Munich\\
		85748 Garching, Germany}

	\begin{abstract}
		Achieving sufficient near-wall resolution remains challenging
		for wall-bounded turbulence simulations using weakly compressible
		smoothed particle hydrodynamics (WCSPH), as local particle
		refinement imposes restrictive time-step requirements.
		This work proposes a multi-scale near-wall approach within
		the Reynolds-averaged Navier--Stokes (RANS) framework to improve
		near-wall predictions without refining the SPH particle distribution.
		A local one-dimensional sublayer solver based on the simplified
		steady $k$--$\omega$ equations is coupled with each wall-adjacent
		fluid particle. A local flow-rate constraint and a
		friction-velocity-based iteration scheme are developed to close
		and solve the sublayer system, while a two-way shear-stress
		coupling with local feedback improves consistency between the
		two scales.
		The proposed approach is evaluated using turbulent straight-channel
		flows over a wide range of Reynolds numbers and a wavy-channel
		flow involving separation and recirculation.
		The results demonstrate improved near-wall velocity, turbulent kinetic energy and
		friction-coefficient predictions, satisfactory convergence,
		and good agreement with reference solutions.
		In particular, the approach improves the convergence of near-wall
		predictions where conventional wall treatments struggle to provide
		consistent results, and agrees closely with reference solutions
		at Reynolds numbers as high as $8.0 \times 10^{7}$ without
		near-wall particle refinement.
		For separated flow, convergence comparable to that of locally
		refined finite-volume simulations is obtained using an
		approximately uniform particle distribution.
		These improvements are achieved with limited computational
		overhead, supporting the application of particle-based methods
		to engineering flows involving wall-bounded turbulence.
	\end{abstract}


\end{frontmatter}
%
%
\section{Introduction}
\label{sec1}
To predict engineering-relevant quantities, such as the friction coefficient,
sufficient near-wall resolving capability is essential in
Reynolds-averaged Navier--Stokes (RANS) modeling~\cite{nguyen2020dual},
particularly when advanced $k$--$\omega$ models are employed
\cite{ecca2018viscous,wang6271943unified}.
In general, two strategies can be used to improve the near-wall prediction:
one is to locally enhance the near-wall resolution, and the other is to employ
a higher-order reconstruction scheme.
Nevertheless, both approaches are inherently limited when applied to turbulence simulations in Lagrangian, fully particle-based frameworks, including smoothed particle hydrodynamics (SPH) \cite{gingold1977smoothed,lucy1977numerical} and moving particle semi-implicit (MPS) \cite{koshizuka1996moving} methods.

First, locally increasing the near-wall resolution is particularly challenging for the unsteady, explicit, mesh-less schemes, such as the compressible or weakly compressible SPH, because the time step is directly constrained by the smallest discretization unit, limiting any substantial reduction in computational cost.
Moreover, while semi-implicit mesh-less methods, including the MPS\cite{khayyer2019multi,tang2016numerical,tanaka2018multi,shibata2017overlapping} and incompressible SPH (ISPH)\cite{khorasanizade2016dynamic}, can partially relax the time-step limitations, the unsteady, Lagrangian characteristics of these particle-based methods persist, meaning that particle advection imposes additional constraints on the time step.
This time-step constraint becomes especially significant in wall-bounded turbulence simulations, where extremely high refinement ratios are necessary.

Although the aforementioned difficulty persists, current research on partial refinement techniques in the SPH method is largely restricted to spatial refinement—similar to mesh-based approaches—commonly referred to as $h$-refinement\cite{rahmani2025anisotropic}.
Despite their successful applications in free-surface\cite{chiron2018analysis,barcarolo2014adaptive,vacondio2012accurate,reyes2013dynamic,omidvar2012wave,bonet2005hamiltonian,zhang2025multi,yang2021smoothed,zhang2024numerical}, multiphase\cite{yang2019adaptive,sun2019study,ju2023study}, and shear flows\cite{sun2018multi,vacondio2016variable,vacondio2013variable,feldman2007dynamic,xiong2013gpu,hu2017consistent,bian2015multi,muta2022efficient,gao2023multi,gao2022block}, as well as in astrophysical\cite{kitsionas2002smoothed,bate1995modelling,meglicki19933d,monaghan1988dynamics,alimi2003smooth,owen1998adaptive}, shock-capturing\cite{lastiwka2005adaptive,shapiro1996adaptive,nelson1994variable,borve2005regularized,liu2006adaptive} problems, and engineering applications such as damage\cite{spreng2014local}, flooding\cite{vacondio2013shallow}, water entries\cite{oger2006two,lyu20223d}, underwater explosions\cite{liang2023pressure,sun2021accurate,sun2021accurateb,zhuang2020smoothed}, and submarine landslides\cite{yi2026novel}, to the best of our knowledge, none of them have been applied to wall-bounded turbulence simulations.

Second, unlike spatial refinement, which improves the prediction by reducing
discretization errors through a smaller particle spacing, the high-order
reconstruction approach improves the approximation quality using higher-order
polynomials, without introducing the additional time-step restriction
associated with near-wall particle refinement.
This approach, commonly referred to as $p$ (polynomial) refinement, has been applied in both mesh-based and mesh-free methods, including finite element\cite{zienkiewicz1983hierarchical}, Galerkin\cite{hillman2021consistent}, finite point\cite{perazzo2008adaptive}, and SPH\cite{zhang2019weakly,renaut2015high} methods, where the approximation is improved either by increasing the polynomial order\cite{zhang2019weakly,kumar2009partition,schweitzer2009adaptive,renaut2015high} or by adding additional nodes\cite{wu2014adaptive} within the same element.
However, most existing studies focus on globally improving the approximation
quality, which can be computationally inefficient for RANS simulations.
This is because RANS predictions usually require enhanced resolution mainly
in the near-wall region, where steep gradients are present.
Furthermore, to the best of our knowledge, none of these approaches have been specifically designed for SPH-based turbulence simulations.

In this work, we propose a novel multi-scale near-wall approach for the
WCSPH--RANS method to improve near-wall predictions without requiring
near-wall particle refinement.
Different from the aforementioned two approaches, an additional 1D local scale, hereafter referred to as the 1D sublayer model, is assigned to each wall-adjacent fluid particle.
Assuming the near-wall flow in the sublayer model has attained a steady state, the simplified $k$-$\omega$ RANS equations are solved locally.
A multi-scale scheme is developed to integrate the 1D sublayer model into the rigorously validated WCSPH--RANS framework\cite{wang2025weakly,wang6271943unified,wang2026effective,wang2025zero}.
Since this approach does not alter the discretization of the WCSPH--RANS method, it does not introduce the additional time-step restriction associated with near-wall particle refinement.

The remainder of this manuscript is organized as follows.
Section \ref{section-coupling} presents the multi-scale coupling scheme.
Numerical examples are provided and discussed in Section \ref{section-numerical-examples}, and concluding remarks are given in Section \ref{section-conclusion}.
The computational code of this work is released in the open-source SPHinXsys repository at https://github.com/Xiangyu-Hu/SPHinXsys.

%
%
\section{WCSPH--RANS method with a 1D sublayer model}
\label{section-coupling}
This section first introduces the multi-scale coupling scheme, then details the formulation and discretization of the 1D sublayer model, and finally describes the modifications to the WCSPH--RANS method.

\subsection{Multi-scale coupling scheme}
\subsubsection{Geometrical coupling}
\begin{figure}[htb!]
	\centering
	\includegraphics[trim = 3.82cm 0cm 0cm 6.89cm, clip,width=1.0\textwidth]{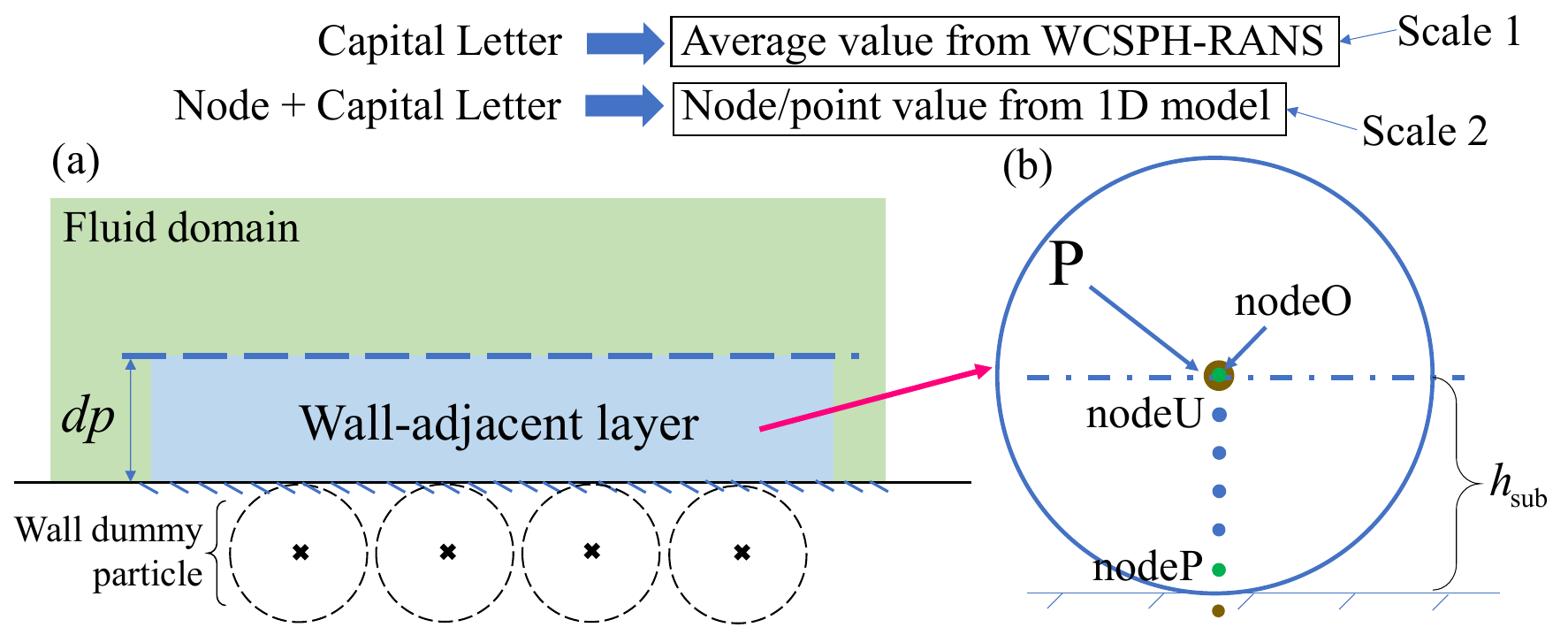}
	\caption{
		Multi-scale coupling framework between the SPH and 1D sublayer scales for a flat-wall configuration: (a) the wall-adjacent SPH layer and (b) the local sublayer model with five computational nodes, where node O coincides with the particle center P. Only one layer of wall dummy particles is shown for clarity, although at least three layers are generally used.
	}
	\label{fig-concept-two-scale-hang}
\end{figure}

Figure \ref{fig-concept-two-scale-hang} illustrates the proposed multi-scale coupling strategy using a flat-wall configuration as an example.
The term multi-scale here specifically refers to two distinct scales:
(1) the SPH scale based on continuum mechanics, in which each particle represents a finite fluid volume and the particle center merely corresponds to the averaged quantities over this volume; and
(2) the local 1D scale based on node discretization.
To clearly distinguish the two scales, a naming convention is introduced in the figure.
As shown in the sub-figure (a), and following our previous work\cite{wang2025weakly}, a wall-adjacent layer is defined to account for the dynamic behavior of fluid particles.
The thickness of this layer is set to the particle spacing, $dp$, and the fluid particles within it are identified as wall-adjacent (or wall-nearest) particles.
The height of the sublayer domain is denoted by $h_\mathrm{sub}$.

More importantly, Figure \ref{fig-concept-two-scale-hang} (b) shows that, to achieve local prediction enhancement in the near-wall region, each wall-adjacent fluid particle independently incorporates a 1D sublayer model, which is embedded at half the particle spacing.
Here, the capital letter P denotes the particle center in the SPH method, and physical quantities defined at P, such as velocity, represent averaged values over the corresponding kernel support domain.
The node P and node U (uppermost) denote the first and last computational nodes of the 1D sublayer model starting from the wall, respectively.
The node O is the boundary node corresponding to node U, and is co-located with the particle center P to ensure coupling between the two scales.

\begin{figure}[htb!]
	\centering
	\includegraphics[trim = 9.53cm 0cm 0cm 3.49cm, clip,width=0.9\textwidth]{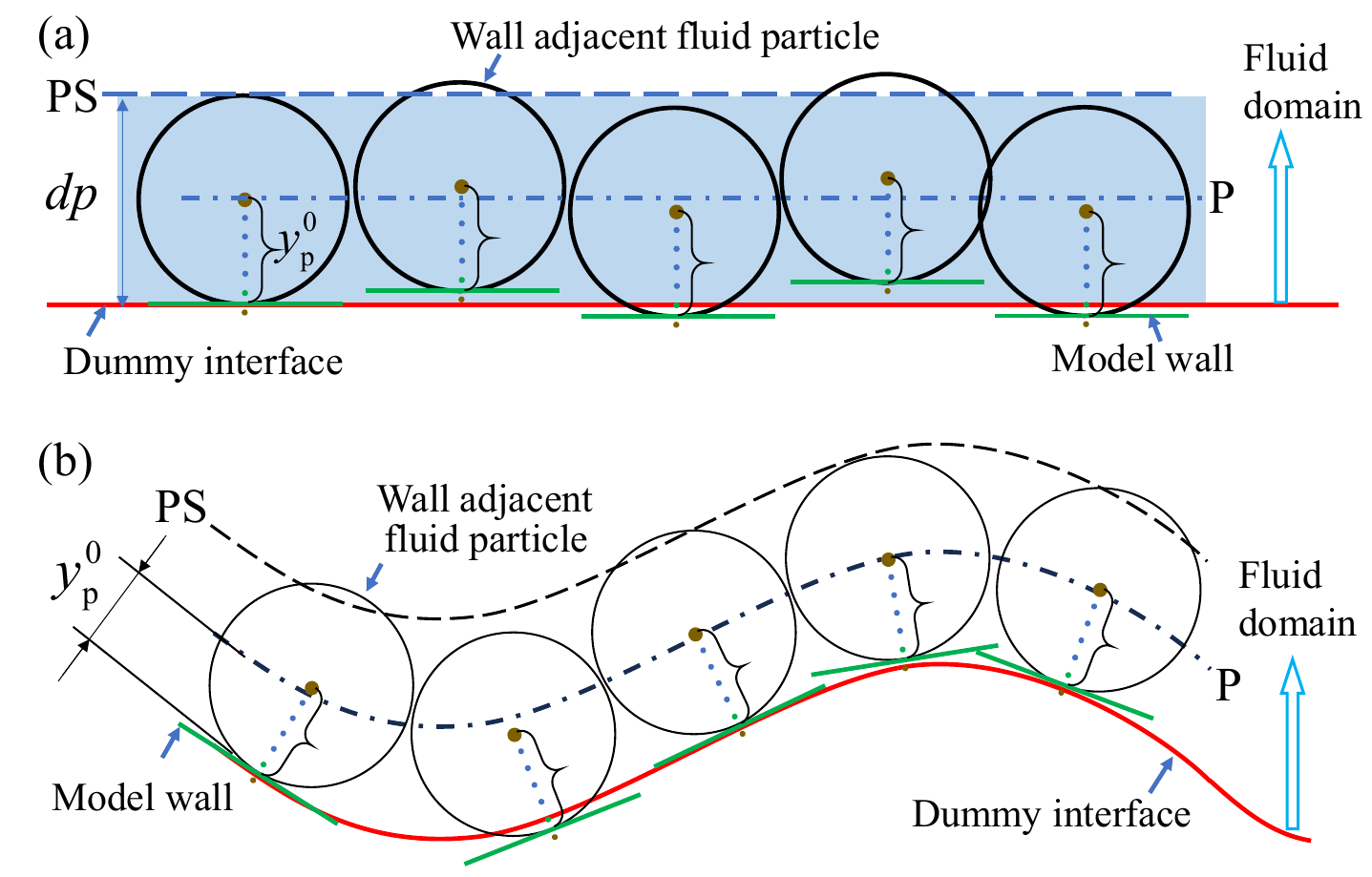}
	\caption{
		Constant $y_p$ strategy for the multi-scale coupling framework, illustrated for (a) near-wall flow over a flat plate and (b) wavy channel flow, using five computational nodes.
	}
	\label{fig-concept-constant-y-p}
\end{figure}

Additionally, to effectively suppress instabilities caused by the migration of wall-adjacent particles, the constant $y_\mathrm{p}$ strategy\cite{wang2025weakly} is adopted.
Figure \ref{fig-concept-constant-y-p} (a) shows that the model wall, serving as the physical wall in the wall-model framework, is separated from the dummy interface.
The 1D computational domain extends from the particle center P to the model wall, with a constant height of $y_\mathrm{p}^0=dp/2$, independent of the distance to the dummy interface.
This treatment significantly improves numerical stability, particularly for
complex geometries where the distance to the dummy interface may vary sharply
because of geometric irregularities.
Meanwhile, it introduces only minor errors, which decrease with increasing
resolution~\cite{wang2025weakly}.

It is also worth noting that, as shown in Fig. \ref{fig-concept-constant-y-p} (b), the 1D model is locally attached to the wall, so that the 1D computational domain adapts to the body-fitted direction, thereby improving near-wall prediction for arbitrary complex geometries, which constitutes an additional advantage of the present approach.
The wall-normal direction is computed using the level-set method\cite{yu2023level,zhu2021cad}, where the geometry surface is defined as the zero level-set of a signed-distance function,
$\Gamma = \{(x,y,z)\mid \phi(x,y,z,t)=0\}$.
The corresponding wall-normal vector is given by
$\nabla \phi / \lvert \nabla \phi \rvert.$

\subsubsection{Physical coupling}
A fundamental assumption underlying the present multi-scale coupling is that the velocity remains continuous across the SPH scale and the 1D sublayer scale at the coupling interface, i.e., the particle center P (node O), given by
\begin{equation}
	U_\mathrm{nodeO} = U_\mathrm{p}^\mathrm{SPH}.
	\label{eq-fundamental-assumption}
\end{equation}

Here, $U$ denotes the wall-tangential velocity magnitude.
The sublayer solver operates in a local wall-aligned coordinate
system using velocity magnitudes.
The corresponding velocity vector is reconstructed as
$\boldsymbol{u}=U\boldsymbol{t}$, where $\boldsymbol{t}$
is the local unit tangent oriented along the wall-tangential
SPH velocity.
Projection onto the global coordinate axes then yields signed
velocity components, including negative streamwise velocities
in reversed-flow regions.

This assumption, which relies on particle-center velocities, may introduce numerical errors, since in SPH the velocity is a kernel-averaged quantity over the support domain, although it is stored at the particle center.
However, with the correction matrix\cite{zhang2025towards} employed in the WCSPH--RANS framework~\cite{wang2025weakly}, second-order accuracy is recovered\cite{wang2025weakly, zhang2025towards,ren2023efficient}.
Additionally, an implicit feedback system (introduced in Section \ref{sec-ISS-feedbackSys}) further reduces numerical errors in the shear stress correction.

\begin{figure}[htb!]
	\centering
	\includegraphics[trim = 7.85cm 0cm 0cm 10.23cm, clip,width=1.0\textwidth]{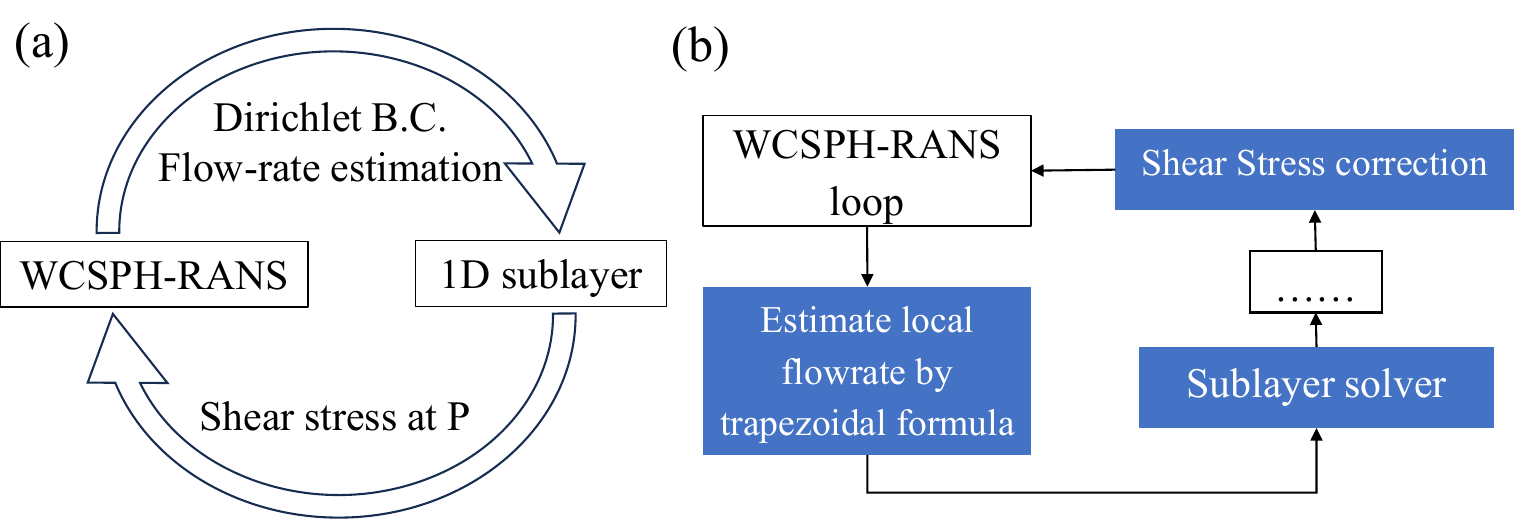}
	\caption{
		Illustration of the two-way coupling scheme: (a) information transfer between the two scales and (b) the overall computational flowchart.
	}
	\label{fig-concept-physical-coupling}
\end{figure}

The two-way coupling concept and the corresponding flowchart are illustrated in Fig. \ref{fig-concept-physical-coupling}(a) and (b), respectively.
For the WCSPH-RANS method, the shear stress at wall-adjacent particles, including the wall shear stress (WSS) and inner shear stress (ISS), is corrected using the corresponding quantities from the 1D sublayer model to maintain continuous approximation across the coupled scales.
The correction strategy follows a wall-function-like treatment, where the correction is applied only within the near-wall region while leaving the overall computational framework unchanged, as illustrated in Fig. \ref{fig-concept-physical-coupling}(b).

In contrast, for the 1D sublayer solver, the Dirichlet boundary condition (B.C.) at node O is provided by the WCSPH--RANS method.
Specifically, the velocity, turbulent kinetic energy and turbulent specific dissipation rate of particle P are mapped to node O, which serves as the boundary node of the computational node U.
Additionally, the local flow rate required for the 1D sublayer model is estimated from the coupled WCSPH--RANS solution.
Details of the estimation procedure are provided in Section \ref{sec-local-flow}.

%

\subsection{The 1D sublayer model}

This section first introduces the simplified one-dimensional $k$-$\omega$ equations, followed by their numerical discretization. Finally, an additional constraint on the local flow rate and the friction-based fixed-point iteration scheme are developed.

\subsubsection{Governing equations}
\label{sec-sub-sub-1D-governing}
In the present wall-treatment model, the sublayer flow is assumed to be locally incompressible, predominantly one-dimensional in the wall-normal direction, and steady.
This approximation is reasonable because the local near-wall variation is primarily governed by the wall-normal gradient.
Under these assumptions, the $k$-$\omega$ RANS model\cite{wilcox2008formulation} can be reduced to
\begin{equation}
	\begin{cases}
		\dfrac{ d}{dy} \! \left[ ({\nu_m} + \nu_t) \, \dfrac{du}{dy}\right]
		=
		\dfrac{1}{\rho} \dfrac{dP}{dx}, \\[8pt]

		\nu_t\left(\dfrac{du}{dy}\right)^2
		- \beta^{*} k \omega
		+ \dfrac{d}{dy} \! \left[\left(\nu_m + \sigma^{*}\dfrac{k}{\omega}\right)  \dfrac{dk}{dy}\right]
		= 0,                            \\[8pt]
		\alpha\,\dfrac{\omega}{k}\,\nu_t\left(\dfrac{du}{dy}\right)^2
		- \beta\,\omega^2
		+ \dfrac{d}{dy} \! \left[\left(\nu_m + \sigma\dfrac{k}{\omega}\right)\dfrac{d\omega}{dy}\right]
		+ \,\dfrac{\sigma_d}{\omega}\,\dfrac{dk}{dy}\,\dfrac{d\omega}{dy}
		= 0,
	\end{cases}
	\label{eq-ODE_straight_kw}
\end{equation}
where $y$ denotes the wall-normal coordinate, and $\nu_m$ and $\nu_t$ denote the kinematic molecular and eddy viscosity, respectively.
The density is denoted by $\rho$, and $u$, $P$, $k$, $\omega$ denote the local wall-tangential velocity magnitude, pressure, turbulent kinetic energy, and turbulent specific dissipation rate, respectively.

The kinematic eddy viscosity is computed as $\nu_t = k / \tilde{\omega}$, where $\tilde{\omega}$ is the limited specific dissipation rate defined by $\tilde{\omega} = \max\left( \omega, C_{\text{lim}} |S| / \sqrt{\beta^*} \right)$ \cite{wilcox2008formulation}. Here, $|S|$ denotes the magnitude of the strain-rate tensor, which reduces to $|S| = |du/dy|$ in the present one-dimensional model.

Here, $\beta^*$, $\sigma^{*}$, $\alpha$, $\beta$, $\sigma$, $\sigma_d$, $C_{\text{lim}}$ are model parameters.
Their values are identical to those in the original formulation of Wilcox \cite{wilcox2008formulation}, are used throughout this work, and are summarized in Table \ref{tab-coeff-kw} in the Appendix.

\subsubsection{Numerical discretization and boundary condition}
\begin{figure}[htb!]
	\centering
	\includegraphics[trim = 11.15cm 0cm 0cm 7.47cm, clip,width=1.0\textwidth]{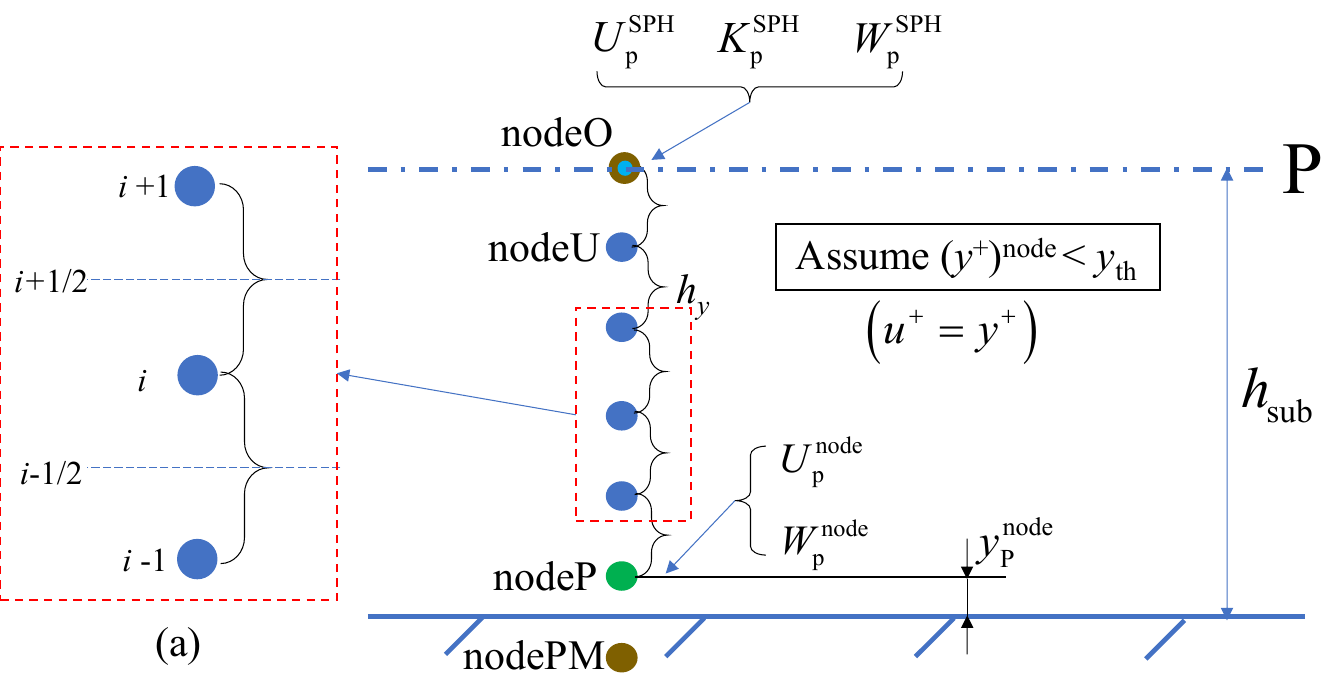}
	\caption{
		Illustration of the near-wall node arrangement, boundary condition treatment, and discretization scheme for the one-dimensional sublayer domain, using five computational nodes as an example.
	}
	\label{fig-concept-5node-BC}
\end{figure}

Figure \ref{fig-concept-5node-BC} (a) demonstrates the node arrangement and discretization.
The first-order derivative in Equation~\eqref{eq-ODE_straight_kw} is discretized using the central difference scheme.
The second-order derivative is discretized in a conservative flux form to enhance numerical stability, as
$\big[D_{i+1/2} (\varphi_{i+1}-\varphi_i)/h_y - D_{i-1/2} (\varphi_i-\varphi_{i-1})/h_y\big]/h_y$,
where the subscript \(_i\) denotes the target node, \(\varphi\) represents an arbitrary variable, whose diffusion coefficient is \(D\), and \(h_y\) is the node spacing.
The interfacial diffusion coefficient, $D_{i-1/2}$ or $D_{i+1/2}$, is computed by the harmonic formulation.

The general discretized form of the three formulations in Equation \eqref{eq-ODE_straight_kw} can be written as,
\begin{equation}
	a_{i-1} \varphi_{i-1}+a_{i} \varphi_{i}+a_{i+1} \varphi_{i+1} = b_i,
	\label{eq-ODE_straight_kw_general}
\end{equation}
in which $a$, $b$ are corresponding coefficients, and their expressions are summarized in Table \ref{tab-unified-discretized-coefficients}.
In this table, $\nu_{eff} =\nu_{m} + \nu_{t} $ is the effective kinematic viscosity.
Additionally, for simplicity, the diffusion coefficients of $k$ and $\omega$ are denoted by $D_k = {\nu_m}+\sigma^* k/\omega$ and $D_\omega = {\nu_m}+\sigma k/\omega$, respectively, and the central difference operator is represented by $\delta_y \varphi_i = (\varphi_{i+1} - \varphi_{i-1}) / (2 h_y) $.

\begin{table}
	\centering
	\small
	\caption{Coefficients of the discretized governing equations
		\eqref{eq-ODE_straight_kw}, written in the form of \eqref{eq-ODE_straight_kw_general}.}
	\renewcommand{\arraystretch}{2.0}
	\setlength{\tabcolsep}{6pt}
	\begin{tabular}{c c c c c}
		\toprule
		\addlinespace[-3pt]
		$\varphi$
		 & $a_{i-1}$
		 & $a_i$
		 & $a_{i+1}$
		 & $b_i/h_y^2$                   \\
		\midrule
		$u$
		 & $-\nu_{\mathrm{eff},(i-1/2)}$
		 & $\nu_{\mathrm{eff},(i-1/2)}
			+\nu_{\mathrm{eff},(i+1/2)}$
		 & $-\nu_{\mathrm{eff},(i+1/2)}$
		 & $\displaystyle
			-\frac{1}{\rho_i}
			\left(\frac{dP}{dx}\right)_i$
		\\
		$k$
		 & $-D_{k,(i-1/2)}$

		 &
		\makecell{$\displaystyle
			D_{k,(i-1/2)}
			+D_{k,(i+1/2)}$
		\\
		$\displaystyle
			+\beta^* \omega_i h_y^2$}

		 & $-D_{k,(i+1/2)}$

		 &
		$\displaystyle
			\nu_{t,i} (\delta_y u_i)^2
		$
		\\[6pt]
		$\omega$
		 & $-D_{\omega,(i-1/2)}$

		 &
		\makecell{$D_{\omega,(i-1/2)}
			+D_{\omega,(i+1/2)}$
		\\
		$+ \beta \omega_i h_y^2$}

		 & $-D_{\omega,(i+1/2)}$
		 &
		\makecell{$\displaystyle
				\alpha \frac{\omega_i}{k_i}\nu_{t,i}
				(\delta_y u_i)^2$
		\\
			$\displaystyle
				+\frac{\sigma_d}{\omega_i}
				(\delta_y k_i)
				(\delta_y \omega_i)$}
		\\
		\bottomrule
	\end{tabular}
	\label{tab-unified-discretized-coefficients}
\end{table}

The boundary condition treatment and near-wall modeling strategy are also illustrated in Fig. \ref{fig-concept-5node-BC}.
At one end of the one-dimensional sublayer domain, the model is coupled with the WCSPH--RANS method through node O.
A Dirichlet boundary condition is imposed at this boundary node, where the values of $u$, $k$, and $\omega$ are provided according to the coupling scheme shown in Fig. \ref{fig-concept-physical-coupling}.

For the boundary condition at the wall-adjacent computational node P, the node is assumed to be located within, or sufficiently close to, the viscous sublayer in order to resolve the near-wall region.
This requirement is expressed as $y^+ < y_\mathrm{th}$, where $y_\mathrm{th} = 5.0$ is a widely adopted empirical threshold in engineering practice \cite{wilcox1998turbulence,wang2025weakly}.
Therefore, the linear law of the wall, $u^+ = y^+$, can be applied, and the boundary values of the velocity and $\omega$ \cite{wilcox1998turbulence} at node P are determined as
\begin{equation}
	\begin{cases}
		U_\mathrm{p}^\mathrm{node} = \dfrac{u_\tau^2 y_\mathrm{p} }{\nu_m}, \\
		W_\mathrm{p}^\mathrm{node}=\dfrac{6\nu_m}{\beta_i y_\mathrm{p}^2}  ,
	\end{cases}
\end{equation}
where $\beta_i=0.075$\cite{wilcox1998turbulence}.
For the turbulent kinetic energy, the wall boundary condition $k=0$ is imposed using a mirror-node treatment. Specifically, the value at the mirror node of node P, denoted as node PM, is assigned as the opposite of that at node P, such that the interpolated value of $k$ at the wall becomes zero.
In practice, however, using a zero-gradient boundary condition leads to only minor differences, since the value of $k$ at node P is already very small due to the strong near-wall dissipation.

It should be noted that, similar to the constant-$y_\mathrm{p}$ strategy \cite{wang2025weakly}, the nodal distance $y_\mathrm{p}^\mathrm{node}$ associated with node P is a model parameter and allowed to be specified manually to ensure the $y^+$ constraint.
The value of $y_\mathrm{p}$ can be initially set to $h_y/2$, followed by a preliminary simulation to examine whether the resulting $y^+$ satisfies the prescribed requirement. If this requirement is not met, $y_\mathrm{p}$ can be adjusted accordingly, following a trial-and-adjustment procedure commonly adopted in engineering practice.

\subsubsection{Additional constraint on local flow rate}
\label{sec-local-flow}
The resulting algebraic system obtained from the discretization of Eq. \eqref{eq-ODE_straight_kw} is under-determined because the pressure gradient remains unknown.
In conventional incompressible flow solvers, the pressure field is determined by coupling the momentum equation with the mass conservation equation, typically through a segregated pressure--velocity coupling algorithm \cite{patankar2018numerical}.

However, in the present one-dimensional formulation, the mass conservation equation does not provide an additional independent constraint, because it reduces to
\begin{equation}
	\frac{d u}{d x} \equiv 0,
\end{equation}
which is automatically satisfied under the one-dimensional wall-normal assumption and therefore cannot be used to determine the unknown pressure gradient.

Since the mass conservation equation cannot provide an independent constraint in the present one-dimensional formulation, an additional constraint on the local flow rate is introduced to close the under-determined system.
Specifically, the current local flow rate within the sublayer is required to approach the target local flow rate, expressed as
\begin{equation}
	\sum_{i=1}^{N_{\mathrm{node}}} u_i A_i
	+ \dfrac{U_{\mathrm{nodeO}} A_\mathrm{nodeO}}{2}
	= Q_{\mathrm{tar}},
	\label{eq-flowrate-constraint}
\end{equation}
where $N_\mathrm{node}$ is the number of computational nodes, and $A_i$ denotes the control area associated with node $i$, extending from $(i-1/2)$ to $(i+1/2)$, as illustrated in Fig. \ref{fig-concept-5node-BC}. For the 2D cases considered in this work, $A_i = h_y$ and $A_\mathrm{nodeO} = h_y$. The local flow rate is evaluated using the midpoint rule; therefore, the contribution from node O is included with a half weight.

\begin{figure}[htb!]
	\centering
	\includegraphics[trim = 9.41cm 0cm 0cm 4.87cm, clip,width=1.0\textwidth]{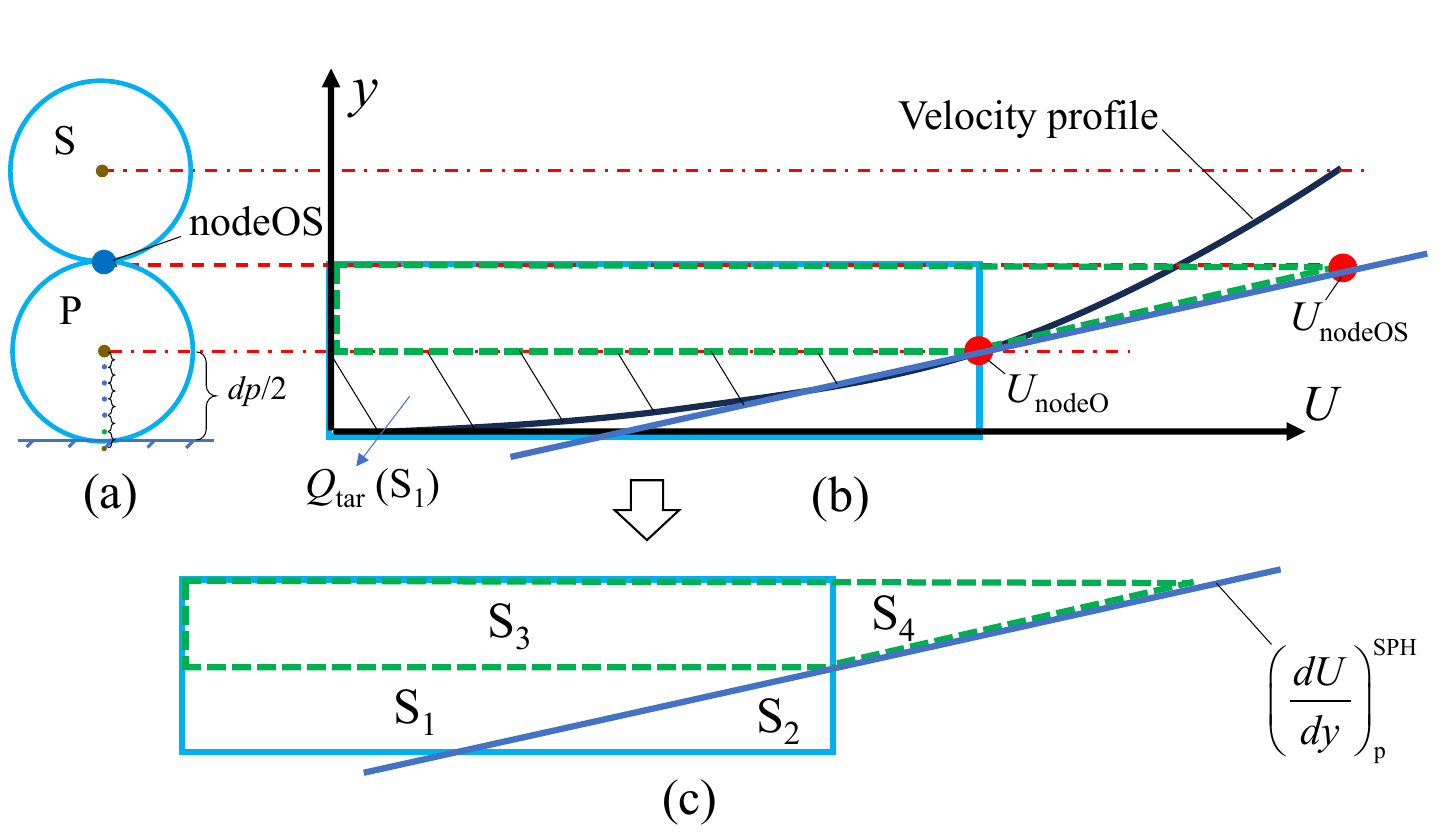}
	\caption{
		Geometrical estimation of the local flow rate for a flat-plate near-wall flow configuration: (a) particle distribution, (b) corresponding flow-rate representation, and (c) definition of the decomposed areas.
	}
	\label{fig-concept-local-flowrate}
\end{figure}

The target local flow rate, $Q_\mathrm{tar}$, is estimated using a geometrical construction.
As illustrated in Fig. \ref{fig-concept-local-flowrate}, the procedure is demonstrated using a cross-sectional near-wall velocity profile of channel flow as an example.
The area in this figure stands for the flow rate, and the tangent line to this curve represents the velocity gradient.

Assuming $S_2 = S_4$, which is valid when the local velocity gradient is sufficiently accurate, the target flow-rate contribution $S_1$, namely $Q_{\mathrm{tar}}$, can be approximated as the difference between $S_1+S_2+S_3$ and $S_3+S_4$. Here, $S_1+S_2+S_3$ corresponds to $Q^\mathrm{SPH}_\mathrm{p}$, while $S_3+S_4$ corresponds to $Q_\mathrm{half}$.
That means, the target local flow rate can be calculated by
\begin{equation}
	Q_{\mathrm{tar}} = Q^\mathrm{SPH}_\mathrm{p}-Q_\mathrm{half},
	\label{eq-flowrate-whole-minus-half}
\end{equation}
where $Q^\mathrm{SPH}_\mathrm{p}$ denotes the total flow-rate contribution associated with the wall-adjacent particle and is computed as
\begin{equation}
	Q^\mathrm{SPH}_\mathrm{p} = U_\mathrm{p} dp,
\end{equation}
where $dp$ is the particle size.
$Q_\mathrm{half}$ is the half flow rate, and is calculated by using the trapezoid formulation, given by
\begin{equation}
	Q_\mathrm{half} = \frac{{(U_\mathrm{nodeO} + U_\mathrm{nodeOS})}dp}{4}.
\end{equation}

Here, $U_\mathrm{nodeOS}$ denotes the velocity at the auxiliary node OS, which is located at the notional interface between the wall-adjacent particle P and the sub-wall-nearest particle S.
Its value is obtained by extrapolation as
\begin{equation}
	U_\mathrm{nodeOS} = U_\mathrm{nodeO} + \frac{dp}{2} \left(\frac{dU}{dy}\right)_\mathrm{p}^\mathrm{SPH},
\end{equation}
in which the velocity gradient magnitude at P is obtained from the WCSPH--RANS method by
\begin{equation}
	\left(\frac{dU}{dy}\right)_p^{\mathrm{SPH}}
	=
	\left\|
	(\boldsymbol{I}-\boldsymbol{n}\otimes\boldsymbol{n})
	(\nabla\boldsymbol{v}_p^{\mathrm{SPH}})
	\boldsymbol{n}
	\right\|,
	\label{eq-flowrate-tan-vel-grad}
\end{equation}
where $\boldsymbol{I}$ is the identity tensor and
$\boldsymbol{n}$ is the unit wall-normal vector.
Since the velocity information on the wall side is not available for the wall-adjacent SPH particle, only the contribution from the inner-fluid side is considered in the velocity-gradient estimation.

Apart from the Dirichlet boundary condition imposed using information from the WCSPH--RANS method, the proposed estimation procedure provides an additional pathway for multi-scale coupling, as both the total flow-rate contribution and the local velocity gradient are evaluated from the SPH-scale solution.
It should be noted that, because the velocity gradient obtained from the SPH method may contain numerical errors, the local flow rate should be regarded as an estimated quantity.
The resulting estimation error is subsequently corrected through the local feedback system introduced in Section \ref{sec-ISS-feedbackSys}.
It should also be noted that the entire estimation procedure is performed locally and independently for each wall-adjacent fluid particle. Although only two-dimensional cases are considered in this work, this local formulation suggests a straightforward extension to three-dimensional configurations by incorporating the corresponding face areas.

\subsubsection{Friction-velocity-based fixed-point iteration scheme}
\begin{figure}[htb!]
	\centering
	\includegraphics[trim = 9.59cm 0cm 0cm 5.65cm, clip,width=1.0\textwidth]{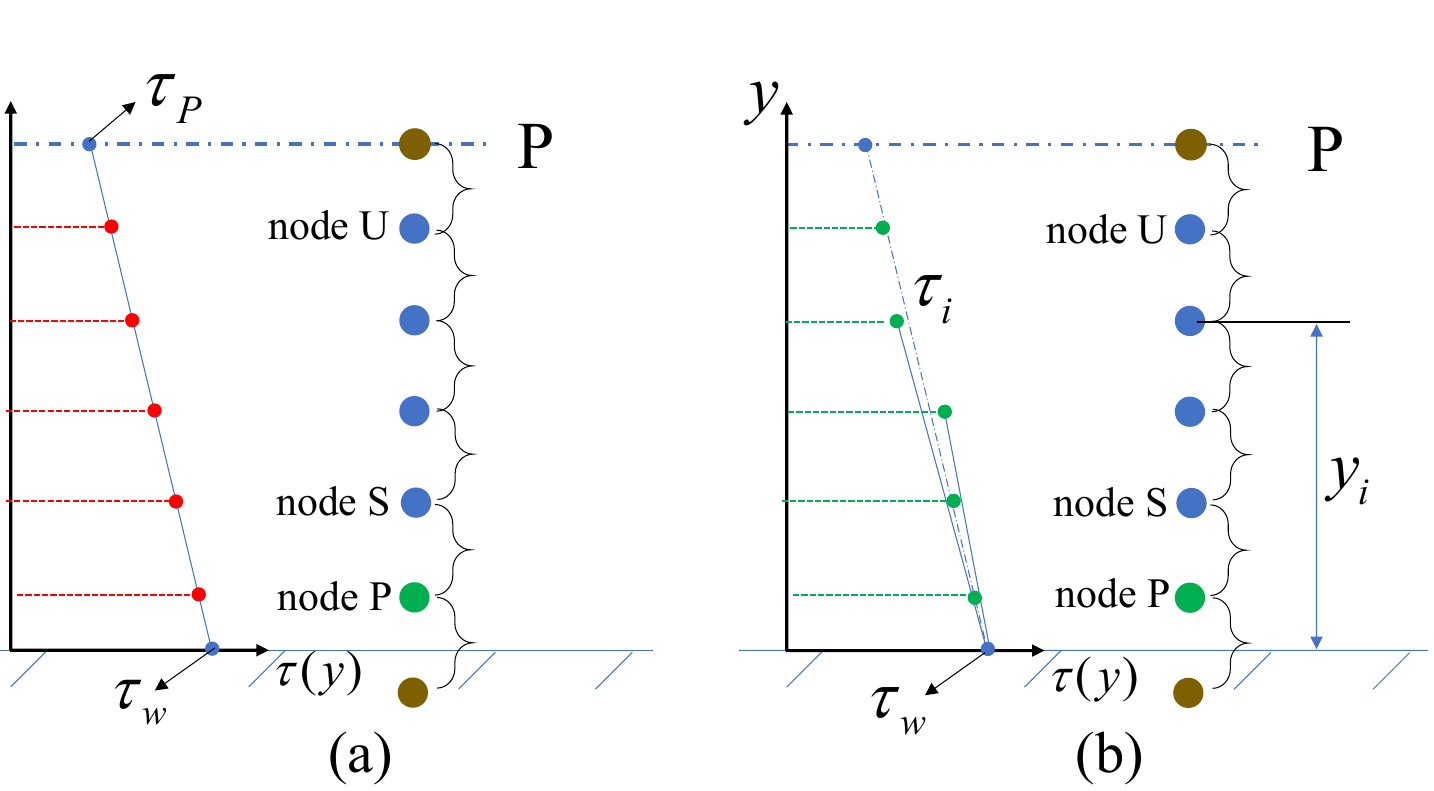}
	\caption{
		Illustration of the local balance between the pressure gradient and shear stress under two treatments: (a) constant local pressure-gradient treatment and (b) adaptive local pressure-gradient treatment.
	}
	\label{fig-concept-pg-difference}
\end{figure}

To satisfy the local flow-rate constraint in Eq. \eqref{eq-flowrate-constraint}, a fixed-point iteration scheme is adopted.
In the present local sublayer model, the sublayer domain is assumed to be sufficiently thin such that the pressure gradient can be regarded as constant within this region.
Under this assumption, a straightforward choice would be to use the constant pressure gradient as the iteration variable.
That means the pressure gradient would be iterated until the current flow rate approaches the target flow rate.
Then the wall shear stress could be calculated according to the balance between the pressure gradient and shear stress gradient in the sublayer, as
\begin{equation}
	\frac{dP}{dx}   = \frac{d \tau}{d y},
\end{equation}
where $\tau$ denotes the shear stress.
As shown in Fig. \ref{fig-concept-pg-difference}(a), integrating this momentum balance across the sublayer domain yields the wall shear stress as
\begin{equation}
	\frac{dP}{dx}   = \frac{\tau_\mathrm{p} - \tau_\mathrm{w}}{h_\mathrm{sub}},
	\label{eq-constant-pg-shear-stress-gradient}
\end{equation}

The friction velocity is then determined from the definition of the wall shear stress, $\tau_\mathrm{w} = \rho u_\tau^2$.
However, this constant pressure-gradient strategy may lead to two main
drawbacks: one related to numerical stability and the other to the evaluation
of the friction velocity, or equivalently the wall shear stress.

First, as shown in Fig.~\ref{fig-concept-pg-difference}(a),
the sublayer system must satisfy both the SPH-imposed Dirichlet
boundary conditions at node O and the estimated local flow-rate
constraint under a spatially constant pressure gradient.
Both inputs are obtained from the instantaneous SPH solution
and may therefore fluctuate even when the time-averaged flow
is smooth.
These fluctuations can disturb the local sublayer iteration
and impair its convergence.

Second, when the pressure gradient is used as the iteration variable, the friction velocity will be back-calculated from Eq. \eqref{eq-constant-pg-shear-stress-gradient}. This procedure introduces discretization errors into the evaluation of the friction velocity, or equivalently the wall shear stress.
Since the friction velocity, through its direct relation to the wall shear
stress, is a key quantity exchanged between the two scales, any discretization
error in its evaluation may contaminate the wall-shear information transferred
to the SPH-scale solution. Although this error is expected to be small due to the small sublayer thickness $h_\mathrm{sub}$, it is preferable to avoid introducing it into the inter-scale coupling.

To avoid the two aforementioned drawbacks, a friction-velocity-based fixed-point iteration scheme is developed to adaptively determine the pressure gradient.
This scheme generally consists of two steps.
First, to improve numerical robustness against fluctuations
in the SPH-derived inputs, the requirement of a spatially
constant pressure gradient is relaxed, allowing the pressure
gradient to vary among computational nodes, as illustrated
in Fig.~\ref{fig-concept-pg-difference}(b).
Instead, at each computational node, a local pressure gradient
is evaluated by integrating the shear-stress balance between
that node and the wall:
\begin{equation}
	\left(\frac{dP}{dx}\right)_i   = \frac{\tau_i - \tau_\mathrm{w}}{y_i},
\end{equation}
where, $y_i$ is the distance from node $i$ to the wall.
The shear stress at node $i$ is also approximated using a central difference scheme. The resulting expression for the local pressure gradient at node $i$ is then given by
\begin{equation}
	\left(\dfrac{dP}{dx}\right)_i = \rho_i \left(  \dfrac{\nu_{\mathrm{eff},i} \delta_y u_i - u_{\tau}^2}{y_i} \right).
	\label{eq-pg-discretized}
\end{equation}

Second, since the friction velocity is a key quantity exchanged between the two scales, it is directly selected as the iteration variable to avoid additional discretization errors in its evaluation.
The local pressure gradient at each node is then computed from Eq. \eqref{eq-pg-discretized}.
This strategy is consistent with the sublayer model: the small sublayer thickness provides a basis for approximating
the local shear-stress gradient by its wall-to-node average,
with the approximation expected to improve as the sublayer
thickness decreases under SPH particle refinement.
Additionally, this strategy is physically more consistent for wall-bounded
flows, because the friction velocity, through its direct relation to the wall
shear stress, represents the key near-wall quantity transferred from the
sublayer model to the SPH-resolved flow field.

\begin{figure}[htb!]
	\centering
	\includegraphics[trim = 6.28cm 0cm 0cm 10.18cm, clip,width=0.9\textwidth]{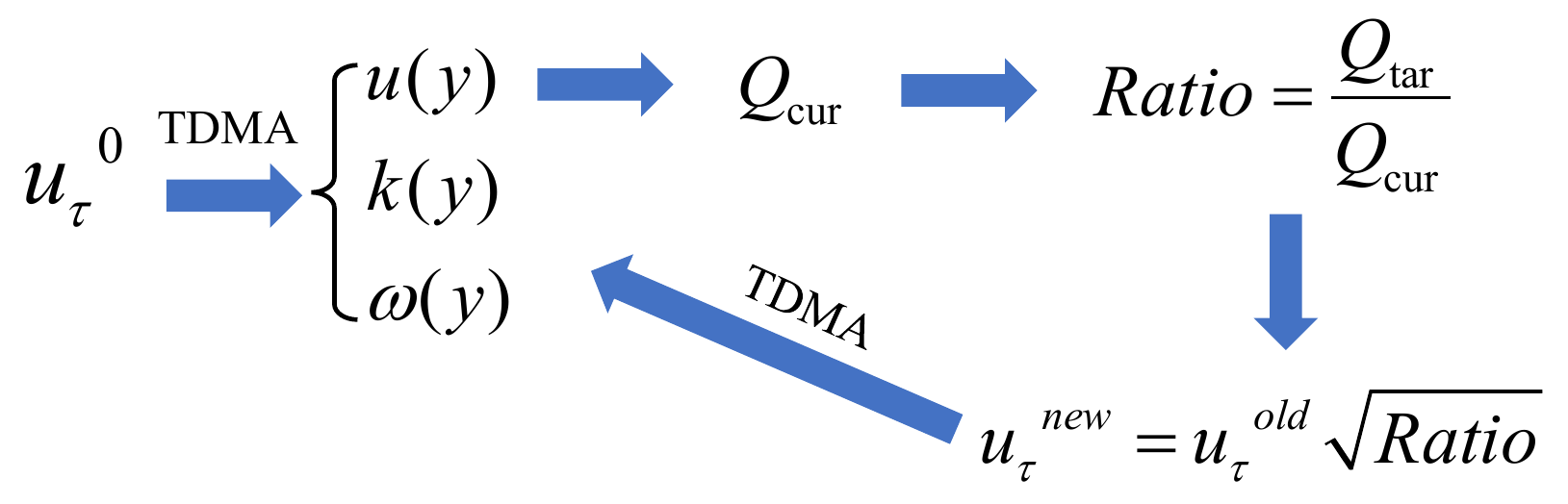}
	\caption{
		Schematic of the fixed-point iterative procedure for determining the friction velocity from the prescribed flow-rate constraint.
	}
	\label{fig-concept-iter-flowchart}
\end{figure}

In summary, by incorporating the friction-velocity-based adaptive pressure gradient approximation and collecting like terms, the coefficients of the momentum equation originally listed in Table \ref{tab-unified-discretized-coefficients} are modified, as summarized in Table \ref{tab-coefficients-u}.
Consequently, once the friction velocity is determined, the one-dimensional system is fully closed and can be solved.
The proportional feedback scheme is illustrated in Fig. \ref{fig-concept-iter-flowchart}.
Here, $Q_\mathrm{cur}$ denotes the current local flow rate evaluated from the left-hand side of Eq. \eqref{eq-flowrate-constraint}, while $u_\tau^0$ denotes the initial guess of the friction velocity.
Additionally, the resulting tri-diagonal linear system is solved using the tri-diagonal matrix algorithm (TDMA), and under-relaxation is applied to further improve numerical stability and convergence.

\begin{table}
	\centering
	\small
	\caption{Unified coefficient form of the discretized momentum governing equation with the friction-velocity-based adaptive pressure gradient approximation method.}
	\renewcommand{\arraystretch}{2.0}
	\setlength{\tabcolsep}{4pt}
	\begin{tabular}{c c c c c}
		\toprule
		\addlinespace[-10pt]
		$\varphi$
		 & $a_{i-1}$
		 & $a_i$
		 & $a_{i+1}$
		 & $b_i$                                                                  \\
		\midrule
		$u$
		 & $-\nu_{\mathrm{eff},(i-1/2)}-\dfrac{h_y \nu_{\mathrm{eff},i} }{2 y_i}$
		 & $\nu_{\mathrm{eff},(i-1/2)}
			+\nu_{\mathrm{eff},(i+1/2)}$
		 & $-\nu_{\mathrm{eff},(i+1/2)}+\dfrac{h_y \nu_{\mathrm{eff},i} }{2 y_i}$
		 & $\displaystyle
			\frac{h_y^2}{y_i} u_{\tau}^2$
		\\[6pt]
		\bottomrule
	\end{tabular}
	\label{tab-coefficients-u}
\end{table}

%

\subsection{SPH-scale WCSPH--RANS formulation}
The present approach is developed within the $k$--$\omega$-based
WCSPH--RANS framework established, verified, and validated
in our previous works
\cite{wang6271943unified,wang2025weakly,wang2026effective,wang2025zero}.
Several established numerical techniques are retained, adapted, or
reformulated in the present formulation because of their essential roles in
numerical robustness, convergence, and computational efficiency.

Within this framework, the wall shear stress (WSS) and inner shear stress
(ISS) correction schemes are developed, together with a feedback mechanism
for maintaining the local balance between the WCSPH--RANS and sublayer
systems.

\subsubsection{Essential supporting numerical techniques}

The adaptive Riemann--eddy dissipation (ARD) scheme was originally proposed
to address the inconsistency between the Lagrangian nature of SPH and RANS
turbulence modeling~\cite{wang2025weakly}.
The present multi-scale approach substantially enhances the near-wall
resolving capability through coupling with the one-dimensional sublayer
solver, without increasing the SPH particle resolution. Nevertheless, this enhanced
resolving capability also makes the coupled solution more sensitive to
numerical disturbances, particularly in the wall-adjacent region. A non-limited one-sided Riemann problem is therefore
introduced for fluid--wall interactions, leading to the enhanced ARD scheme,
which plays a crucial role in maintaining near-wall numerical stability.

The transport-velocity formulation (TVF)~\cite{adami2013transport} is
employed to maintain a regular particle distribution. Within the
WCSPH--RANS framework, however, it may introduce weak spurious velocity
disturbance that hinders the convergence of the turbulent kinetic energy
profiles. The de-noised TVF~\cite{wang2025weakly} is therefore adopted to
suppress the fluctuation and ensure stable convergence.
Additionally, the dual-criteria time-stepping scheme
\cite{zhang2020dual} is employed to improve computational efficiency.
To accommodate the two-equation RANS formulation, the original scheme is
further extended by introducing a term-splitting strategy and accounting
for the eddy viscosity in the time-step criterion
\cite{wang6271943unified}.

\subsubsection{Enhanced boundary offset technique}
Following the concept of the previously proposed boundary-offset
technique (BOT)~\cite{wang2025weakly}, the wall-normal distance employed
in the near-wall RANS treatment is regarded as a prescribed numerical
parameter rather than being strictly determined by the SPH particle
spacing. In the original BOT, the effective boundary and the
corresponding wall dummy particles are adaptively offset to maintain
the prescribed wall distance and avoid the formation of a numerical
gap. In the present multi-scale approach, this concept is transferred
to the one-dimensional sublayer solver. Specifically, the
wall-adjacent node (node P) is assigned a prescribed wall-normal
position $y_P^{\mathrm{node}}$, and only its mirror node PM is offset
accordingly, while the SPH particles and the dummy boundary remain
unchanged.

\begin{figure}[htb!]
	\centering
	\includegraphics[trim = 11.81cm 0cm 0cm 6.05cm, clip,width=1.0\textwidth]{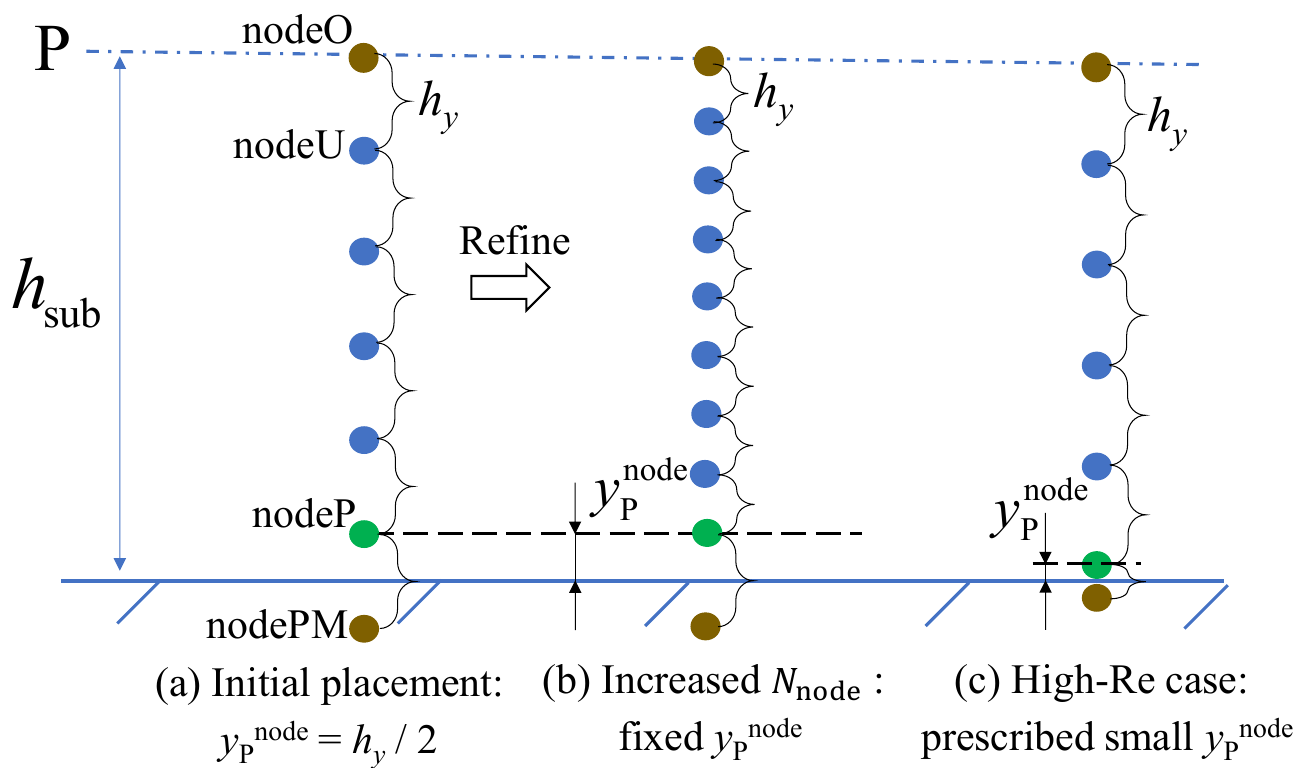}
	\caption{Illustration of the boundary-offset technique (BOT$^{+}$)
	incorporated into the one-dimensional sublayer solver. The mirror node
	PM is offset according to the prescribed $y_P^{\mathrm{node}}$,
	independently of the nominal nodal spacing $h_y$.}
	\label{fig-bot-plus}
\end{figure}

As illustrated in Fig.~\ref{fig-bot-plus}(a), an initial computation is
performed using the conventional midpoint placement
$y_P^{\mathrm{node}}=h_y/2$, where $h_y$ denotes the nominal nodal
spacing of the sublayer discretization. This midpoint placement is
retained as the default choice whenever it already provides sufficient
near-wall resolution, since it preserves the regular nodal arrangement
and avoids the additional numerical errors potentially associated with
the boundary offset. The resulting value of $y^{+}$ at node P is
then examined, and $y_P^{\mathrm{node}}$ is adjusted when the
default placement does not provide an appropriate near-wall resolution.
When a prescribed offset is required, the selected
$y_P^{\mathrm{node}}$ is kept fixed throughout the subsequent
simulation and when varying the number of sublayer nodes.
Otherwise, the default midpoint placement
$y_P^{\mathrm{node}}=h_y/2$ is retained.
For the prescribed-offset treatment, the location of node P
therefore remains unchanged when the sublayer resolution is
increased, as shown in Fig.~\ref{fig-bot-plus}(b).
It is worth noting that, when the sublayer computational domain
lies within the viscous sublayer, further refinement of its
nodal discretization is generally unnecessary.
A five-node system is sufficient for most of the cases
considered herein, as demonstrated by the numerical tests below.

The present technique is particularly
useful at extremely high Reynolds numbers, for which the viscous
sublayer becomes exceedingly thin. As illustrated in
Fig.~\ref{fig-bot-plus}(c), a sufficiently small
$y_P^{\mathrm{node}}$ can be prescribed and retained, allowing node
P to resolve the viscous sublayer without increasing the SPH
particle resolution. This treatment is employed in the subsequent
straight-channel case at $Re=8.0\times10^{7}$.

The prescribed offset may introduce two potential sources of numerical
error. First, when $y_P^{\mathrm{node}}\neq h_y/2$, the resulting
nonuniform spacing may introduce a
small error into the discrete evaluation of the local flow rate.
However, because the velocity is small in the immediate vicinity of
the wall, the corresponding contribution to the total flow rate is
expected to be minor. Second, the offset of node PM may affect the
imposition of the wall boundary conditions. To minimize this influence,
the boundary conditions for velocity and specific dissipation rate
are imposed directly at node P through equation replacement, as demonstrated in Fig. \ref{fig-concept-5node-BC}, such
that only the boundary treatment of turbulent kinetic energy involves
node PM. Since the turbulent kinetic energy approaches zero toward
the wall and remains small in this region, the associated error is
also expected to be limited.

Overall, by confining the boundary offset to the mirror node PM, the
present node-based implementation avoids any modification of the SPH
particles or the dummy-boundary configuration and therefore introduces
substantially less perturbation into the SPH particle system than the
original BOT. This enhanced formulation is hereafter referred to as
BOT$^{+}$.

\subsubsection{Correction on shear stress}
Similar to the wall function method, the wall shear stress of the wall-adjacent fluid particle is corrected by that from the 1D sublayer solver as
\begin{equation}
	\boldsymbol{\tau}_{w} = \rho \left(u_\tau^{\mathrm{sub}} \right)^2 (\mathbf{t} \otimes \mathbf{n} ) ,
	\label{eq-wss-from-sublayer}
\end{equation}
in which $u_\tau^{\mathrm{sub}}$ is the friction velocity from the 1D sublayer solver, and $\boldsymbol{t}$ is the local unit tangent defined in
Section~2.1.2.
For the wall adjacent fluid particles, namely the particles in layer P, the acceleration in the momentum equation induced from viscosity can be calculated by two parts, one is from wall, the other is from the inner neighboring fluid particles, as
\begin{equation}
	\left(\frac{\text{d} \mathbf v}{\text{d} t}\right)^\nu _\mathrm{p}
	=
	\left(\frac{\text{d} \mathbf v}{\text{d} t}\right)^\nu _\mathrm{p, inner} +
	\left(\frac{\text{d} \mathbf v}{\text{d} t}\right)^\nu _\mathrm{p, wall}
\end{equation}

The formulation of the viscous acceleration contributed by wall is
\begin{equation}
	\left(\frac{\text{d} \mathbf v}{\text{d} t}\right)^\nu _\mathrm{p, wall}=
	- \frac{2}{\rho_i} \sum_{j}^{N_\mathrm{w}} \boldsymbol{\tau}_{w} \cdot  {\nabla} W_{ij} V_j ,
	\label{eq-wss-sph}
\end{equation}
where $V$ refers to the particle volume, $N_\mathrm{w}$ is the number of the wall dummy particle near the particle $i$.

To achieve a continuous approximation of the viscous force at the wall-adjacent fluid particle, the formulation of the viscous acceleration contributed by the inner fluid particles is corrected as
\begin{equation}
	\left(\frac{\text{d} \mathbf v}{\text{d} t}\right)^\nu _\mathrm{p, inner}=
	- \frac{2}{\rho_i} \sum_{j}^{N_\mathrm{in}} \boldsymbol{\tau}_\mathrm{msc} \cdot  {\nabla} W_{ij} V_j ,
	\label{eq-iss-p-sph}
\end{equation}
in which $N_\mathrm{in}$ denotes the number of inner-fluid particles located
within the kernel support of the target particle $i$.
The shear stress obtained from the multi-scale coupling (MSC) scheme between the sublayer solver and SPH solver is calculated by
\begin{equation}
	\boldsymbol{\tau}_\mathrm{msc} = \mu_\mathrm{eff}\left( \nabla \mathbf{v}_\mathrm{msc} + \left( \nabla \mathbf{v}_\mathrm{msc} \right)^\mathrm{T} \right).
\end{equation}

It is worth noting that the velocity gradient here is quite different from that in Equation \eqref{eq-flowrate-tan-vel-grad}, because the velocity gradient here is from the multi-scale coupling scheme, and is calculated by
\begin{equation}
	\nabla \mathbf{v}_\mathrm{msc} = \left( \dfrac{du}{dy} \right)_\mathrm{msc} (\mathbf{t} \otimes \mathbf{n} ),
\end{equation}
in which
\begin{equation}
	\left( \dfrac{du}{dy} \right)_\mathrm{msc} = \dfrac{\lvert U_\mathrm{p} - U_\mathrm{nodeU} \rvert }{{h_y}}.
	\label{eq-vel-grad-msc}
\end{equation}

Additionally, to ensure conservation of viscous force between particle-pairs, for the fluid particles that interact with the wall-adjacent fluid particles, the formulation is accordingly corrected as Equation \eqref{eq-iss-p-sph}.

\subsubsection{The local feedback system}
\label{sec-ISS-feedbackSys}
\begin{figure}[htb!]
	\centering
	\includegraphics[trim = 8.77cm 0cm 0cm 8.02cm, clip,width=1.0\textwidth]{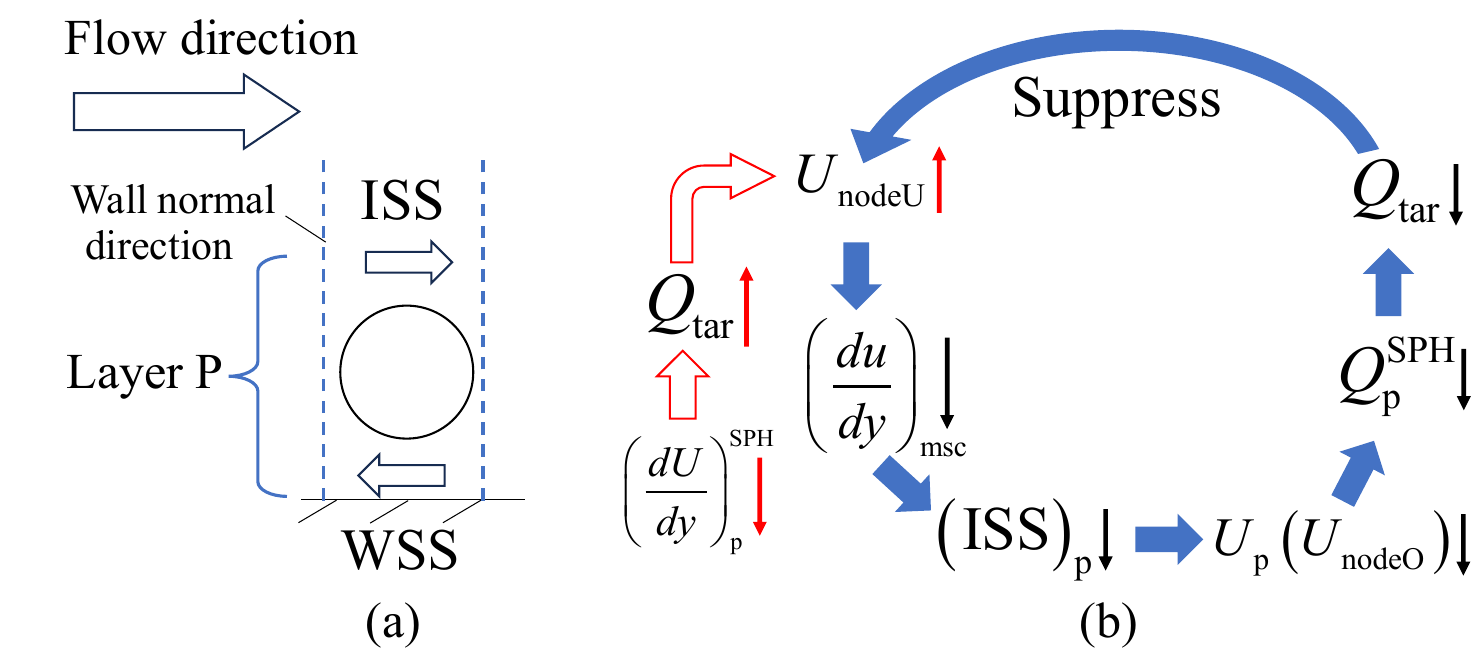}
	\caption{
		Schematic illustration of the error-absorption feedback mechanism: (a) local balance between the inner shear stress and wall shear stress, and (b) error-absorption procedure when an error occurs in the velocity gradient.
	}
	\label{fig-concept-feedback-vel-grad}
\end{figure}

The discretization of the velocity gradient in Eq. \eqref{eq-vel-grad-msc}
is deliberately designed to enable the feedback mechanism and improve the
coupling consistency between the WCSPH--RANS and sublayer systems.
As shown in Fig. \ref{fig-concept-feedback-vel-grad}(a), the feedback
mechanism is constructed based on the local balance between the wall shear
stress (WSS) and the inner shear stress (ISS) associated with the
wall-adjacent fluid particle.
Any departure from the local momentum balance drives an
adjustment of the wall-adjacent particle velocity, which
feeds back into the sublayer system and promotes a
self-correcting response through the coupled ISS and WSS.

Figure~\ref{fig-concept-feedback-vel-grad}(b) illustrates how
the proposed feedback mechanism suppresses errors in the
estimated velocity gradient.
At low SPH resolutions, the velocity gradient tends to be
underestimated~\cite{wang6271943unified}.
Through the local flow-rate estimation, this leads to an
overestimated target flow rate and consequently an
overestimated velocity at node U.
For the illustrated local profile, where
$U_\mathrm{p}>U_\mathrm{nodeU}$, an increase in
$U_\mathrm{nodeU}$ at a temporarily fixed $U_\mathrm{p}$
reduces $(du/dy)_\mathrm{msc}$ and hence the ISS.
With the WSS temporarily regarded as unchanged, the reduced
ISS decreases the momentum supplied by the inner fluid,
causing the velocity at particle P, or equivalently node O,
to decrease.
This reduces the particle-associated flow rate
$Q_\mathrm{p}^{\mathrm{SPH}}$ and, through
Eq.~\eqref{eq-flowrate-whole-minus-half}, lowers the target
local flow rate.
The resulting feedback counteracts the initial overestimation
of $U_\mathrm{nodeU}$ and thereby suppresses the error.

For the same local velocity ordering, the feedback acts in
the opposite direction when the estimated velocity gradient
is overestimated rather than underestimated.
This velocity ordering is typical of near-wall flow in
stationary-wall pipes and channels and in attached boundary layers.
The feedback behavior for profiles with
$U_\mathrm{p}\leq U_\mathrm{nodeU}$ has not been assessed
here and will be investigated in future work.

The local construction allows this feedback mechanism to operate
in complex geometries under the same local velocity ordering.
Although errors may be introduced in the estimation of the local flow rate,
they can be partly absorbed through the feedback correction, which enforces
the local balance between the WCSPH--RANS and sublayer systems.
However, this error-absorption capability is not unlimited.
As demonstrated by the validation cases presented below, when the SPH
particle resolution is too coarse, the resulting discretization error may
be too large to be fully compensated by the local feedback mechanism.
Therefore, the reliability of the coupled solution relies primarily on the
local shear-stress balance, the numerical consistency of the near-wall
coupling, and a sufficiently resolved SPH-scale solution.

\section{Numerical examples}
\label{section-numerical-examples}
This section evaluates the proposed multi-scale near-wall approach for
WCSPH--RANS simulations through a set of representative benchmark tests.
The numerical examples include turbulent straight-channel flows at three
typical and challenging Reynolds numbers, followed by a turbulent wavy-channel
flow involving flow separation and complex near-wall geometry.

It should be noted that the present convergence assessment should be
interpreted as a practical resolution-convergence study rather than a strict
mathematical proof of SPH convergence.
As discussed by Quinlan et al.~\cite{quinlan2006truncation}, strict
convergence of SPH operators generally requires the smoothing length and
particle spacing to be refined in a consistent manner, typically with an
increasing number of neighboring particles.
The results reported here are obtained using the Wendland C2 kernel with
$h=1.3dp$, the same CFL numbers, and the same remaining numerical parameters
as those used in Ref.~\cite{wang2025weakly}.
All line plots are extracted after the flow reaches a statistically steady
state and are based on time-averaged results.
The reported computational costs are averaged wall-clock times obtained from
repeated runs under the same hardware and software environment.

\subsection{Turbulent flow in a straight channel}
The numerical setup, including the boundary conditions, follows Refs. \cite{wang2025weakly,wang6271943unified,wang2026effective,wang2025zero}.
The Reynolds number is defined based on the channel height and the inflow bulk velocity.
The resolution of each test is characterized by the number of fluid particles (nodes or cells) across the channel height, denoted by $N_f$, and the observation position is set at the fully-developed segment.

\subsubsection{Case 1: baseline case at Re = 5714}
\label{case-straight-Re5714}
Although the Reynolds number is relatively low, the flow is already fully
turbulent~\cite{lee2015direct}. In this case, the near-wall length scale
remains sufficiently large in physical space, such that a uniformly
discretized WCSPH--RANS simulation with $y^+$ approaching unity remains
computationally affordable.
Therefore, the original WCSPH--RANS $k$--$\omega$ method without the
multi-scale approach can still provide a uniformly resolved reference
solution for this case, as previously verified and validated in
Ref.~\cite{wang6271943unified}.
This makes the present case suitable for assessing the proposed multi-scale
approach against the conventional uniformly resolved WCSPH--RANS result.

\begin{figure}[htb!]
	\centering
	\includegraphics[trim = 10.61cm 0cm 0cm 1.95cm, clip,width=1.0\textwidth]{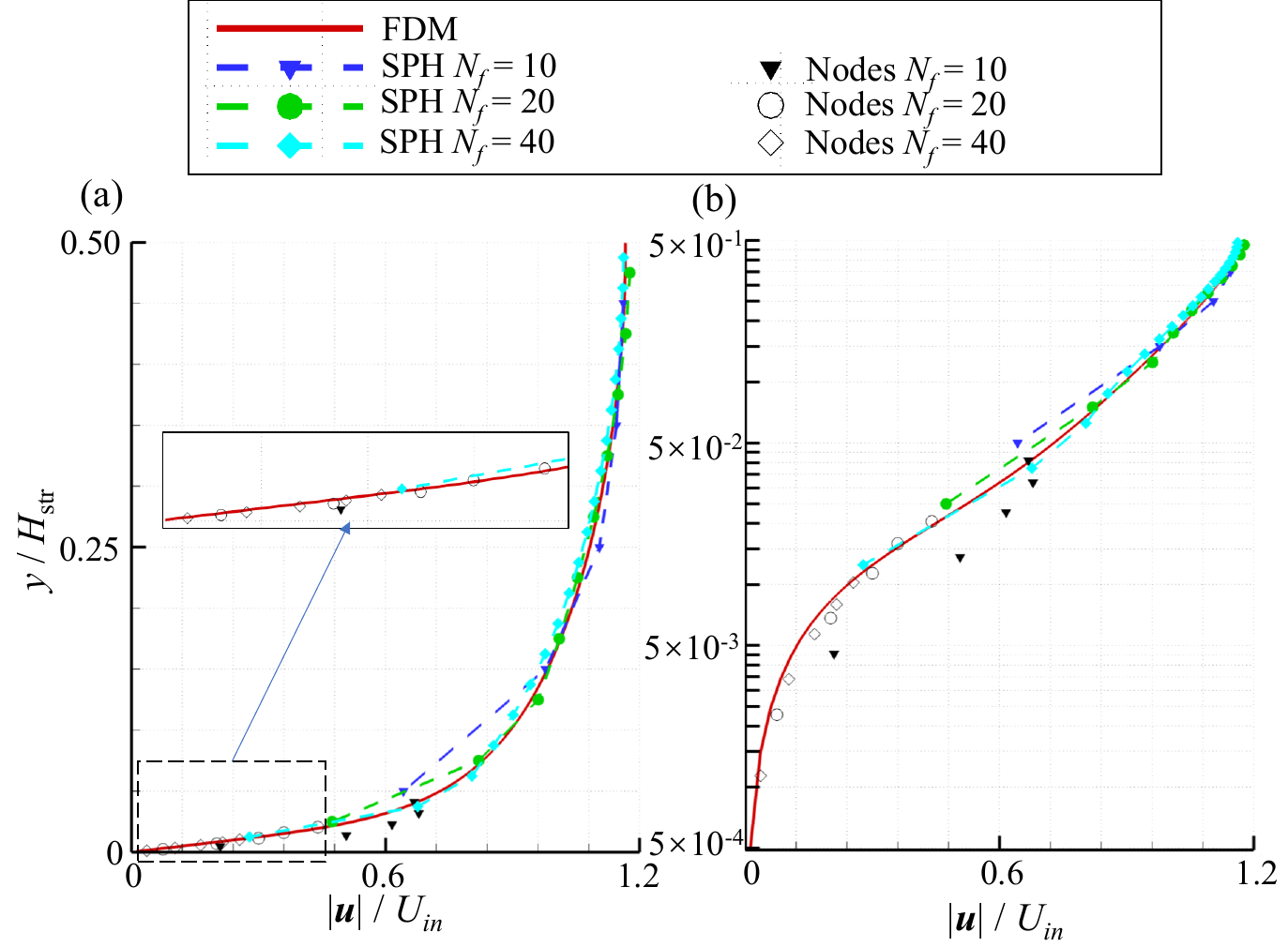}
	\caption{
		Case 1: comparison of cross-sectional velocity profiles obtained using the
		multi-scale approach at three SPH particle resolutions, denoted by $N_f$,
		with the finite-difference reference solution: (a) linear axis and
		(b) logarithmic axis.
		Line-symbol curves denote the SPH particle values, while the isolated symbols
		denote the sublayer-node values computed by the sublayer solver.
	}
	\label{fig-straight-re5714-vel-converge-vg-inner}
\end{figure}

First, to test the convergence of the proposed approach, the velocity profiles calculated by the three resolutions are shown in Fig.~\ref{fig-straight-re5714-vel-converge-vg-inner}.
The converged reference profile is taken from Ref.~\cite{wang6271943unified}, where the simplified one-dimensional model was solved over the half-channel height by the finite difference method (FDM), and validated with the DNS result~\cite{lee2015direct}.
A five-node sublayer system is employed in the present test, and satisfactory convergence is observed at $N_f=40$. At this resolution, the present approach shows good agreement with the converged one-dimensional reference solution, particularly in the near-wall region.

Additionally, at the coarsest resolution, $N_f=10$, a localized kink is observed at the
multi-scale coupling position.
This is mainly caused by the insufficient SPH-scale resolution, because the
local flow-rate constraint depends on the velocity-gradient information
estimated from the SPH particle field.
Although the proposed feedback mechanism can partially absorb such estimation
errors, its correction capability is not unlimited; when the SPH resolution is
extremely coarse, the SPH-scale discretization error may still be too large
to be fully compensated.
This is consistent with the general observation that even laminar channel-flow
simulations usually require approximately 20 particles~\cite{zhang2025dynamical} across the channel
height to obtain a reasonably resolved velocity profile.
With increasing resolution, the kink rapidly disappears and a smooth
connection between the SPH particle values and the sublayer-node values is
recovered, demonstrating the resolution consistency of the proposed
multi-scale coupling strategy.

\begin{figure}[htb!]
	\centering
	\includegraphics[trim = 10.66cm 0cm 0cm 4.02cm, clip,width=1.0\textwidth]{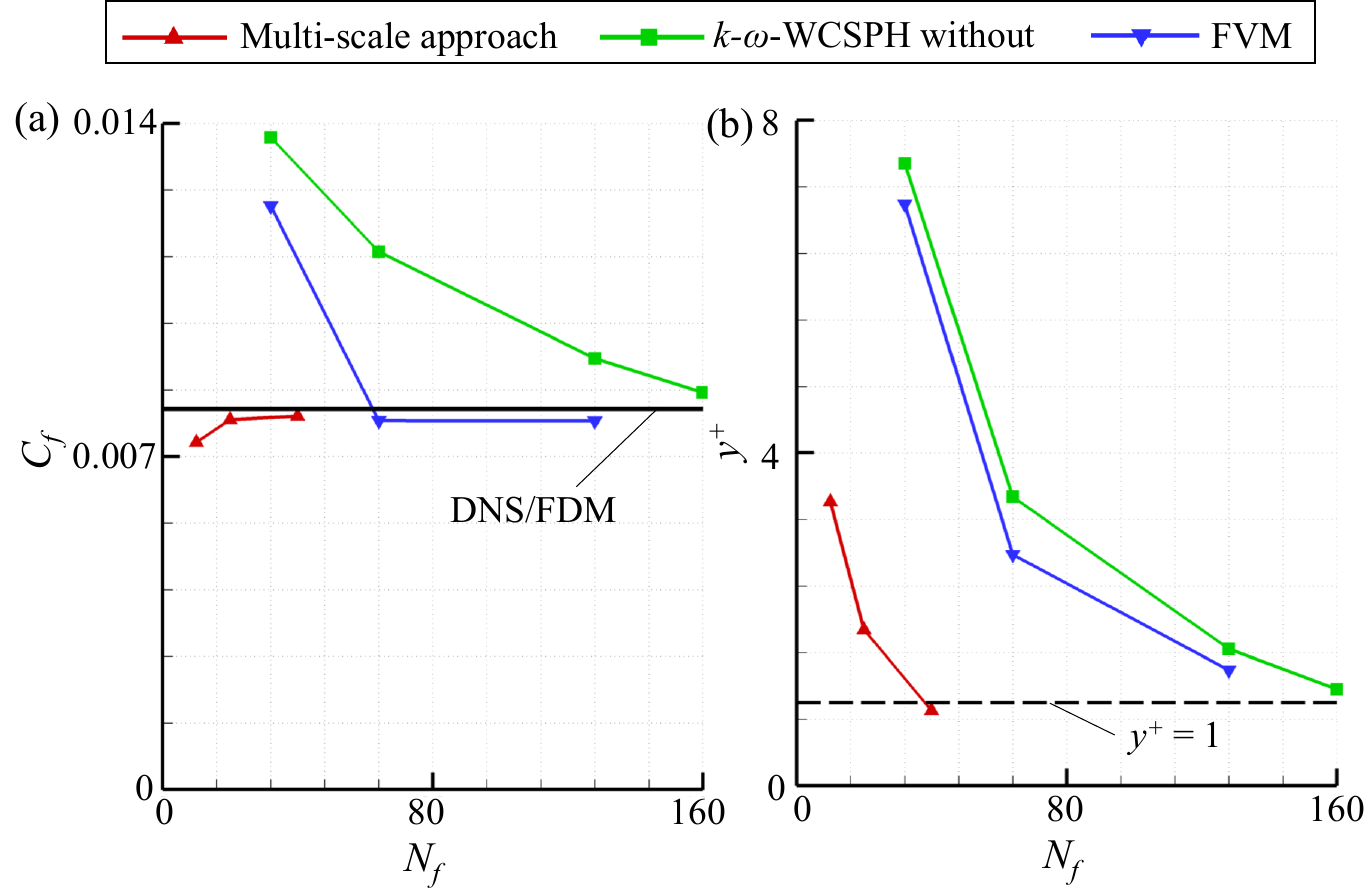}
	\caption{
		Case 1: comparison of the (a) friction coefficient, and (b) $y^+$ with the different methods, in which the FVM results are from Ref.~\cite{wang6271943unified}.
	}
	\label{fig-straight-re5714-vel-comp-cf}
\end{figure}

Second, the friction coefficient~\cite{wang6271943unified} and $y^+$ are compared at different resolutions in Fig.~\ref{fig-straight-re5714-vel-comp-cf}, in which $C_f = \tau_w / (0.5\rho U_{\mathrm{ref}}^2) $, $\tau_w = \rho (u_\tau)^2 $, and reference velocity $U_\mathrm{ref} = 1$.
For the multi-scale approach, the reported $y^+$ value is evaluated at node P.
With the proposed multi-scale approach, the predicted friction coefficient
approaches the DNS/FDM reference value already at $N_f=20$ and can be regarded
as converged at $N_f=40$.
At the same time, the corresponding $y^+$ value is also reduced to the
order of unity, indicating that the near-wall condition is properly
represented.
In contrast, the $k$--$\omega$--WCSPH method without the proposed near-wall
treatment still exhibits noticeable resolution dependence even up to
$N_f=160$.
Therefore, for this case, the proposed approach achieves a converged
friction-coefficient prediction using at least four times lower wall-normal
particle resolution than the conventional uniformly resolved WCSPH--RANS
simulation.

\begin{figure}[htb!]
	\centering
	\includegraphics[trim = 9.07cm 0cm 0cm 2.81cm, clip,width=1.0\textwidth]{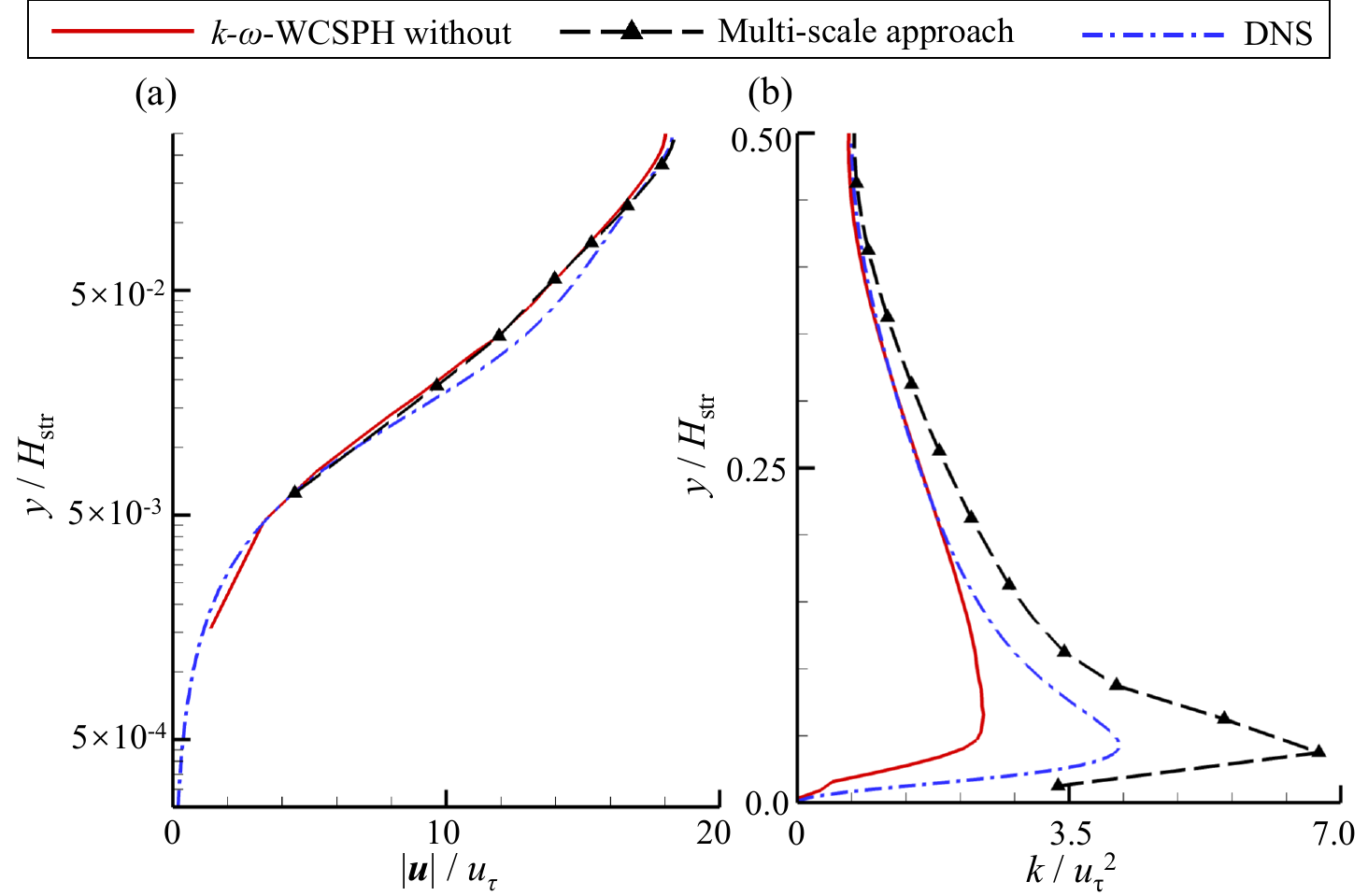}
	\caption{
		Case 1: comparison of the converged cross-sectional velocity and turbulent kinetic energy profiles obtained by the different methods, and the DNS results are obtained from Ref. \cite{lee2015direct}.
	}
	\label{fig-straight-re5714-vel-k-comp-converged}
\end{figure}

Third, the cross-sectional velocity and turbulent kinetic energy profiles are
compared with the DNS data of Lee and Moser~\cite{lee2015direct}, as shown in
Fig.~\ref{fig-straight-re5714-vel-k-comp-converged}.
Here, the proposed multi-scale approach is evaluated at $N_f=40$, where the
friction coefficient has already reached a practically converged state,
whereas the conventional $k$--$\omega$--WCSPH result without the multi-scale
treatment is shown at $N_f=160$.
This comparison therefore highlights that the proposed approach can achieve a
comparable velocity prediction using a much lower wall-normal particle
resolution.
Moreover, the turbulent kinetic energy predicted by the proposed approach shows improved agreement with the DNS trend, although $k$ is still over-predicted, possibly due to the relatively low resolution, while the baseline simulation underestimates it.

Lastly, the computational cost is compared in Table~\ref{tab-straight-cpu-time}.
the relative computational cost remains
close to unity for all tested resolutions. Therefore, the additional local
sublayer solution does not introduce a noticeable computational overhead for this case.
It is also expected that the additional computational cost introduced by the sublayer model might become less significant as the resolution increases.
This is because the sublayer model is solved only for the wall-adjacent particles, whose proportion decreases relative to the total number of particles at higher resolutions.
As a result, the overall computational cost remains dominated by the global particle interaction, neighbor-search, configuration updating procedures.

\begin{table}[htbp]
	\centering
	\small
	\setlength{\tabcolsep}{10pt}
	\caption{
		Wall-clock time (s) for simulating the fully developed turbulent flow in
		the straight channel at three spatial resolutions with and without the sublayer model.
		The wall-clock time is measured for parallel computations up to a non-dimensional time of 100.
		All simulations are performed on a desktop computer equipped with an
		AMD Ryzen 7 5700G processor (8 cores, 16 threads), 48~GiB RAM.
	}
	\label{tab-straight-cpu-time}
	\begin{tabular}{lccc}
		\toprule
		Resolution $N_f$                           & 20    & 40     & 80      \\
		\midrule
		Without sublayer model                     & 46.67 & 537.10 & 4520.01 \\
		With sublayer model                        & 45.26 & 518.17 & 4566.20 \\
		Relative computational cost (With/Without) & 0.97  & 0.96   & 1.01    \\
		\bottomrule
	\end{tabular}
\end{table}

\subsubsection{Case 2: practical case at Re = 40,000}
\label{case-straight-Re40000}
The second case considers a more practically relevant Reynolds number. It is more challenging than the baseline case because, at the adopted particle or mesh resolution ($N_f = 40$), the wall-adjacent particle or cell falls within the buffer layer ($y^+$ is around 25 theoretically), where conventional near-wall modeling is generally subject to increased uncertainty.

\begin{figure}[htb!]
	\centering
	\includegraphics[trim = 8.74cm 0cm 0cm 6.61cm, clip,width=1.0\textwidth]{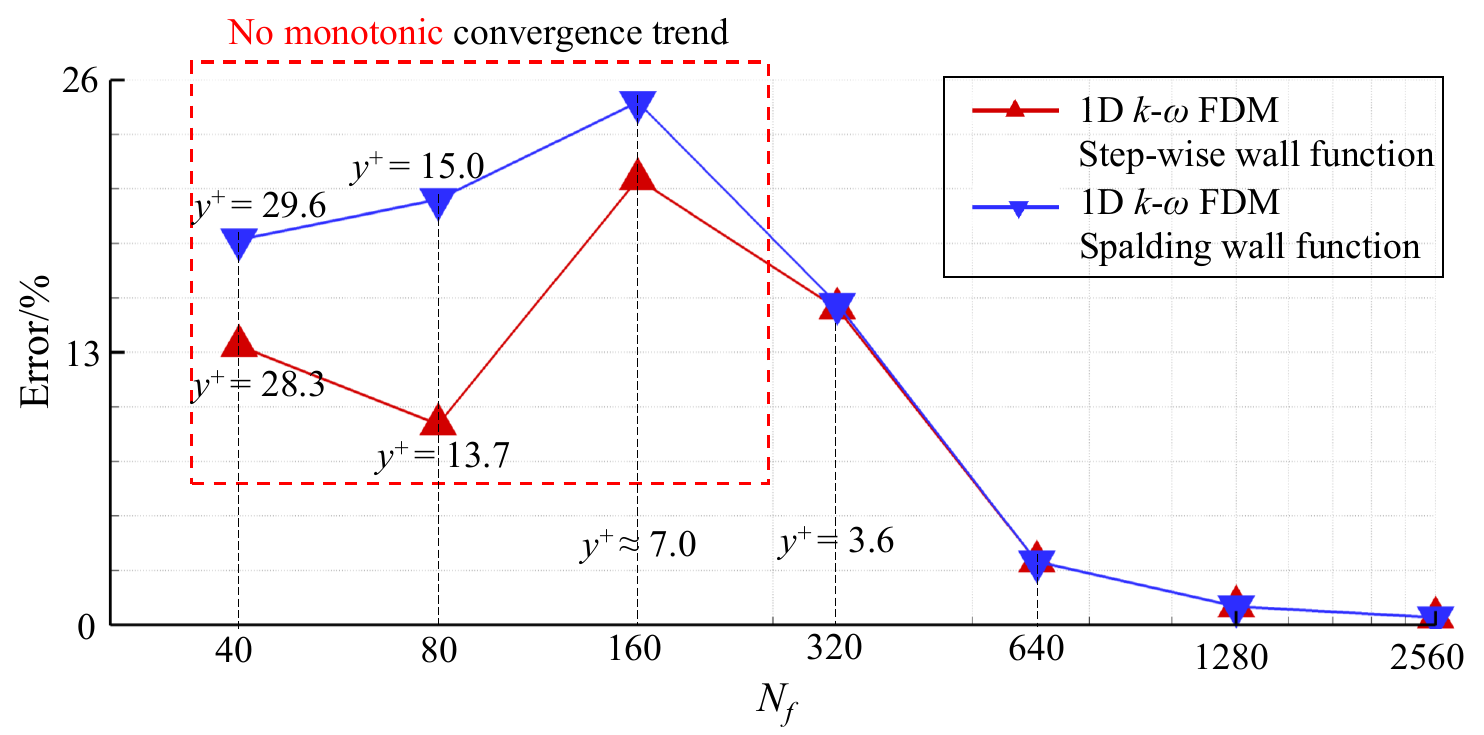}
	\caption{
		Case 2: Friction-coefficient error with respect to the DNS result~\cite{lee2015direct} predicted by the one-dimensional $k$--$\omega$ model using the step-wise and Spalding wall functions at different resolutions, with the horizontal axis plotted on a logarithmic scale.
	}
	\label{fig-straight-re40k-1D_cf-yplus}
\end{figure}

First, a reference solution is obtained by solving the simplified 1D $k$--$\omega$ model using the finite difference method (FDM). This calculation also provides a preliminary indication of the resolution sensitivity discussed above, particularly when the near-wall node falls within the buffer layer.
The relative errors of the friction coefficient predicted by the one-dimensional model at different resolutions are evaluated with respect to the DNS reference value $C_f = 5.00 \times 10^{-3}$ \cite{lee2015direct}, as shown in Fig. \ref{fig-straight-re40k-1D_cf-yplus}. Two wall-function treatments, namely the step-wise wall function\cite{wang2025weakly} and the Spalding wall function\cite{wang6271943unified}, are considered for comparison.

As shown in Fig. \ref{fig-straight-re40k-1D_cf-yplus}, neither the step-wise nor the Spalding wall function exhibits a monotonic convergence trend when the near-wall node falls within the buffer layer.
Although the Spalding wall function exhibits a simpler resolution
dependence, it requires an additional iterative procedure
and thus incurs additional computational cost.
As the resolution is further increased and the near-wall node approaches the viscous sublayer, the predictions gradually converge toward the DNS reference value.

\begin{figure}[htb!]
	\centering
	\includegraphics[trim = 3.81cm 0cm 0cm 2.9cm, clip,width=1.0\textwidth]{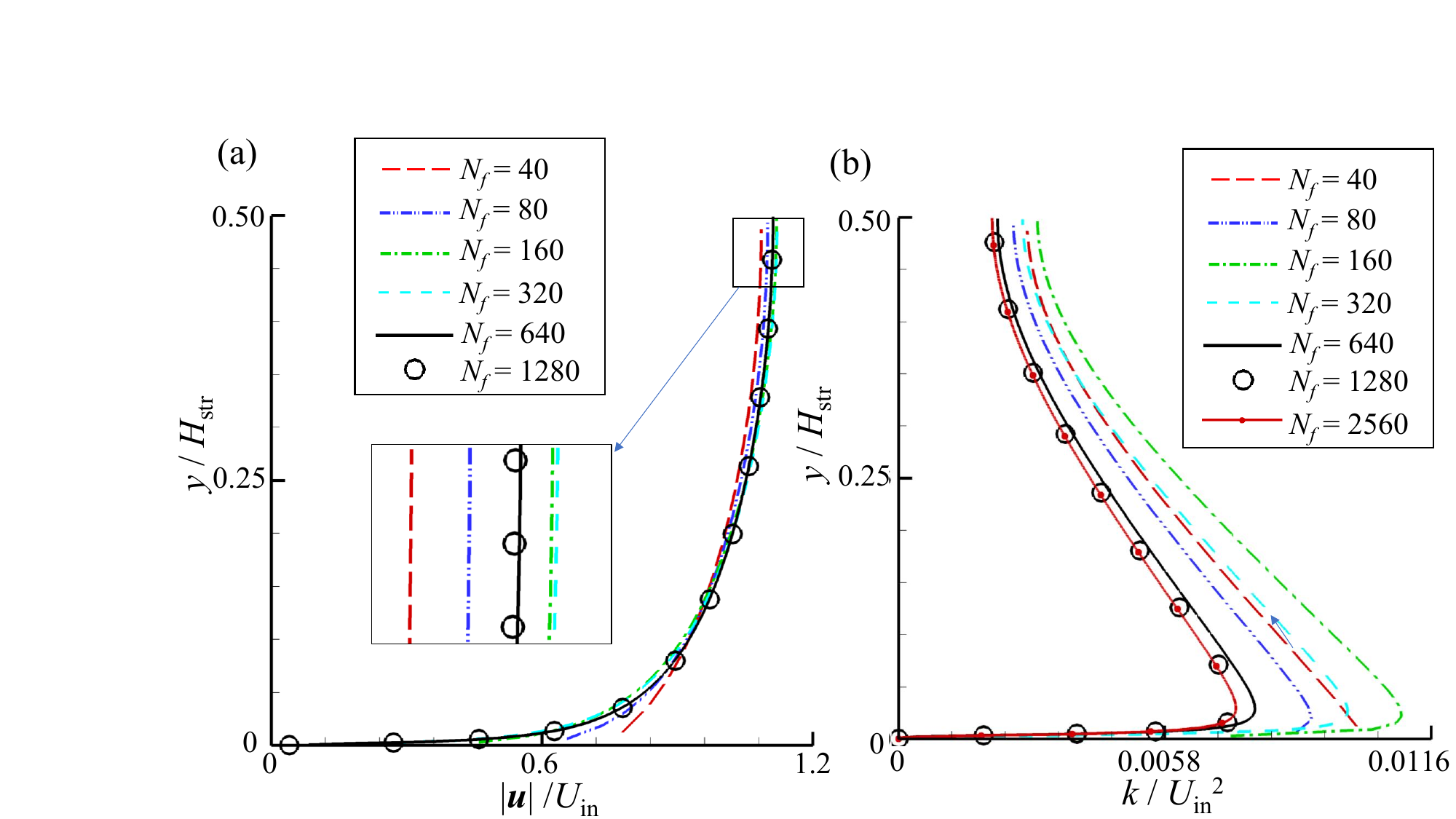}
	\caption{
		Case 2: resolution dependence of the velocity and turbulent kinetic energy profiles predicted by the one-dimensional $k$--$\omega$ model using the step-wise wall function: (a) velocity profile with an enlarged view near the channel centerline and (b) turbulent kinetic energy profile.
	}
	\label{fig-straight-re40k-1D_vel-tke}
\end{figure}

The corresponding velocity and turbulent kinetic energy ($k$) profiles are shown in Fig. \ref{fig-straight-re40k-1D_vel-tke}. Similar to the friction-coefficient results, the velocity profiles also exhibit non-monotonic convergence at relatively low resolutions. This behavior is particularly visible in the enlarged channel centerline region, where the maximum velocity varies non-monotonically as the resolution is increased. The $k$ profiles show similar resolution sensitivity, as the results first approach the reference solution, then deviate from it, and finally converge again with increasing resolution. This non-monotonic behavior further indicates the uncertainty associated with wall-normal resolutions corresponding to the buffer layer.

\begin{figure}[htb!]
	\centering
	\includegraphics[trim = 8.74cm 0cm 0cm 7.42cm, clip,width=1.0\textwidth]{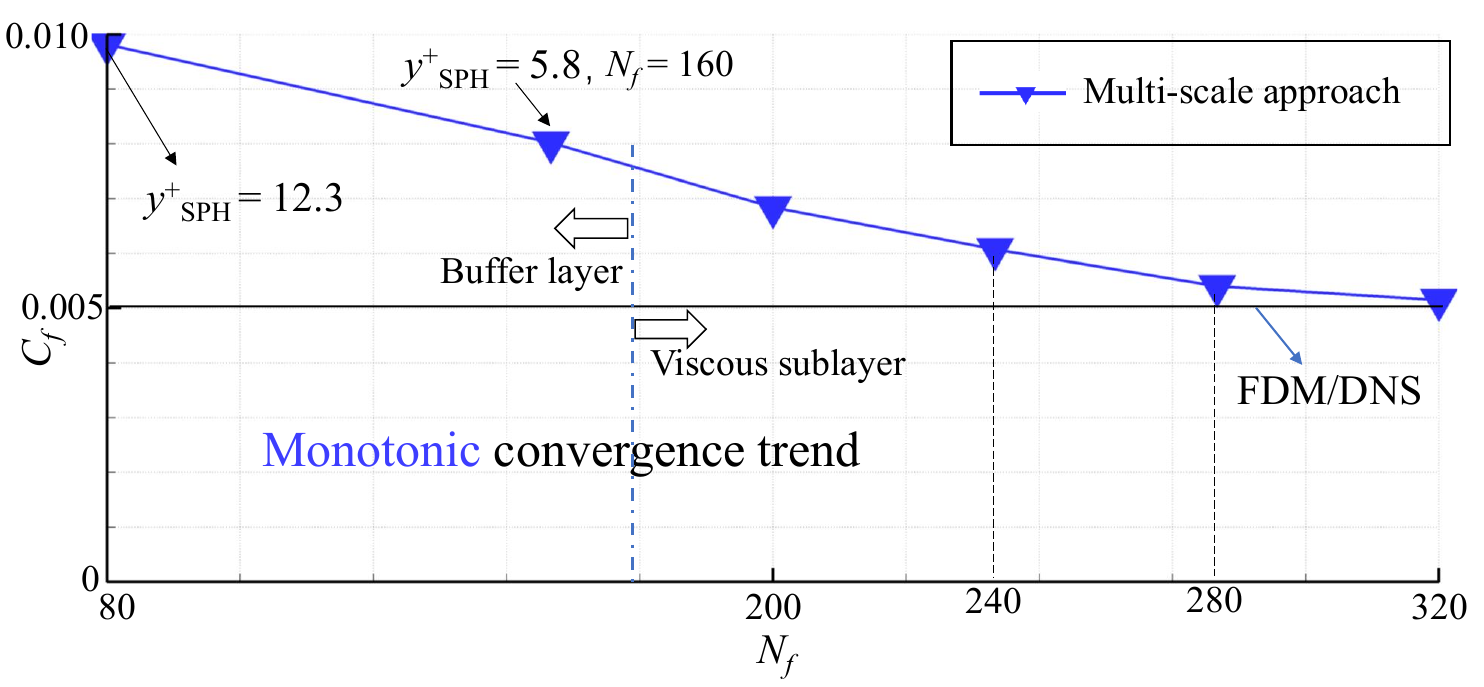}
	\caption{
	Case 2: friction-coefficient values of the SPH method with the multi-scale approach at different resolutions, compared with the DNS reference result~\cite{lee2015direct}, which is also approached by the converged FDM solution. The blue dotted line schematically indicates the approximate transition of the wall-adjacent SPH particles from the buffer layer to the viscous sublayer, and the corresponding $y^+_{\mathrm{SPH}}$ values of the wall-adjacent SPH particles are also indicated.
	}
	\label{fig-straight-re40k-SPH_cf}
\end{figure}

In contrast, as shown in Fig. \ref{fig-straight-re40k-SPH_cf}, the SPH results obtained with the proposed multi-scale approach exhibit a clear monotonic convergence trend in the friction-coefficient prediction. The error decreases continuously as the resolution increases, even when the wall-adjacent SPH particles are located in the buffer layer, as indicated by the corresponding $y^+_{\mathrm{SPH}}$ values. This behavior contrasts with the non-monotonic convergence observed in the simplified one-dimensional tests, indicating that the proposed approach mitigates the resolution sensitivity associated with the buffer layer. The reduction in error becomes much less pronounced beyond $N_f=280$, and the result at $N_f=320$ can be regarded as approaching convergence.

\begin{figure}[htb!]
	\centering
	\includegraphics[trim = 3.81cm 0cm 0cm 2.9cm, clip,width=1.0\textwidth]{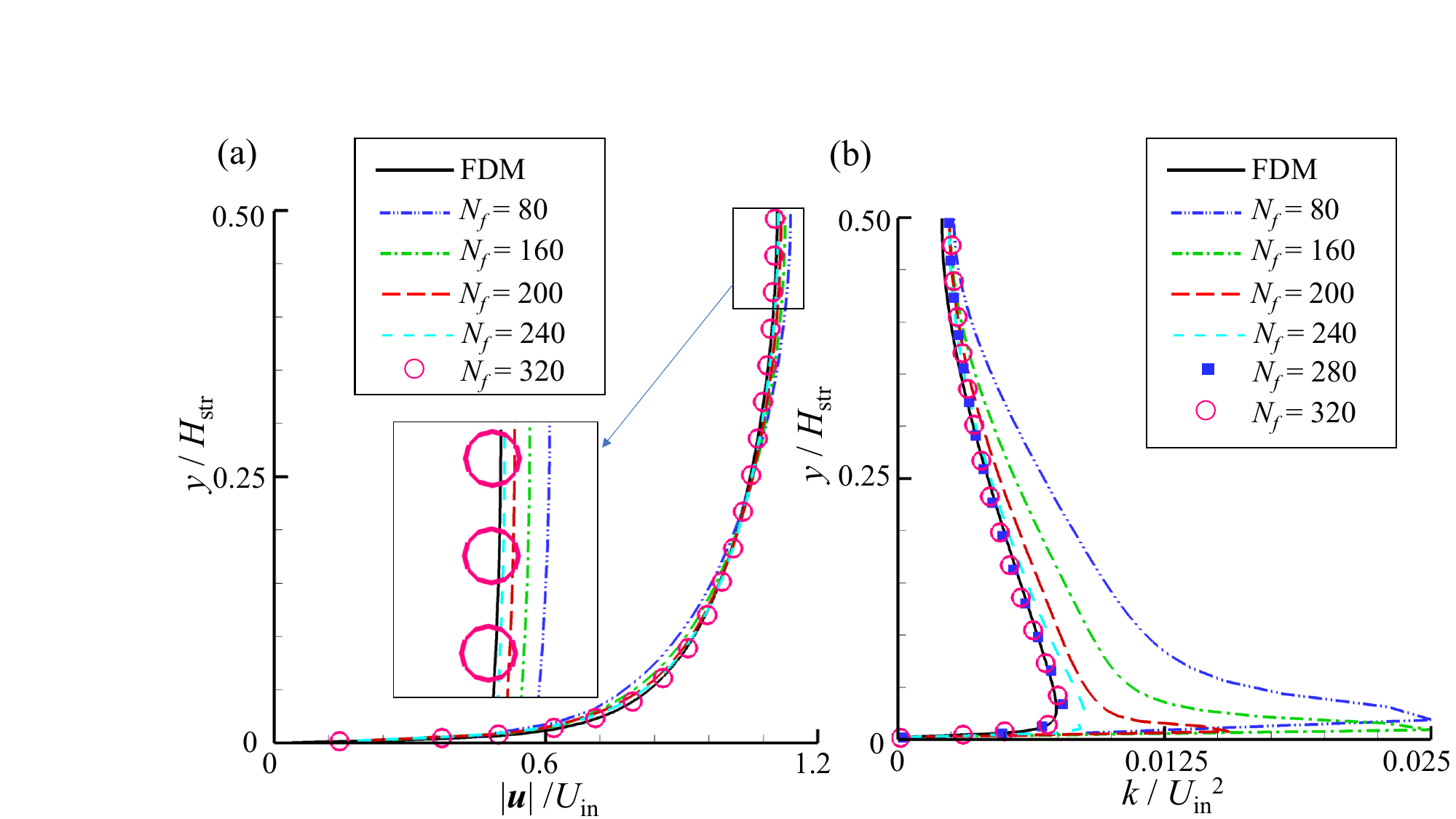}
	\caption{
		Case 2: velocity and turbulent kinetic energy profiles predicted by the SPH method with the multi-scale approach at different resolutions, compared with the one-dimensional FDM reference solution: (a) velocity profile with an enlarged view near the channel centerline and (b) turbulent kinetic energy profile.
	}
	\label{fig-straight-re40k-SPH-vel-tke}
\end{figure}

This improved convergence behavior is further supported by the cross-sectional velocity and $k$ profiles shown in Fig. \ref{fig-straight-re40k-SPH-vel-tke}.
The velocity profile shows clear convergence at $N_f=240$, whereas the $k$ profile requires a higher resolution, approximately $N_f=280$, to achieve comparable convergence in the near-wall region.
Both the velocity and $k$ profiles achieve the monotonic convergence and the converged results agree well with the reference values.

\begin{figure}[htb!]
	\centering
	\includegraphics[trim = 3.81cm 0cm 0cm 2.9cm, clip,width=1.0\textwidth]{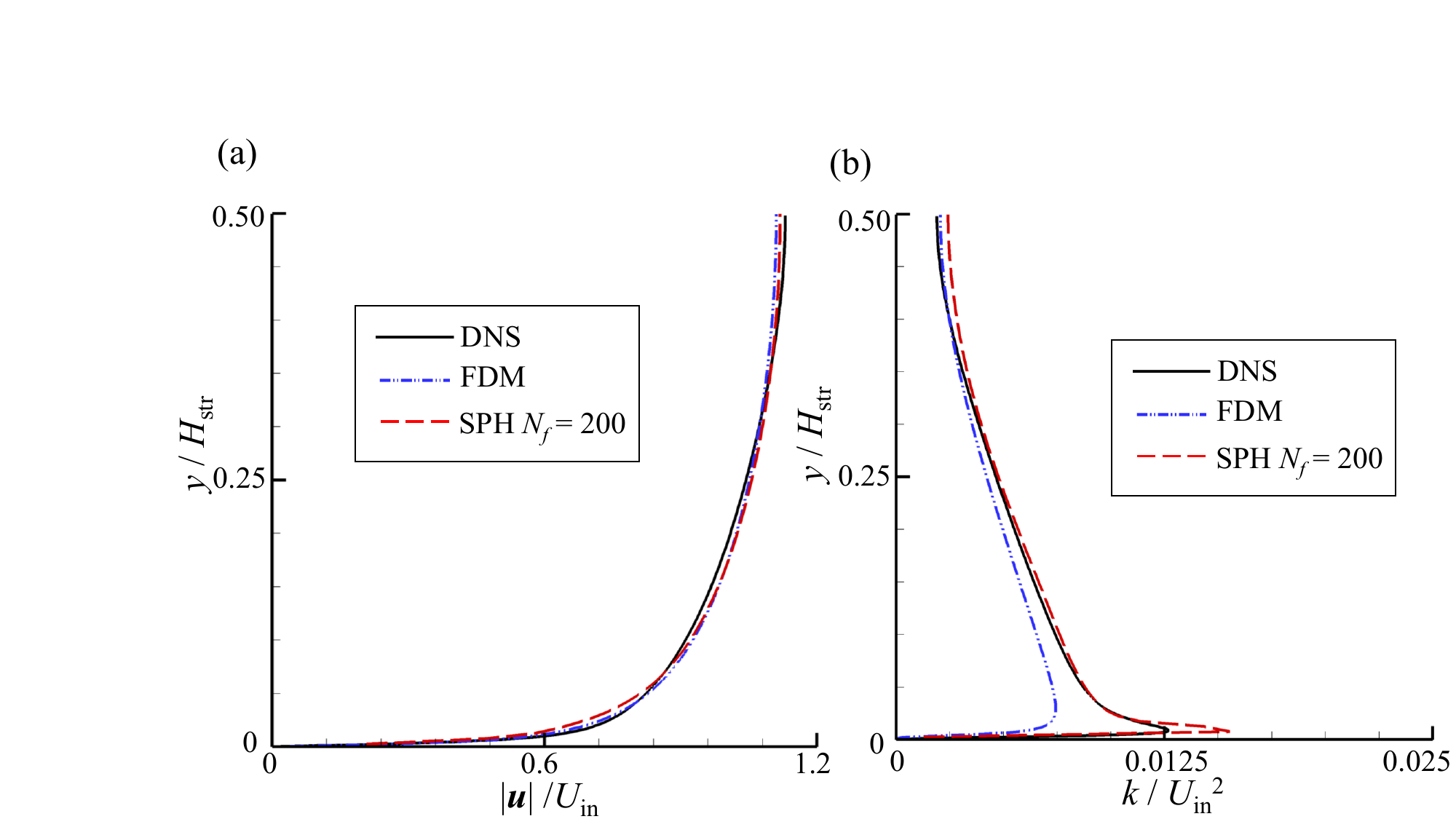}
	\caption{
		Case 2: velocity and turbulent kinetic energy profiles predicted by the SPH method with the multi-scale approach, compared with the one-dimensional FDM reference solution and DNS result~\cite{lee2015direct}: (a) velocity profile and (b) turbulent kinetic energy profile.
	}
	\label{fig-straight-re40k-SPH_comp-dns}
\end{figure}

Additionally, the SPH result at a representative resolution, $N_f=200$, is
compared with the converged one-dimensional FDM solution and the DNS data, as
shown in Fig. \ref{fig-straight-re40k-SPH_comp-dns}.
The comparison between the FDM and DNS results indicates that the
one-dimensional $k$--$\omega$ model tends to produce an overly smoothed
near-wall $k$ profile.
In contrast, the WCSPH--RANS result obtained with the proposed multi-scale
approach yields a sharper near-wall $k$ distribution and shows better
agreement with the DNS data in this region.
This behavior is consistent with the observation in
Fig. \ref{fig-straight-re5714-vel-k-comp-converged} for the baseline case,
suggesting that, at a relatively low resolution, the Lagrangian nature of the
SPH discretization may help preserve the near-wall variation of turbulent
kinetic energy.

This observation suggests a potential advantage of the
Lagrangian WCSPH--RANS formulation.
In the present comparison, the converged one-dimensional
FDM solution exhibits a smoother near-wall turbulent
kinetic energy profile than the DNS data.
By contrast, the Lagrangian particle discretization appears to retain a
sharper near-wall $k$ distribution at the tested resolution, leading to closer
agreement with the DNS data in this region.
Although this effect may diminish as the solution further
converges, it suggests that the present WCSPH--RANS formulation may provide a
useful advantage in representing near-wall turbulent kinetic energy
distributions.

\begin{figure}[htb!]
	\centering
	\includegraphics[trim = 3.89cm 0cm 0cm 3.06cm, clip,width=1.0\textwidth]{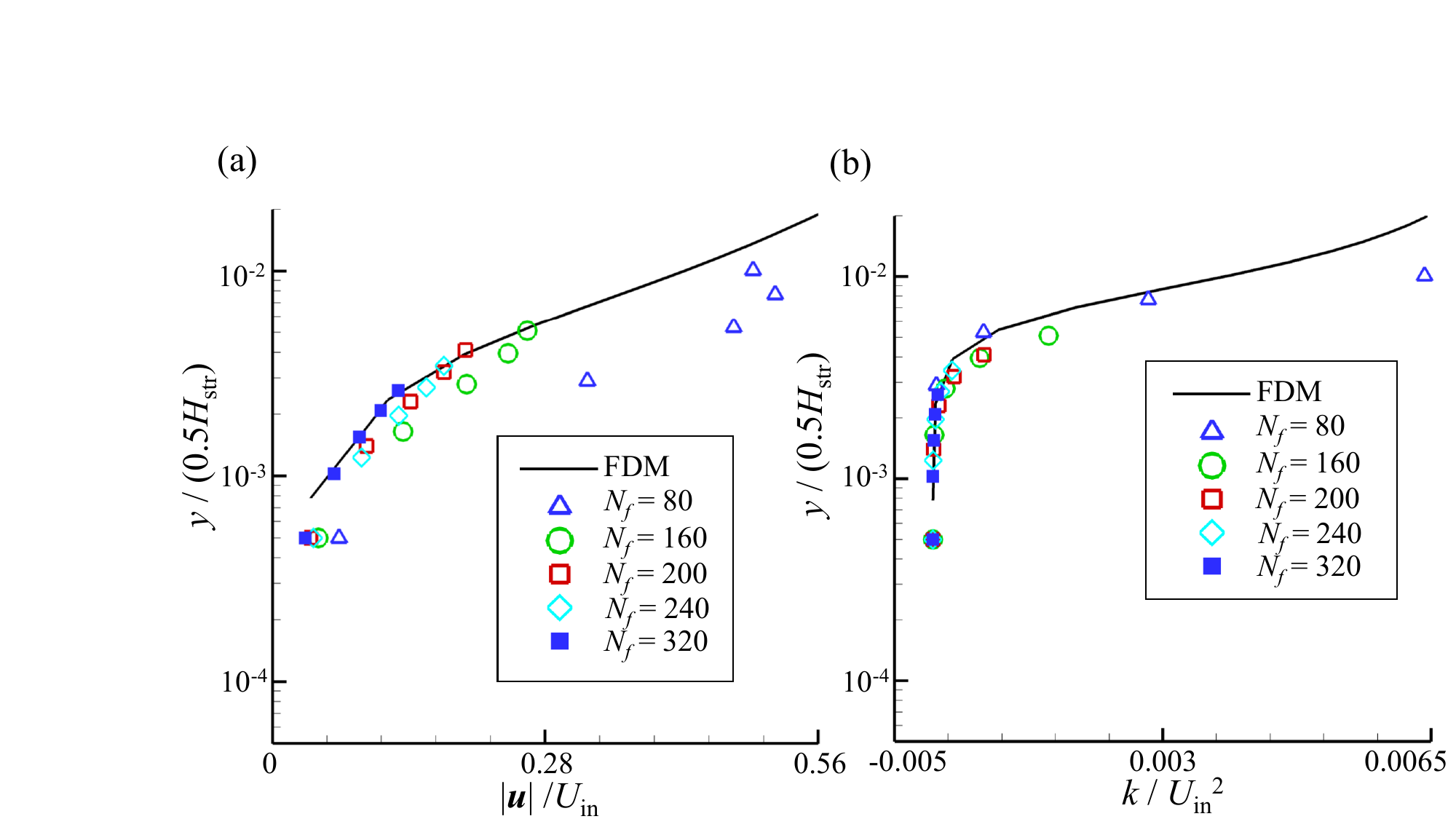}
	\caption{
		Case 2: sublayer-node profiles of velocity and turbulent kinetic energy obtained from the SPH simulation with the multi-scale approach: (a) velocity profile and (b) turbulent kinetic energy profile. The vertical axis is plotted on a logarithmic scale.
	}
	\label{fig-straight-re40k-SPH-vel-tke-node}
\end{figure}

The velocity and turbulent kinetic energy values computed by the sublayer solver are presented in Fig. \ref{fig-straight-re40k-SPH-vel-tke-node}.
At the lowest resolution, $N_f=80$, the velocity profile exhibits a localized kink, which disappears when the resolution is increased to $N_f=160$ and beyond.
Both the velocity and $k$ profiles show good agreement with the reference results and exhibit satisfactory convergence behavior. It is also worth noting that, owing to the use of the BOT$^+$ technique, the height of the first node adjacent to the wall remains fixed at $5.0 \times 10^{-4}$, although the influence of this treatment is limited in the present case.

\begin{figure}[htb!]
	\centering
	\includegraphics[trim = 3.81cm 0cm 0cm 2.9cm, clip,width=1.0\textwidth]{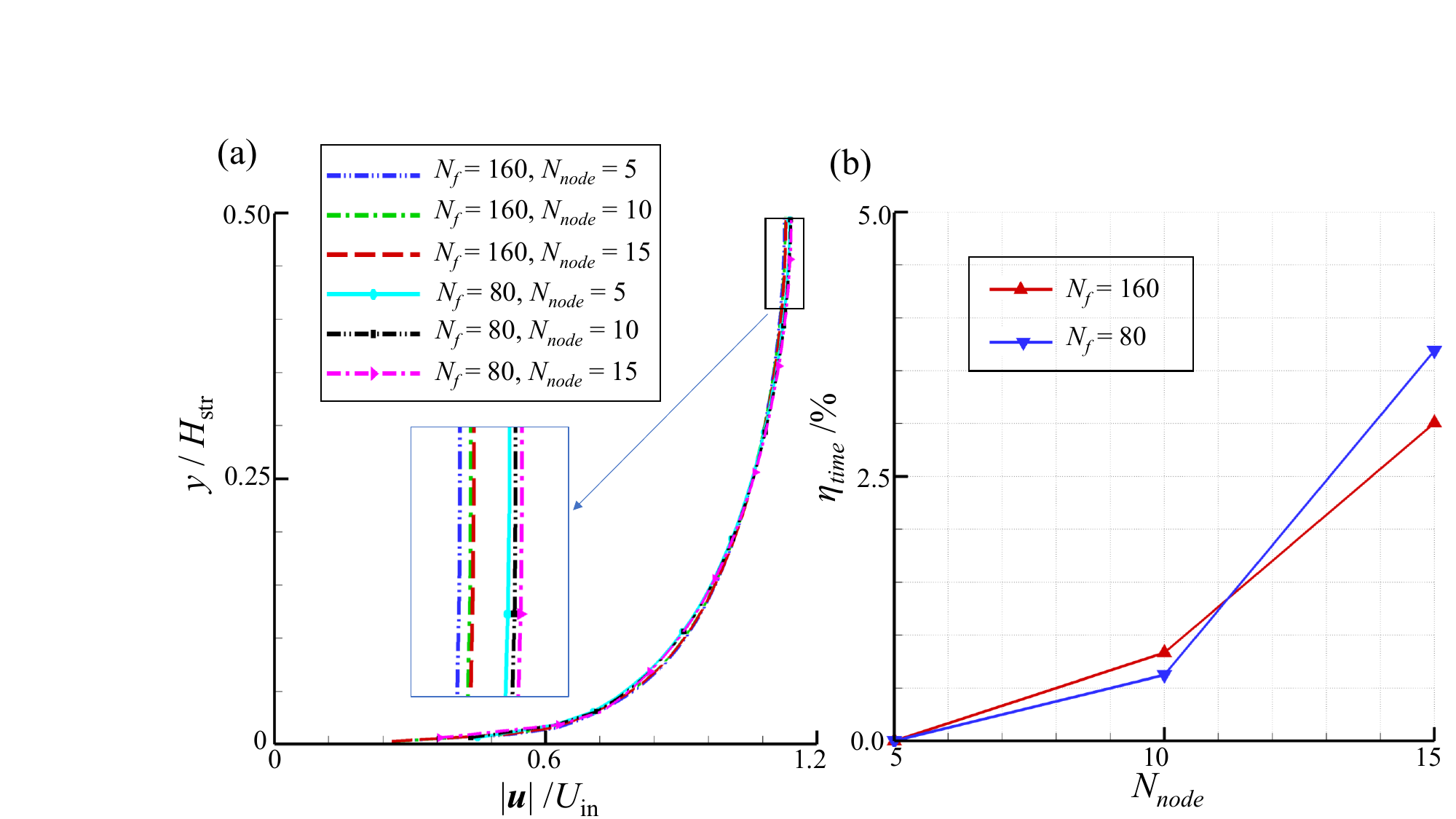}
	\caption{
		Case 2: (a) velocity profiles predicted by the SPH method with the multi-scale approach at the two resolutions with different number of nodes of the sublayer solver, $N_{node}$, and (b) the corresponding relative increase in computational cost.
	}
	\label{fig-straight-re40k-SPH_diff_node_number}
\end{figure}

The influence of the number of sublayer nodes on the predicted profiles and
computational cost is investigated in
Fig. \ref{fig-straight-re40k-SPH_diff_node_number}.
For each SPH resolution, simulations with different numbers of sublayer nodes,
$N_{\mathrm{node}}$, are performed under the same computational device and
software environment.
The relative increase in computational cost is defined as
$\eta_\mathrm{time}=(T_{N_\mathrm{node}}-T_5)/T_5$,
where $T_{N_\mathrm{node}}$ denotes the wall-clock time
for the corresponding sublayer-node count at a fixed SPH
resolution.

At the two representative resolutions considered here, increasing
$N_{\mathrm{node}}$ leads to only minor changes in the velocity profile.
The main difference is observed near the channel centerline, where a larger
number of sublayer nodes results in a slightly increased deviation from the
one-dimensional FDM reference solution.
This limited sensitivity to $N_{\mathrm{node}}$ indicates that,
for this case, a small number of sublayer nodes is
sufficient to capture the local wall-normal variation relevant
to the coupled solution.
Therefore, further increasing the number of sublayer nodes does not
necessarily improve the overall prediction, especially when the SPH particle
resolution remains unchanged.

Regarding the computational time, as shown in Fig. \ref{fig-straight-re40k-SPH_diff_node_number}(b), increasing
the number of sublayer nodes inevitably increases the computational cost.
Nevertheless, even for the 15-node system, the additional cost remains below
5\%, indicating that the overhead introduced by the sublayer solver is limited.
This relative overhead is expected to further decrease as the SPH resolution
increases, because the overall particle operations, such as neighbor searching
and configuration updates, become increasingly dominant in the total
computational cost.

In summary, considering both agreement with the reference solution and computational cost, a five-node
sublayer system is generally preferred.
For more extreme cases with very high Reynolds numbers, such as the case
considered next, a ten-node sublayer system combined with the BOT$^+$
technique is used to resolve the viscous sublayer and provide a more reliable near-wall representation.
The node-count tests in the present case also indicate a limited
cost increase: increasing the number of nodes from 5 to 10
raises the wall-clock time by only 0.62\% at $N_f=80$
and 0.83\% at $N_f=160$.

\begin{figure}[htb!]
	\centering
	\includegraphics[trim = 7.51cm 0cm 0cm 7.35cm, clip,width=1.0\textwidth]{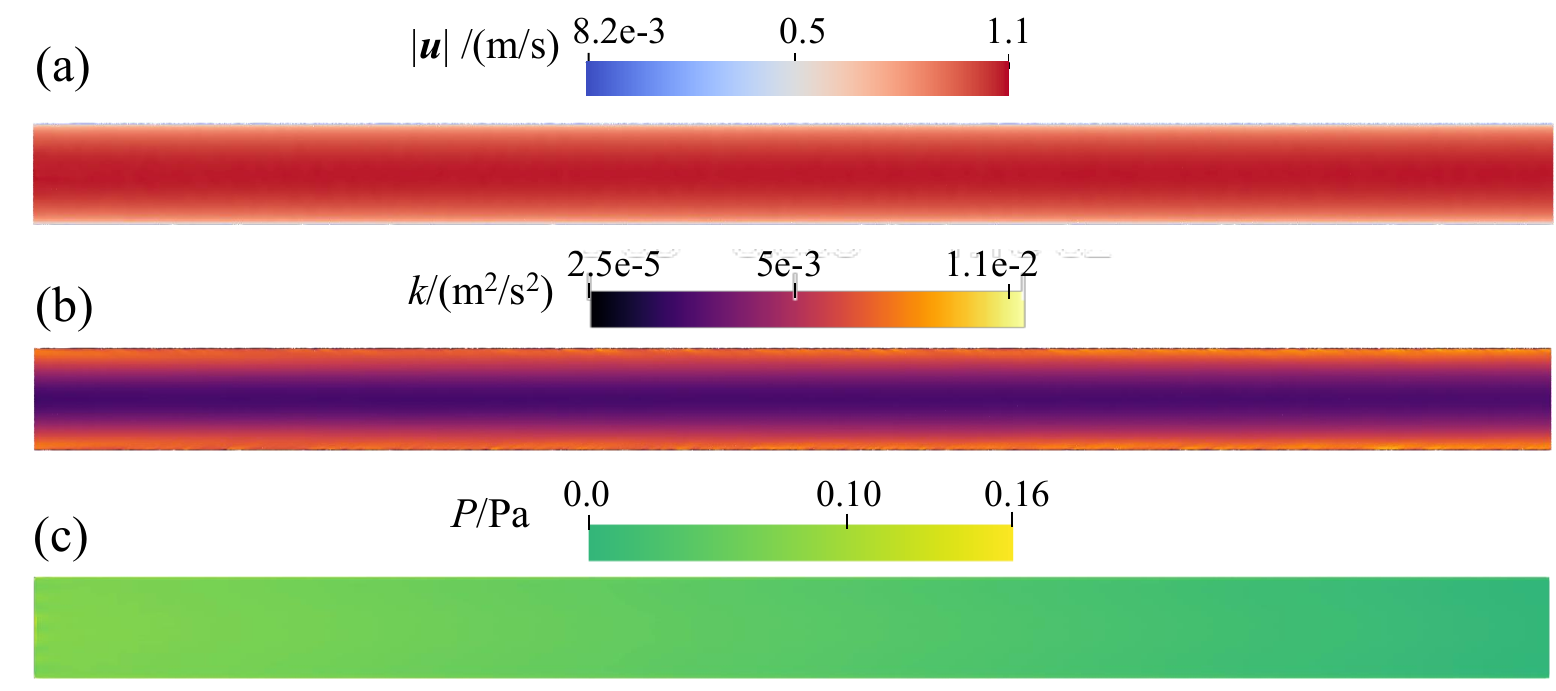}
	\caption{
		Case 2: contour fields of the turbulent channel flow obtained using the SPH method with the proposed multi-scale approach: (a) velocity magnitude, (b) turbulent kinetic energy, and (c) time-averaged pressure.
	}

	\label{fig-straight-re40k-SPH-vel-tke-p-contour}
\end{figure}

The corresponding contour fields at $N_f=280$, are presented in Fig. \ref{fig-straight-re40k-SPH-vel-tke-p-contour}, together with the time-averaged pressure field.
All three fields exhibit smooth distributions, consistent with the expected behavior of a RANS solution.

%

\subsubsection{Case 3: extreme-Reynolds-number case at Re = 80,000,000}
\label{case-straight-Re80000000}

The third case considers a benchmark case documented in the NASA turbulence modeling resource (TMR)~\cite{NASA_TMR_channel}, which provides publicly available turbulence model verification and validation cases, computational grids, and reference data.
Owing to its broad use in the CFD community for turbulence-model benchmarking, this case is adopted here as a well-established reference for assessing the proposed near-wall approach under extreme-Reynolds-number wall-bounded turbulent-flow conditions.
For comparison, the NASA TMR results obtained using the
Wilcox 2006 $k$--$\omega$ model are adopted. The CFL3D reference results are normalized consistently with the present
results, with the reference friction coefficient taken from the fully
developed region near the channel outlet.

\begin{figure}[htb!]
	\centering
	\includegraphics[trim = 3.91cm 0cm 0cm 1.91cm, clip,width=1.0\textwidth]{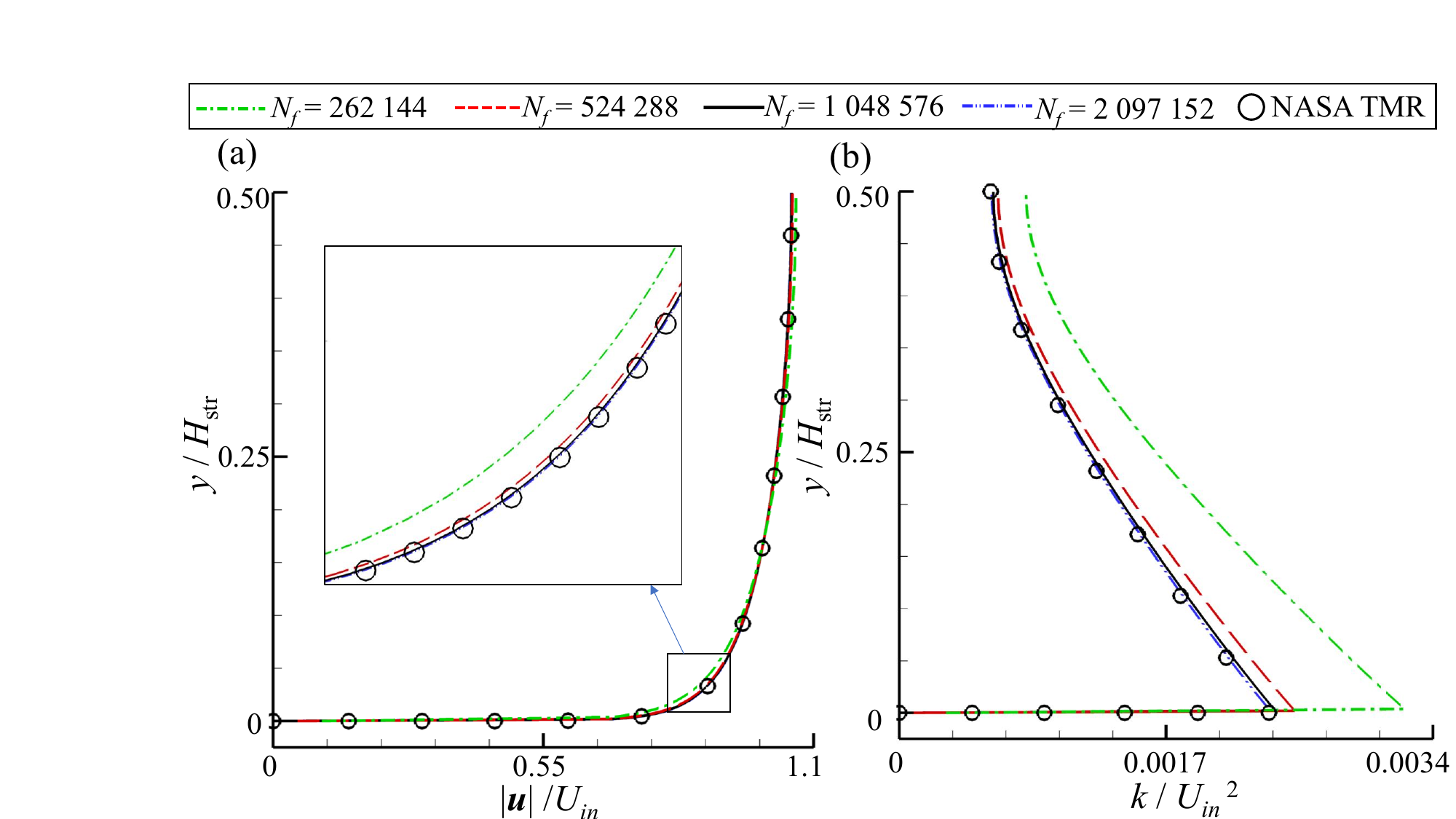}
	\caption{
		Case 3: comparison of the velocity and turbulent kinetic energy profiles obtained from the one-dimensional model and the NASA TMR reference data computed using CFL3D. The NASA TMR data correspond to the converged resolution ($N_f = 512$) with extreme near-wall grid refinement.
	}
	\label{fig-straight-re5714-vel-k-converge-python}
\end{figure}

First, this case is simulated using the simplified one-dimensional $k$--$\omega$ model with uniform discretization to provide a preliminary test and a reference solution, as shown in Fig. \ref{fig-straight-re5714-vel-k-converge-python}.
Due to the extremely high Reynolds number, the viscous sublayer becomes extremely thin in physical space.
In contrast to the baseline case, resolving this layer with a uniform node distribution would require an extremely high resolution, with $N_f$ exceeding one million to obtain a converged solution.
Although the one-dimensional results agree well with those reported by NASA TMR, such a resolution would clearly be computationally prohibitive for an SPH simulation of this benchmark case, thereby highlighting the necessity of the multi-scale approach.

\begin{figure}[htb!]
	\centering
	\includegraphics[trim = 3.91cm 0cm 0cm 2.9cm, clip,width=1.0\textwidth]{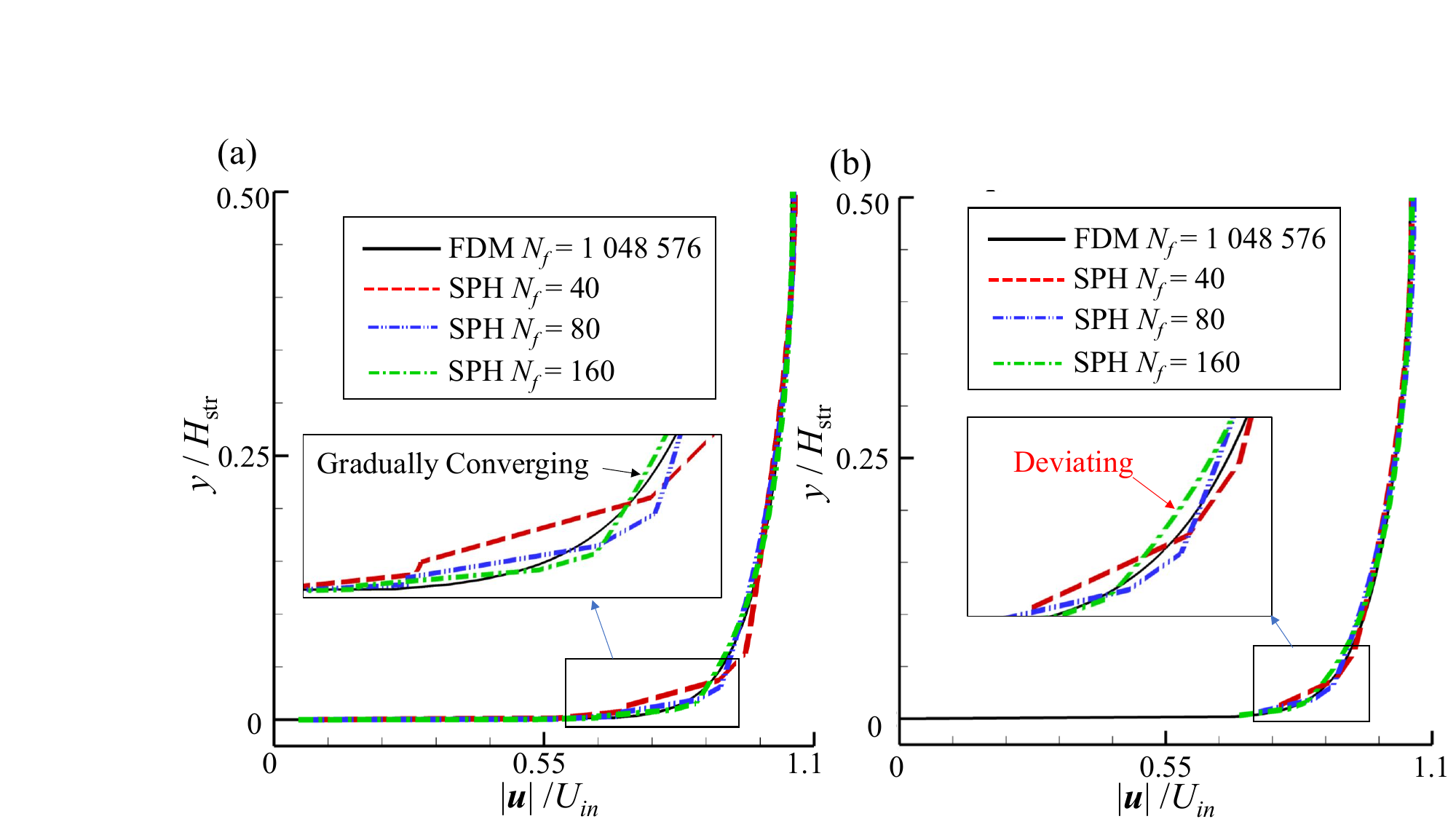}
	\caption{
		Case 3: comparison of cross-sectional velocity profiles obtained at three resolutions using (a) the proposed multi-scale approach and (b) the WCSPH--RANS method without the multi-scale treatment. The sublayer-node values are plotted together with the corresponding SPH particle values.
	}
	\label{fig-straight-re80m-vel-converge-comp-withTrash}
\end{figure}

Second, convergence tests of the cross-sectional velocity profiles are conducted using the proposed multi-scale approach and the $k$--$\omega$ WCSPH method without the multi-scale treatment~\cite{wang6271943unified}.
The results are presented in Fig. \ref{fig-straight-re80m-vel-converge-comp-withTrash}.
For the multi-scale approach, the number of sublayer nodes is set to 10, and $y_p^\mathrm{node}$ is fixed at $2.0 \times 10^{-6}$.
Although both approaches achieve satisfactory agreement near the channel centerline, differences are observed in the near-wall region. The multi-scale approach exhibits a converging trend, particularly in the region with a large velocity gradient, whereas the method without the multi-scale treatment does not show comparable convergence due to insufficient resolution within the viscous sublayer.

\begin{figure}[htb!]
	\centering
	\includegraphics[trim = 7.46cm 0cm 0cm 4.96cm, clip,width=1.0\textwidth]{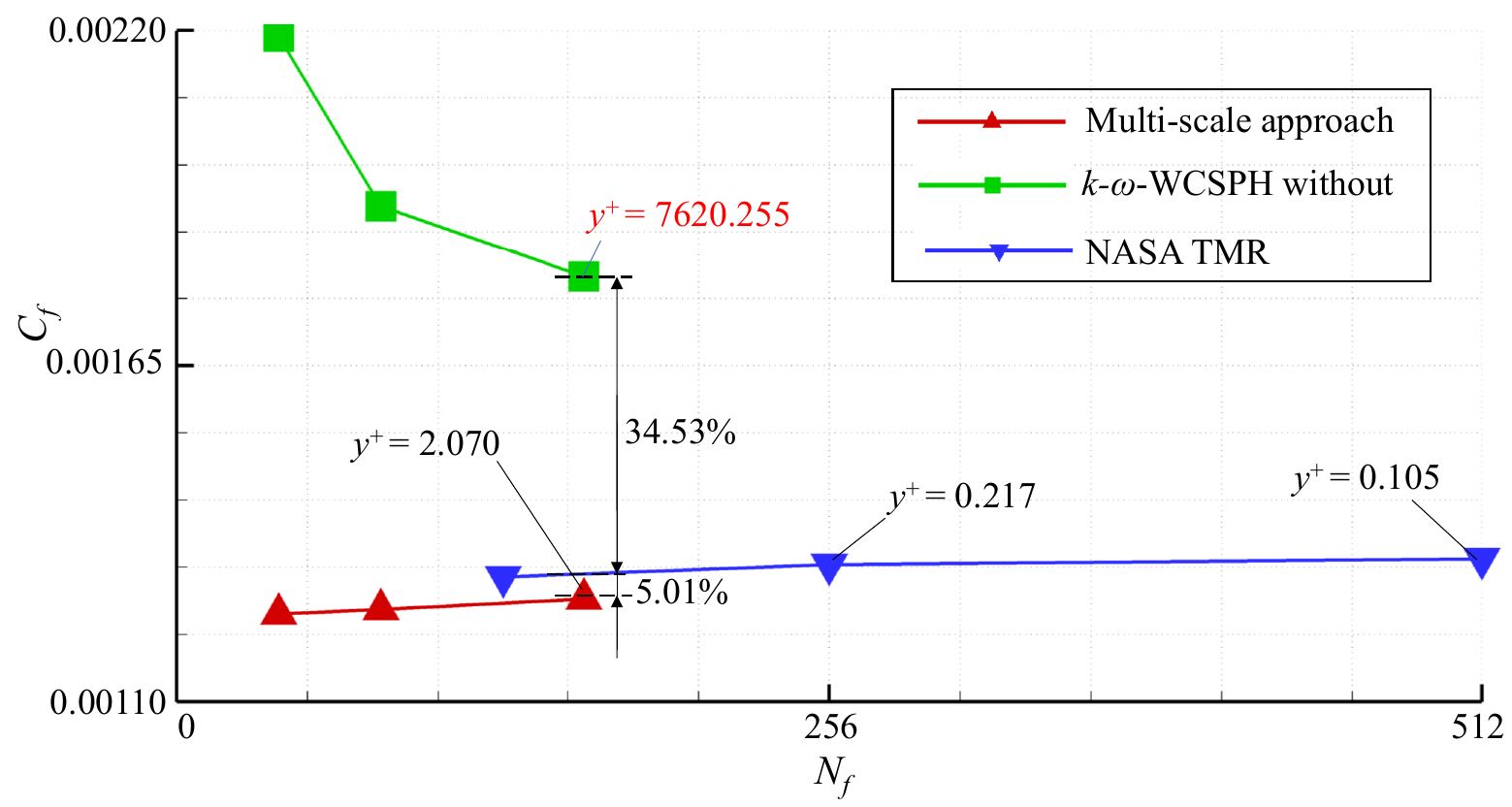}
	\caption{
		Case 3: comparison of the friction coefficient at different resolutions obtained using the proposed multi-scale approach and the WCSPH--RANS method without the multi-scale treatment, together with the NASA TMR reference result~\cite{NASA_TMR_channel} obtained using extreme near-wall grid refinement.
	}
	\label{fig-straight-re80m-vel-comp-cf}
\end{figure}

Finally, Figure \ref{fig-straight-re80m-vel-comp-cf} demonstrates the advantage of the proposed approach in predicting the friction coefficient.
Compared with the TMR result obtained at $N_f = 512$ with extreme near-wall grid refinement, the proposed multi-scale approach achieves comparable results using a relatively coarse particle resolution of $N_f = 160$, without requiring extreme near-wall particle refinement.
Further increasing the resolution is expected to improve the agreement, as a clear convergence trend is observed. In contrast, directly applying the WCSPH--RANS method without the multi-scale treatment leads to a significant over-prediction of $C_f$.

\begin{figure}[htb!]
	\centering
	\includegraphics[trim = 7.32cm 0cm 0cm 3.53cm, clip,width=1.0\textwidth]{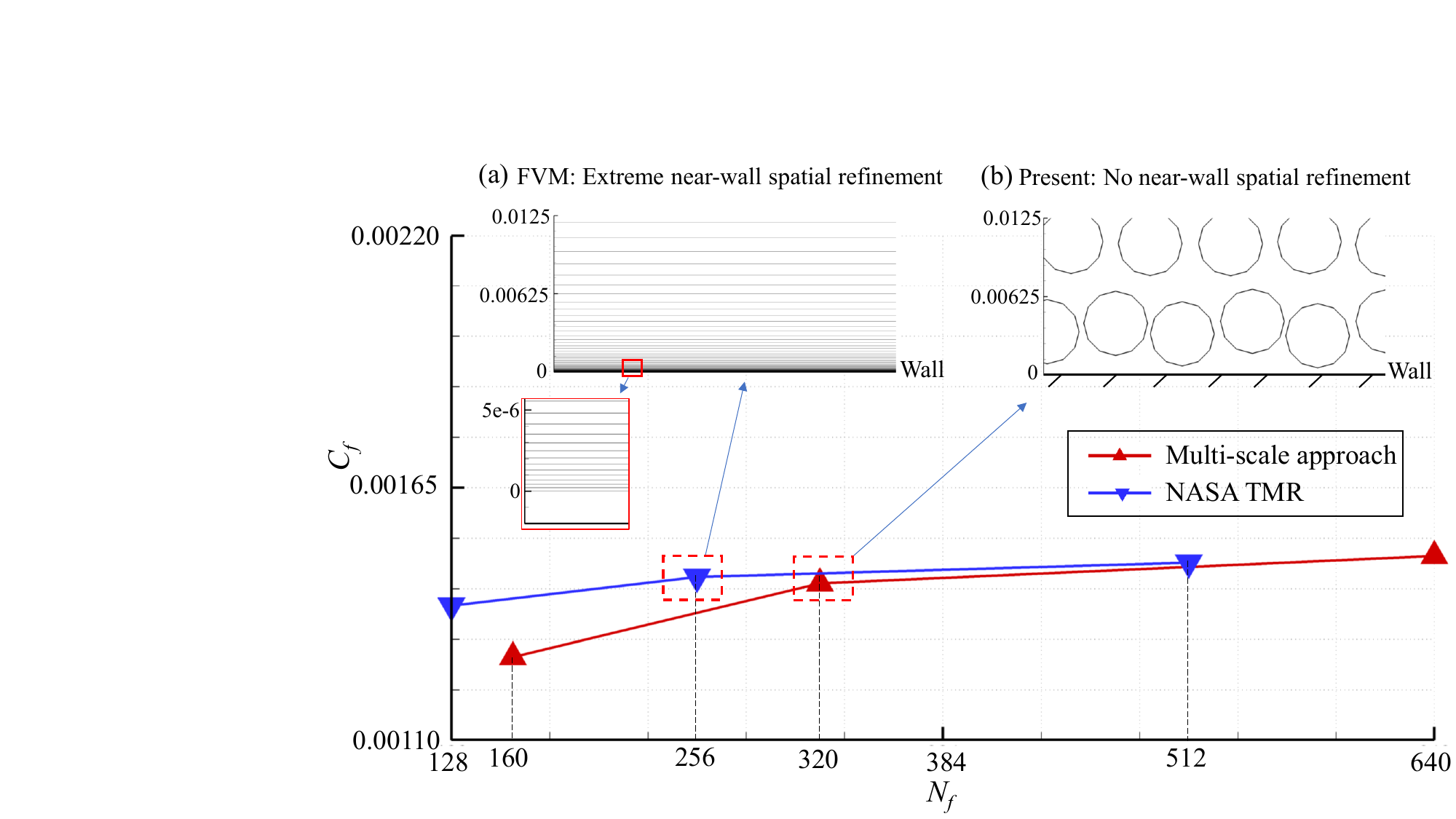}
	\caption{
		Case 3: comparison of friction coefficients at higher resolutions between the NASA TMR reference results~\cite{NASA_TMR_channel} with extreme near-wall grid refinement and the proposed multi-scale SPH results with uniform particle discretization: (a) near-wall grid distribution and (b) particle distribution.
	}
	\label{fig-straight-re80m-further_refined}
\end{figure}

The results obtained at higher resolutions are presented in Fig. \ref{fig-straight-re80m-further_refined}. The relative error of the friction coefficient computed by the proposed approach, which was previously 5.01\% at $N_f=160$, decreases to 0.53\% at $N_f=320$. The solution is considered converged at $N_f=320$, since further increasing the resolution to $N_f=640$ produces only negligible changes.
A similar convergence behavior is observed for the NASA TMR reference data, which reach convergence at $N_f=256$. However, it should be noted that the finite-volume (TMR) results were obtained using extremely strong near-wall grid refinement. As shown in Fig. \ref{fig-straight-re80m-further_refined}(a), the maximum cell size in the wall-normal direction is approximately $2.56 \times 10^{5}$ times larger than the minimum cell size.

In contrast, the proposed approach does not require such extreme near-wall refinement in the SPH discretization, as shown in Fig. \ref{fig-straight-re80m-further_refined}(b). This comparison clearly demonstrates the advantage of the proposed multi-scale approach.
%

\subsection{Turbulent flow in a wavy (sinusoidal) channel}

\begin{figure}[htb!]
	\centering
	\includegraphics[trim = 10.05cm 0cm 0cm 7.61cm, clip,width=1.0\textwidth]{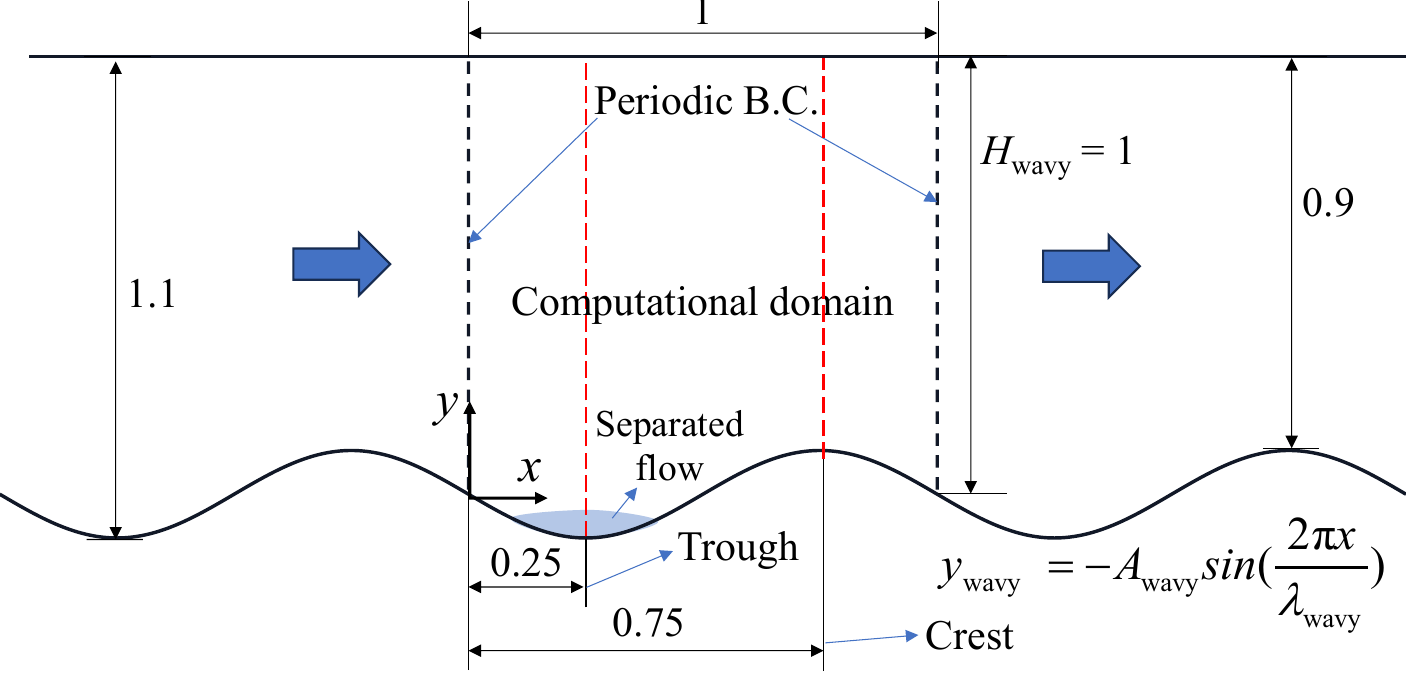}
	\caption{
		Wavy channel: schematic illustration of the geometry and computational domain. The sinusoidal lower wall, periodic boundary conditions, flow direction, separated-flow region, and crest and trough observation locations indicated by red dashed lines are shown. Dimensions are given in meters.
	}
	\label{fig-wavy-geo}
\end{figure}

To test the current approach on predicting more complicated flow problem, particularly when the separation flow exists, the turbulent flow over a wavy channel case is simulated.
This case is a classical benchmark case provided by ANSYS numbered VMFL012~\cite{ansys_vmfl012_wavy_channel}.
Figure \ref{fig-wavy-geo} shows the geometry, and the boundary conditions, quantitative data including the flow parameters are presented in Table~\ref{tab-wavy-para}.
The Reynolds number of this case, $Re_\mathrm{wavy}$, is based on the mean channel height, $H_\mathrm{wavy}$, and mean streamwise velocity, $U_{\mathrm{mean}}$.

\begin{table}[!htb]
	\centering
	\caption{Geometric, flow parameters and boundary conditions of the wavy-channel case.}
	\label{tab-wavy-para}
	\begin{tabular}{@{}lll@{}}
		\toprule
		Category & Parameter                                     & Value            \\
		\midrule

		\multirow{6}{*}{Geometry}
		         & Periodic domain length                        & $1.0~\mathrm{m}$ \\
		         & Mean channel height, $H_\mathrm{wavy}$        & $1.0~\mathrm{m}$ \\
		         & Wave amplitude, $A_\mathrm{wavy}$             & $0.1~\mathrm{m}$ \\
		         & Wavelength, $\lambda_\mathrm{wavy}$           & $1.0~\mathrm{m}$ \\
		         & Channel height at the crest                   & $0.9~\mathrm{m}$ \\
		         & Channel height at the trough                  & $1.1~\mathrm{m}$ \\

		\midrule

		\multirow{5}{*}{Flow parameters}
		         & Fluid density
		         & $1.0~\mathrm{kg\,m^{-3}}$                                        \\

		         & Dynamic viscosity
		         & $1.0\times10^{-4}~\mathrm{Pa\,s}$                                \\

		         & Mean streamwise velocity, $U_{\mathrm{mean}}$
		         & $0.816~\mathrm{m\,s^{-1}}$                                       \\

		         & Mass flow rate
		         & $0.816~\mathrm{kg\,s^{-1}}$                                      \\

		         & Reynolds number, $Re_\mathrm{wavy}$
		         & $8160$                                                           \\

		\midrule

		\multirow{3}{*}{Boundary conditions}
		         & Streamwise boundaries                         & Periodic         \\
		         & Upper straight wall                           & No-slip          \\
		         & Lower sinusoidal wall                         & No-slip          \\

		\bottomrule
	\end{tabular}
\end{table}

To systematically investigate this case, we perform a reference simulation using the well-known open-source software OpenFOAM v12, with the same case setup and the same $k$--$\omega$ RANS model~\cite{wilcox2008formulation} as those employed in the present work.
Additionally, consistent with the SPH simulation, the OpenFOAM calculation was performed within the unsteady RANS framework, with pressure-velocity coupling achieved using the pressure-implicit with splitting of operators (PISO) algorithm. An adaptive time step was employed to maintain the maximum Courant number below unity, and the simulation was advanced to 300 seconds, identical to that used in the SPH simulation.

\begin{figure}[htb!]
	\centering
	\includegraphics[trim = 10.48cm 0cm 0cm 1.21cm, clip,width=1.0\textwidth]{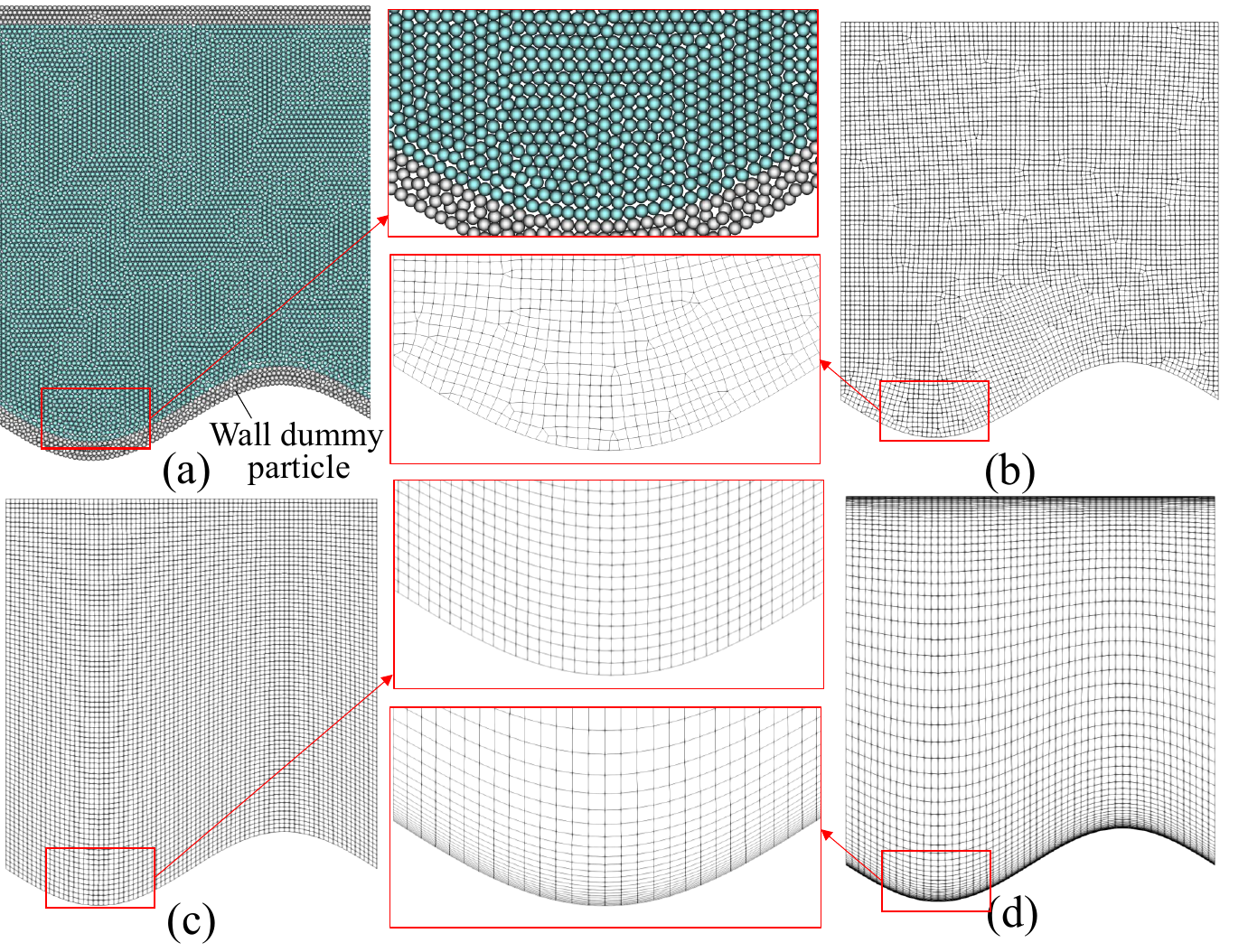}
	\caption{
		Wavy channel: comparison of particle and grid distributions: (a) initial SPH particle distribution, where wall dummy particles are also shown, (b) uniform unstructured grid, (c) uniform structured grid, and (d) locally refined structured grid. The enlarged views highlight the near-wall resolution around the trough region.
	}
	\label{fig-wavy-mesh-particle-comp}
\end{figure}

As shown in Fig. \ref{fig-wavy-mesh-particle-comp}, three different types of mesh are generated using ICEM and adopted in the OpenFOAM simulations.
Following the definition used for the SPH particle discretization in subfigure (a), the resolutions of the uniform grids in subfigures (b) and (c) are characterized by the number of cells across the cross-section, $H_\mathrm{wavy}$, denoted by $N_f$.
As summarized in Table \ref{tab:wavy-discretizations}, three resolution levels, namely coarse, medium, and fine, are considered for both the proposed SPH approach and the FVM simulations.
The fine locally refined mesh shown in subfigure (d) was adopted directly
from the ANSYS official documentation~\cite{ansys_vmfl012_wavy_channel},
whereas the corresponding medium and coarse meshes are generated through
systematic coarsening according to the prescribed refinement criterion.
Details of the three locally-refined meshes are summarized in
Table~\ref{tab:wavy-discretizations}.

\begin{table}[htbp]
	\centering
	\caption{Summary of the spatial discretizations and resolution levels
		employed for the wavy-channel simulations.
		Here, $N_x$ and $N_y$ denote the numbers of cells in the streamwise
		and wall-normal directions, respectively. The first-cell height adjacent
		to the wall, $\Delta y_1$, is reported in units of
		$10^{-4}\,\mathrm{m}$.}
	\label{tab:wavy-discretizations}
	\small
	\setlength{\tabcolsep}{3.5pt}
	\renewcommand{\arraystretch}{1.15}
	\begin{tabularx}{\linewidth}{
		@{}l
		>{\raggedright\arraybackslash}X
		>{\centering\arraybackslash}p{0.17\linewidth}
		>{\centering\arraybackslash}p{0.14\linewidth}
		>{\centering\arraybackslash}p{0.095\linewidth}
		>{\centering\arraybackslash}p{0.095\linewidth}
		>{\centering\arraybackslash}p{0.095\linewidth}
		@{}
		}
		\hline
		Method
		 & Discretization
		 & Distribution
		 & Quantity
		 & Coarse
		 & Medium
		 & Fine                                   \\
		\hline

		SPH
		 & Particle
		 & \multirow{3}{*}{\shortstack{Uniform }}
		 & \multirow{3}{*}{$N_f$}
		 & \multirow{3}{*}{80}
		 & \multirow{3}{*}{160}
		 & \multirow{3}{*}{320}                   \\

		FVM
		 & Structured
		 &
		 &
		 &
		 &
		\\

		FVM
		 & Unstructured
		 &
		 &
		 &
		 &
		\\
		\hline

		\multirow{2}{*}{FVM}
		 & \multirow{2}{*}{Structured}
		 & \multirow{2}{*}{Hyperbolic}
		 & $N_x\times N_y$
		 & $40\times45$
		 & $56\times63$
		 & $80\times90$                           \\

		 &
		 &
		 &
		$\Delta y_1$
		 & 12.53
		 & 8.646
		 & 5.897                                  \\
		\hline
	\end{tabularx}
\end{table}

\begin{figure}[htb!]
	\centering
	\includegraphics[trim = 11.17cm 0cm 0cm 0.4cm, clip,width=1.0\textwidth]{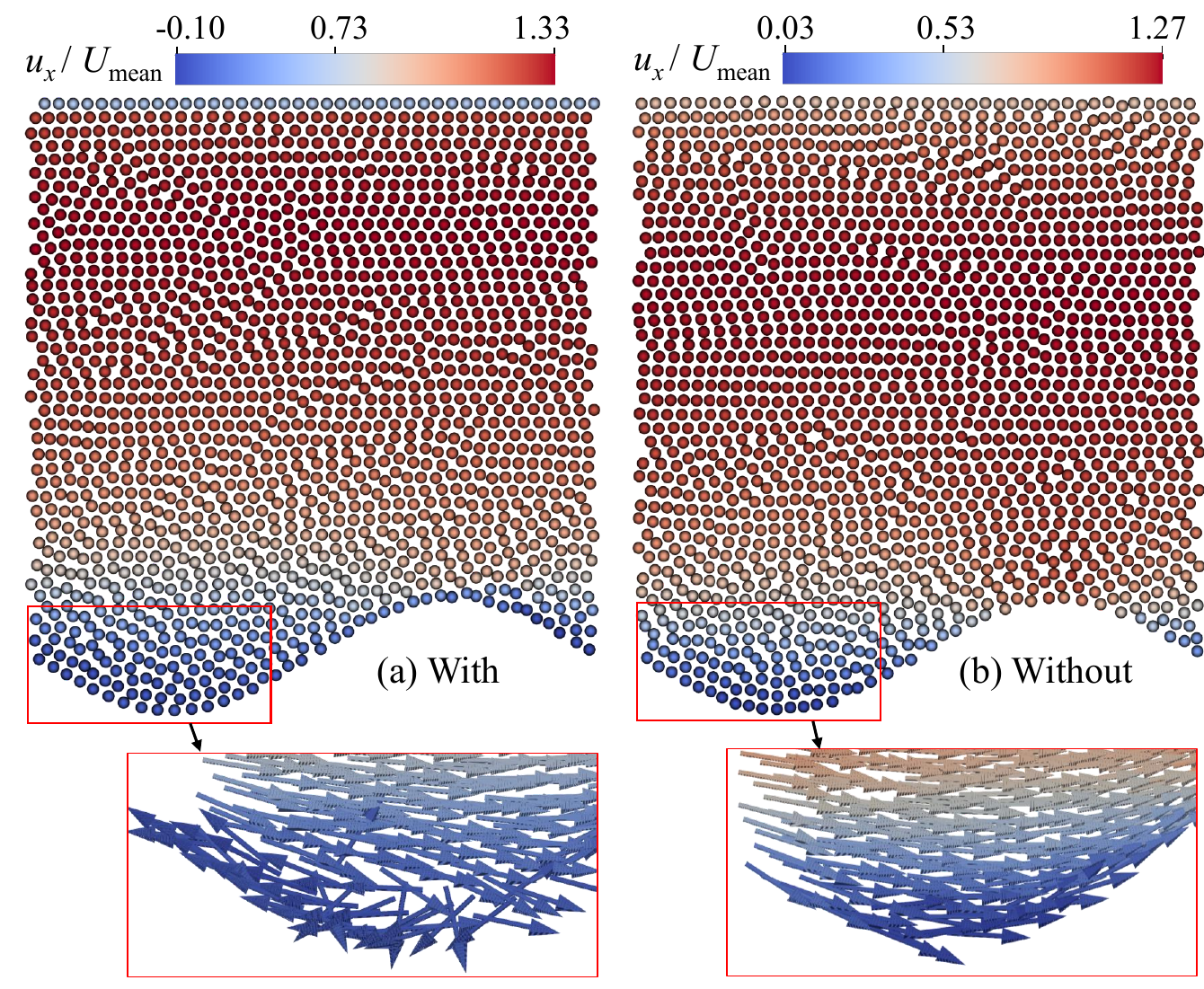}
	\caption{
		Wavy channel: comparison of low-resolution SPH results at $N_f=40$
		computed (a) with and (b) without the proposed multi-scale approach.
		The contours show the normalized streamwise velocity $u_x/U_\mathrm{mean}$,
		and the enlarged views illustrate the velocity-vector distributions in the
		near-wall reversed-flow region around the trough.
	}
	\label{fig-wavy-contour-vel-with-without-comp}
\end{figure}

First of all, Figure \ref{fig-wavy-contour-vel-with-without-comp} compares the low-resolution SPH results obtained with and
without the proposed multi-scale approach at $N_f=40$.
With the multi-scale approach, a coherent recirculation structure is clearly
formed near the trough region, as indicated by both the negative streamwise
velocity and the organized velocity vectors in the enlarged view.
This indicates that the proposed near-wall treatment can still provide an
effective representation of the separated near-wall flow, even when the SPH
particle resolution is very coarse.
This improvement may be attributed to the combined near-wall treatment in the
proposed formulation, including the sublayer solver and the modified
Riemann-based wall interaction.
The former provides additional wall-normal resolution, while the latter
improves the consistency of the wall-shear response, thereby alleviating the
insufficient wall-normal resolution of the coarse SPH particle discretization.

In contrast, without the multi-scale treatment, as presented in Fig. \ref{fig-wavy-contour-vel-with-without-comp} (b), the reversed flow region becomes less
distinct and the near-wall velocity field is more strongly smoothed by the
coarse particle discretization.
As a result, the low-resolution SPH simulation without the multi-scale
approach is less capable of resolving the local velocity reversal and the
associated recirculating motion near the wavy wall.

\begin{figure}[htb!]
	\centering
	\includegraphics[trim = 0.36cm 0cm 0cm 1.4cm, clip,width=1.0\textwidth]{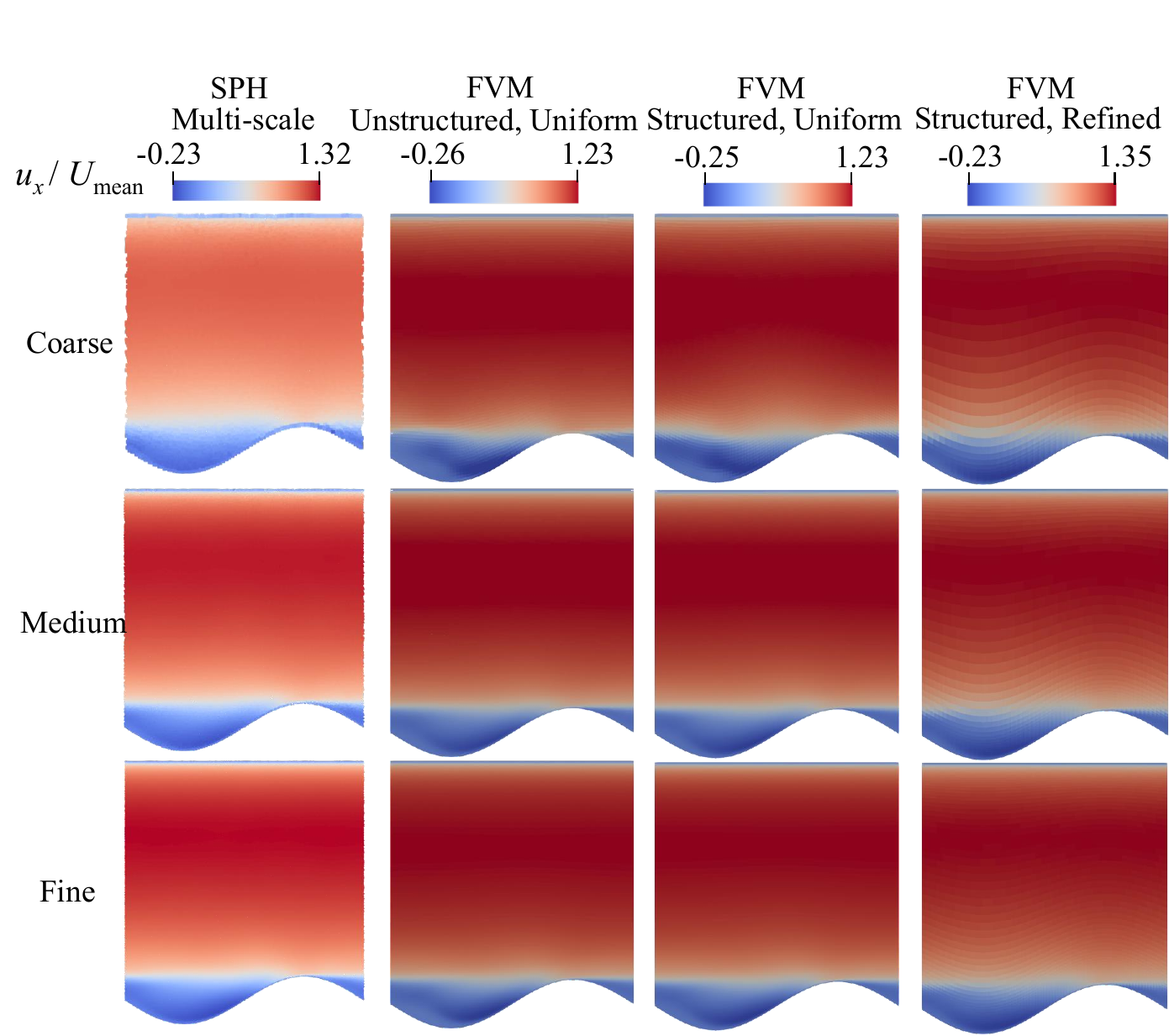}
	\caption{
		Wavy channel: comparison of streamwise velocity ($u_x$) contours at different resolutions. From left to right, the columns correspond to the SPH method with the proposed multi-scale approach, FVM with a uniform unstructured grid, FVM with a uniform structured grid, and FVM with a locally refined structured grid. From top to bottom, the rows correspond to coarse, medium, and fine resolutions.
	}
	\label{fig-wavy-large-vel-contour-comp}
\end{figure}

\begin{figure}[htb!]
	\centering
	\includegraphics[trim = 0.36cm 0cm 0cm 1.51cm, clip,width=1.0\textwidth]{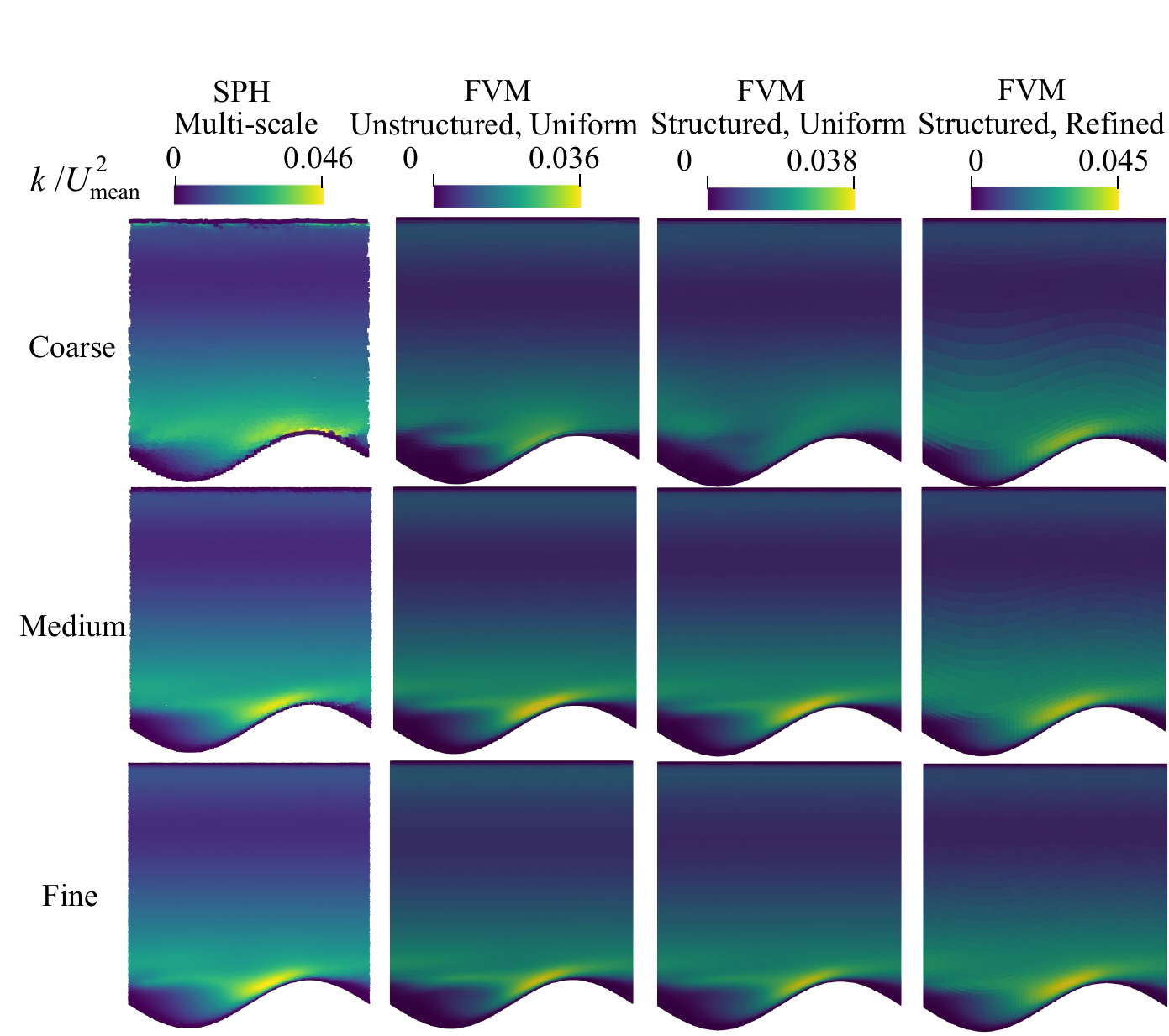}
	\caption{
		Wavy channel: comparison of turbulent kinetic energy ($k$) contours at different resolutions. From left to right, the columns correspond to the SPH method with the proposed multi-scale approach, FVM with a uniform unstructured grid, FVM with a uniform structured grid, and FVM with a locally refined structured grid. From top to bottom, the rows correspond to coarse, medium, and fine resolutions.
	}
	\label{fig-wavy-large-tke-contour-comp}
\end{figure}

Second, the qualitative comparisons of the velocity and turbulent kinetic energy fields are presented in Figs. \ref{fig-wavy-large-vel-contour-comp} and \ref{fig-wavy-large-tke-contour-comp}.
For each simulation series, including the SPH case with the multi-scale approach and the three FVM grid types, the color scale is determined based on the corresponding fine-resolution result.
As the spatial resolution increases, all four discretization approaches yield increasingly consistent flow patterns.
In particular, all simulations capture the separated-flow region characterized by negative streamwise velocity within the trough, as well as the subsequent flow recovery along the ascending wall.

Notably, even at the coarse resolution, the proposed multi-scale SPH method produces smooth velocity and turbulent kinetic energy fields, including within and around the separated-flow region.
This behavior is particularly evident in the $k$ contours, for which stable and spatially coherent distributions remain challenging to obtain using fully particle-based turbulence simulations~\cite{wang2025weakly,wang2022isph,bao2023pof}.

As the resolution increases, the SPH predictions become increasingly close to the FVM results, particularly those obtained using the locally refined structured grid. The similarity is also reflected in the ranges of the fine-resolution solutions.
In comparison, the two uniform FVM grids yield lower peak values, especially for $k$, indicating that the localized near-wall turbulence production is more strongly smoothed at the investigated resolutions.

\begin{figure}[htb!]
	\centering
	\includegraphics[trim = 2.44cm 0cm 0cm 0.67cm, clip,width=1.0\textwidth]{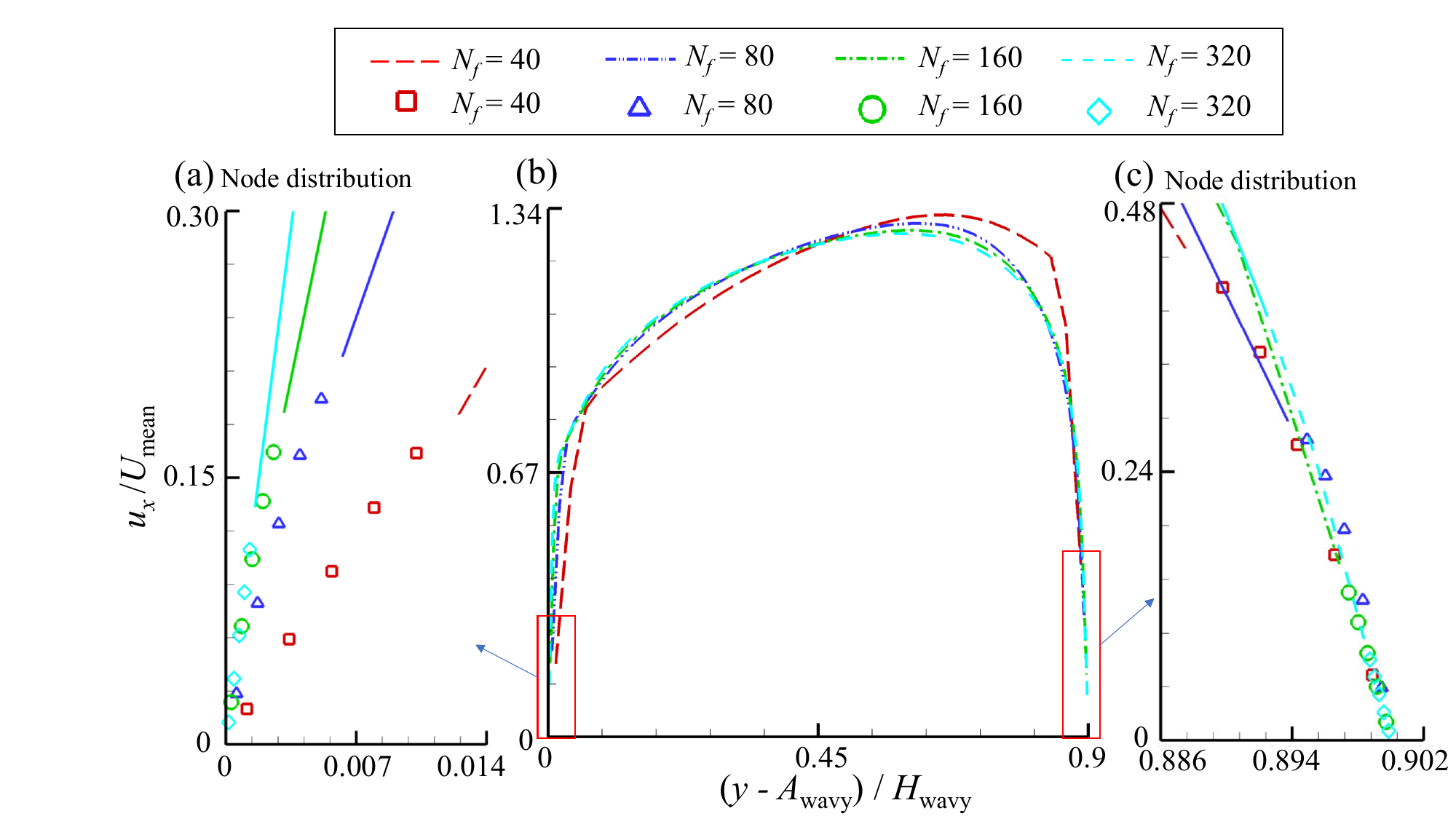}
	\caption{
		Wavy channel: velocity profiles predicted by the SPH method with the proposed multi-scale approach at the crest section for different resolutions: (a) local enlargement of the lower-wall region, (b) cross-sectional velocity profile, and (c) local enlargement of the upper-wall region. Panels (a) and (c) use the same horizontal coordinate as panel (b).
	}
	\label{fig-wavy-sph-crest-converge-vel-node}
\end{figure}

\begin{figure}[htb!]
	\centering
	\includegraphics[trim = 2.2cm 0cm 0cm 0.67cm, clip,width=1.0\textwidth]{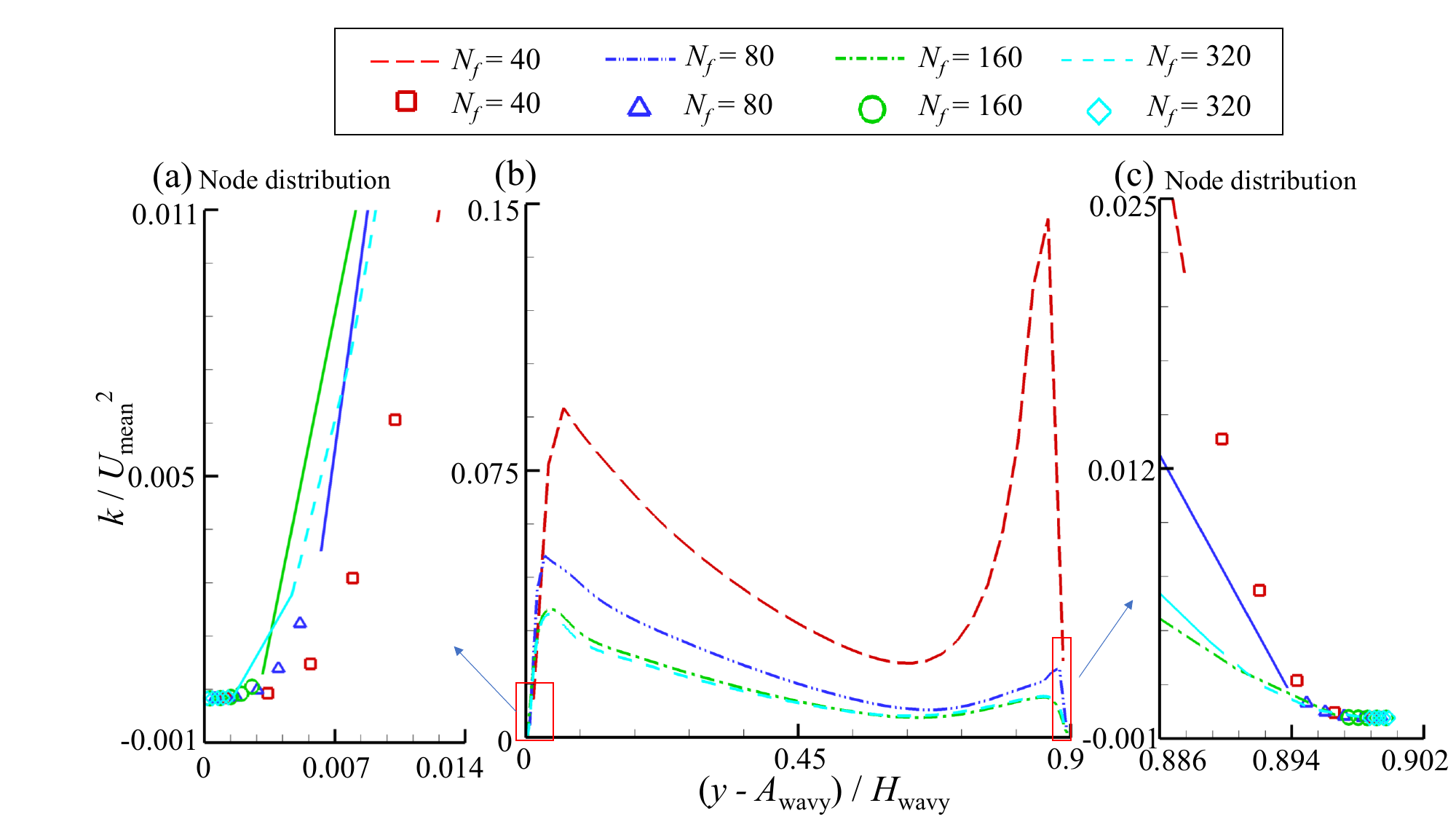}
	\caption{
		Wavy channel: turbulent kinetic energy profiles predicted by the SPH method with the proposed multi-scale approach at the crest section for different resolutions: (a) local enlargement of the lower-wall sublayer-node distribution, (b) cross-sectional turbulent kinetic energy profile, and (c) local enlargement of the upper-wall sublayer-node distribution. Panels (a) and (c) are plotted using the same horizontal coordinate as panel (b).
	}
	\label{fig-wavy-sph-crest-converge-tke-node}
\end{figure}

\begin{figure}[htb!]
	\centering
	\includegraphics[trim = 3.36cm 0cm 0cm 1.76cm, clip,width=1.0\textwidth]{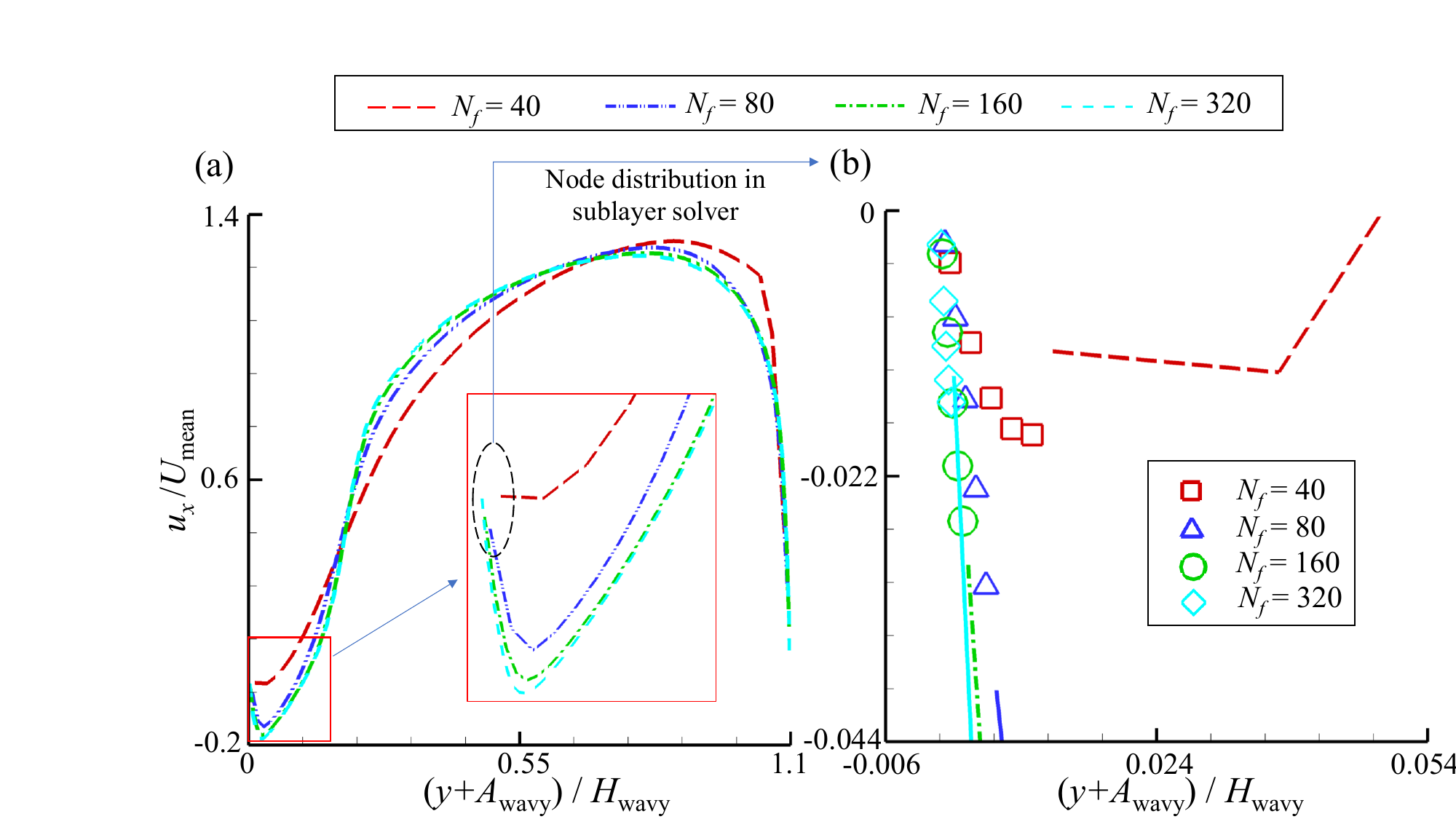}
	\caption{
		Wavy channel: velocity profiles predicted by the SPH method with the proposed multi-scale approach at the trough section for different resolutions: (a) cross-sectional velocity profile and (b) local enlargement of the lower-wall sublayer-node distribution, plotted using the same vertical coordinate as in panel (a).
	}
	\label{fig-wavy-sph-converge-vel-tke}
\end{figure}

\begin{figure}[htb!]
	\centering
	\includegraphics[trim = 3.03cm 0cm 0cm 1.76cm, clip,width=1.0\textwidth]{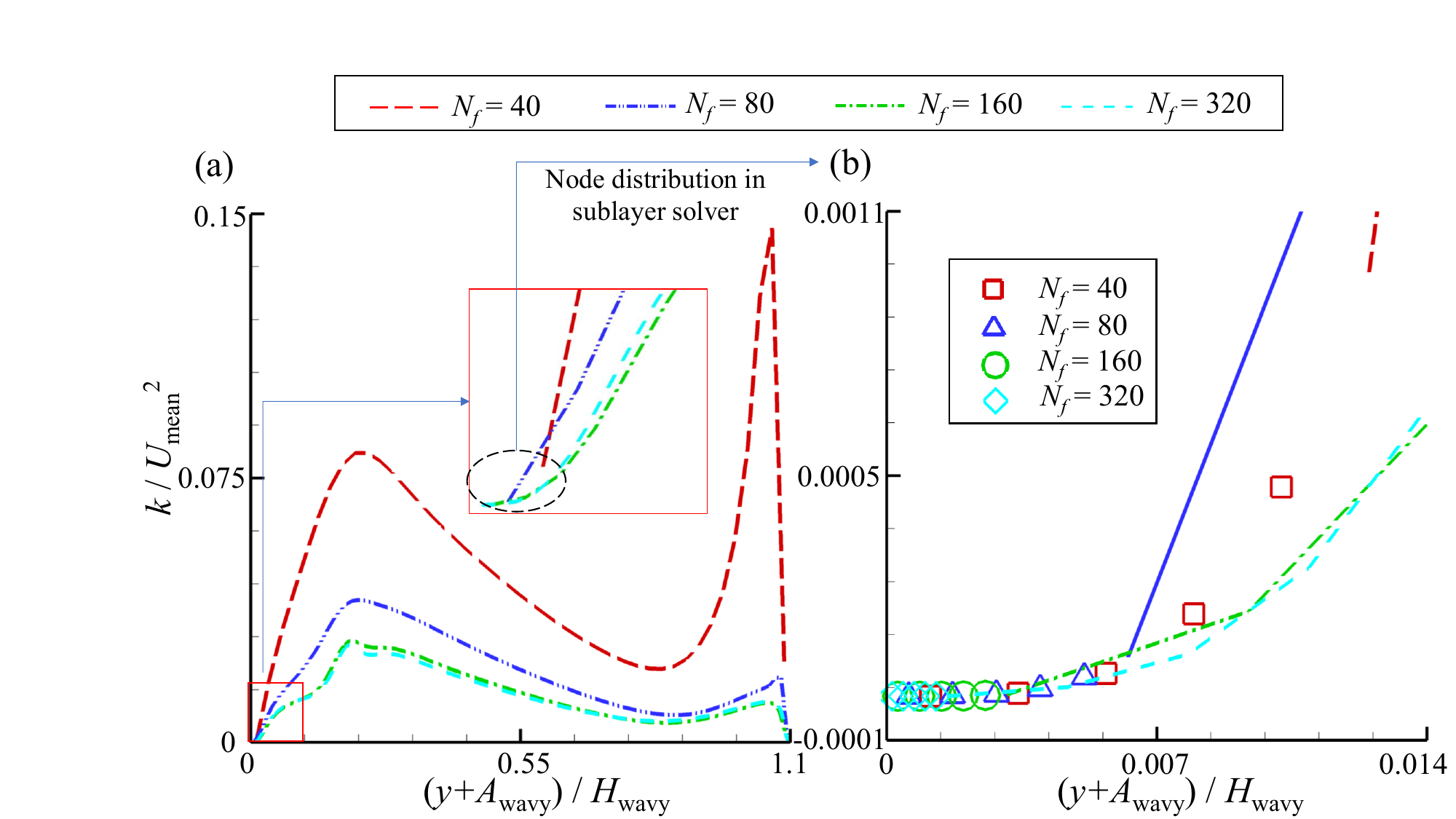}
	\caption{
		Wavy channel: turbulent kinetic energy profiles predicted by the SPH method with the proposed multi-scale approach at the trough section for different resolutions: (a) cross-sectional turbulent kinetic energy profile and (b) local enlargement of the lower-wall sublayer-node distribution, plotted using the same vertical coordinate as in panel (a).
	}
	\label{fig-wavy-sph-converge-tke-node}
\end{figure}

\begin{figure}[htb!]
	\centering
	\includegraphics[trim = 3.36cm 0cm 0cm 1.76cm, clip,width=1.0\textwidth]{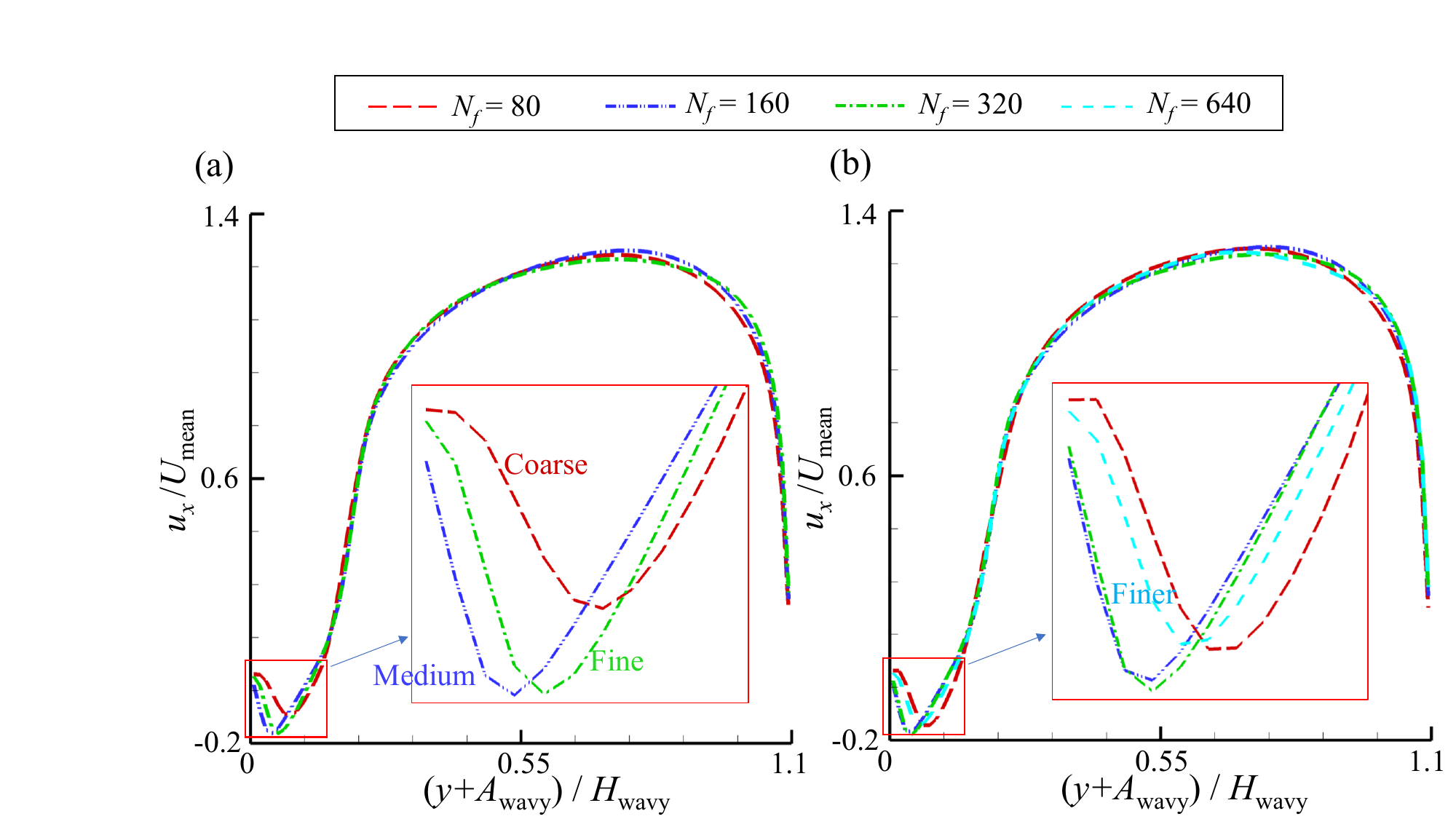}
	\caption{
		Wavy channel: streamwise velocity profiles at the trough section obtained from FVM simulations with uniform grids at different resolutions: (a) unstructured grid and (b) structured grid. The enlarged views highlight the near-wall reversed-flow region.
	}
	\label{fig-wavy-OF-uniform-bad-converge-vel}
\end{figure}

\begin{figure}[htb!]
	\centering
	\includegraphics[trim = 3.36cm 0cm 0cm 1.76cm, clip,width=1.0\textwidth]{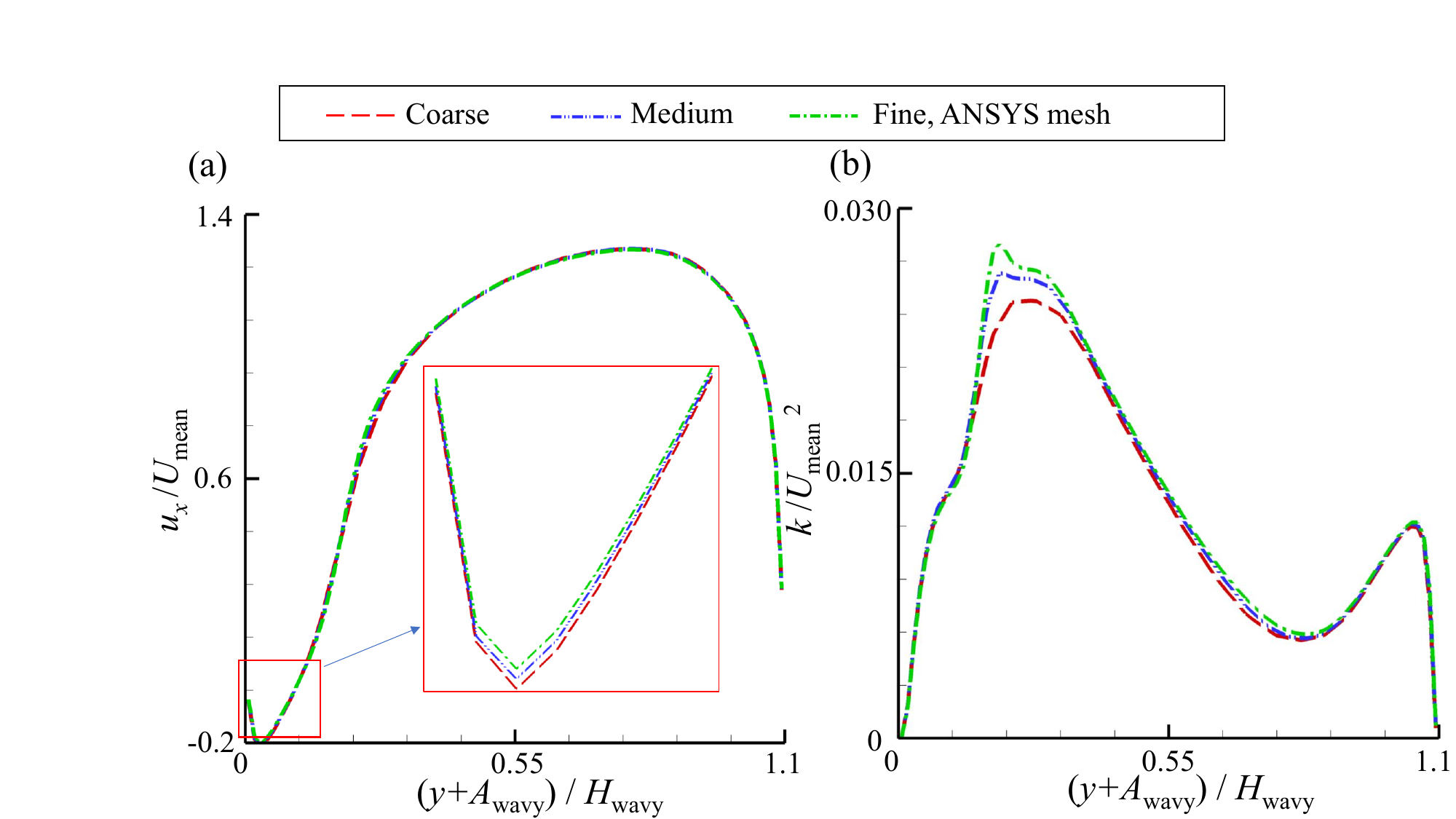}
	\caption{
		Wavy channel: streamwise velocity and turbulent kinetic energy profiles at the trough section obtained from FVM simulations with locally refined structured grids at different refinement levels: (a) streamwise velocity profile and (b) turbulent kinetic energy profile. The fine-grid result is computed using OpenFOAM on the mesh
		provided in the ANSYS benchmark~\cite{ansys_vmfl012_wavy_channel}.
	}
	\label{fig-wavy-OF-partial-converge-vel-tke}
\end{figure}

Third, the quantitative comparisons at different resolutions are presented in
Figs. \ref{fig-wavy-sph-crest-converge-vel-node},
\ref{fig-wavy-sph-crest-converge-tke-node},
\ref{fig-wavy-sph-converge-vel-tke},
\ref{fig-wavy-sph-converge-tke-node},
\ref{fig-wavy-OF-uniform-bad-converge-vel}, and
\ref{fig-wavy-OF-partial-converge-vel-tke}.
These results include the proposed SPH approach and the FVM simulations based
on the three grid types introduced above.
As indicated in Fig. \ref{fig-wavy-geo}, the observation lines
are located at the crest and trough sections, respectively.

To systematically assess the convergence of the proposed SPH method with
the multi-scale approach, four resolutions are considered, and both the SPH
particle values and the sublayer-node values are presented.
At the crest section, both the velocity and $k$ profiles exhibit satisfactory
convergence trends, as shown in
Figs. \ref{fig-wavy-sph-crest-converge-vel-node} and
\ref{fig-wavy-sph-crest-converge-tke-node}.
When the resolution is increased from $N_f=160$ to $N_f=320$,
only minor differences remain in the high-velocity region near the upper wall
and in the near-wall $k$ region close to the lower wall.
The profiles in the remaining regions almost collapse onto each other,
indicating that the solution at $N_f=320$ can be regarded as converged.
In addition, at the crest section, the velocity profile remains free of
kink-like behavior even at the lowest resolution, $N_f=40$.
The sublayer-node values also exhibit good convergence trends for both the
velocity and $k$ profiles, as demonstrated in subfigures (a) and (c).

For the SPH results at the trough section, the presence of flow separation
makes it more challenging to accurately reproduce the profiles than at the
crest section.
In particular, convergence in the near-wall reversed-flow region is more
difficult to achieve.
However, as shown in Fig. \ref{fig-wavy-sph-converge-vel-tke}, with the multi-scale approach, an obvious monotonic converging behavior is observed even in the separated flow area, and the negative velocity is well reproduced.

It is also worth noting that, at the lowest resolution, $N_f=40$,
the reconstructed sublayer velocity has a negative
streamwise component near the wall, indicating the
presence of local reversed flow.
Although this sublayer response provides a correction tendency toward the
reversed-flow solution, the resolved SPH field remains too coarse to fully
capture the corresponding near-wall velocity variation, and the kink-like behavior appears.
As the resolution is increased to $N_f=80$, the kink-like behavior disappears,
and a smooth transition is obtained between the SPH particle values and the
sublayer-node values, as presented in Fig. \ref{fig-wavy-sph-converge-vel-tke} (b).

The $k$ profiles show a similarly satisfactory monotonic convergence trend,
as shown in Fig. \ref{fig-wavy-sph-converge-tke-node}.
Although the $k$ value is substantially overestimated at the lowest resolution,
the profiles become consistent at $N_f=320$.
Moreover, no kink-like behavior is observed for any of the tested resolutions.
Considering the above results, the velocity and $k$ profiles show only minor
changes when the resolution is further increased to $N_f=320$.
Although small differences remain locally, especially in the near-wall
regions, the solution at $N_f=320$ can be regarded as sufficiently converged
for the purpose of the present comparison.

After assessing the convergence behavior of the proposed SPH approach,
the corresponding FVM results are examined for comparison.
To keep the discussion concise, only the trough-section profiles are presented,
since this region contains near-wall reversed flow and is therefore the most
challenging part of the wavy-channel benchmark.
For the uniform unstructured grids shown in
Fig. \ref{fig-wavy-OF-uniform-bad-converge-vel}(a),
the velocity profiles do not exhibit a monotonic convergence trend as the
resolution is increased.
In particular, the prediction of the near-wall reversed-flow region remains
highly sensitive to the grid resolution, suggesting that convergence is
difficult to achieve using uniform unstructured grids for this case.

For the uniform structured grids shown in
Fig. \ref{fig-wavy-OF-uniform-bad-converge-vel}(b),
the profiles obtained at $N_f=160$ and $N_f=320$ appear to be close to each
other, which may suggest an apparent convergence trend.
However, comparison with the experimental data indicates that the
$N_f=320$ result still shows a noticeable discrepancy.
Therefore, an additional refinement test at $N_f=640$ is performed to examine
whether this discrepancy can be reduced by further increasing the uniform
grid resolution.
The $N_f=640$ result does not improve the agreement in a monotonic manner and
instead deviates from the intermediate-resolution results, particularly in
the near-wall reversed-flow region.
This suggests that, for the present separated-flow case, simply increasing
the resolution of a uniform structured grid does not provide a reliable
convergence path.

For the locally refined structured grids, the corresponding results are shown
in Fig. \ref{fig-wavy-OF-partial-converge-vel-tke}.
Compared with the uniform-grid results, the locally refined structured grids
exhibit a much clearer convergence behavior for both the velocity and $k$
profiles at the trough section.
Although small differences remain, especially in the near-wall reversed-flow
region and around the peak of the $k$ profile, the medium-grid result is generally close to the fine-grid
OpenFOAM solution.
Therefore, the locally refined FVM results can be regarded as practically
converged for the present comparison.

Overall, the convergence assessments show that satisfactory convergence can
be obtained by both the proposed multi-scale SPH approach and the locally
refined FVM simulation.
Nevertheless, the locally refined FVM result is strongly dependent on a
problem-specific mesh design, including the refinement region, near-wall
spacing, and grid-stretching distribution.
These choices usually require prior knowledge of the separated-flow structure
and considerable mesh-generation experience.
In contrast, the proposed SPH approach achieves comparable convergence with
an approximately uniform particle discretization, thus reducing the reliance
on case-specific discretization design.

\begin{figure}[htb!]
	\centering
	\includegraphics[trim = 3.36cm 0cm 0cm 1.76cm, clip,width=1.0\textwidth]{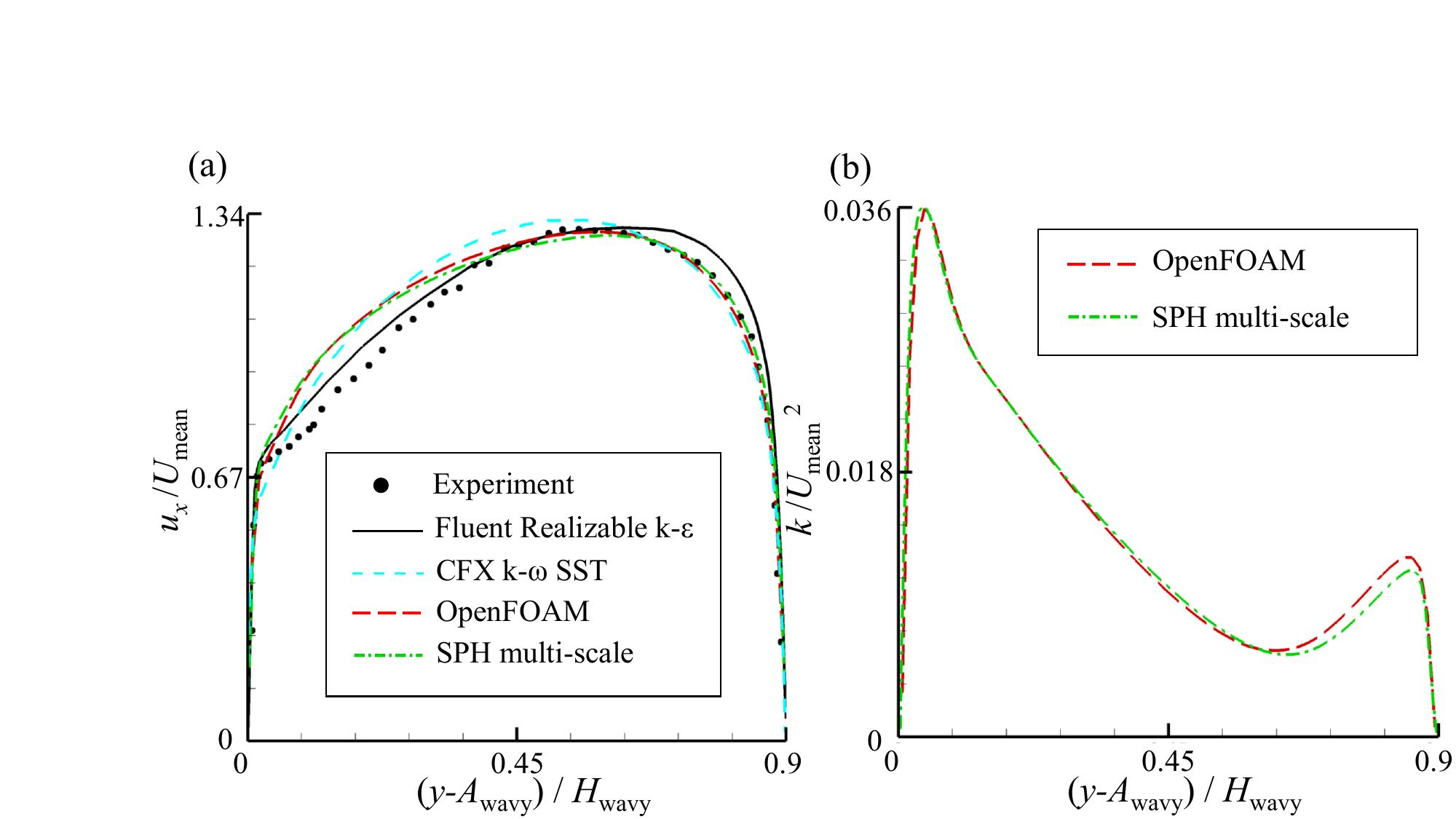}
	\caption{
		Wavy channel: comparison of profiles at the crest section: (a) streamwise velocity profile compared with experimental data~\cite{kuzan1986velocity}, ANSYS Fluent realizable $k$--$\varepsilon$ result~\cite{ansys_vmfl012_wavy_channel}, ANSYS CFX $k$--$\omega$ SST result~\cite{ansys_vmfl012_wavy_channel}, OpenFOAM $k$--$\omega$ 2006 result, and the SPH result with the proposed multi-scale approach; and (b) turbulent kinetic energy profile compared between OpenFOAM and the SPH result with the proposed multi-scale approach.
	}
	\label{fig-wavy-OF-SPH-crest-vel-k-comp}
\end{figure}

Fourth, the converged SPH results obtained using the proposed multi-scale
approach are compared with the OpenFOAM results, the experimental
data~\cite{kuzan1986velocity}, and the ANSYS reference
results~\cite{ansys_vmfl012_wavy_channel}.
As shown in Fig. \ref{fig-wavy-OF-SPH-crest-vel-k-comp}, the proposed SPH
approach and OpenFOAM give very similar predictions at the crest section.
For the streamwise velocity profile, both methods reproduce the overall
experimental trend well, including the velocity increase from the lower wall
and the smooth variation toward the upper wall.
The two methods also show close agreement in the turbulent kinetic energy
profile, although no experimental or ANSYS reference data for $k$ are available
for this comparison.
Near the upper wall, the SPH and OpenFOAM results agree closely with the
experimental data and show a level of agreement comparable to the CFX
$k$--$\omega$ SST reference result.
In contrast, the Fluent realizable $k$--$\varepsilon$ result over-predicts the
velocity in this region.
Near the wavy wall, the numerical results generally over-predict the
velocity compared with the experimental data, while the Fluent realizable $k$--$\varepsilon$ result appears to be closer to the experimental measurements.

\begin{figure}[htb!]
	\centering
	\includegraphics[trim = 3.36cm 0cm 0cm 0.9cm, clip,width=1.0\textwidth]{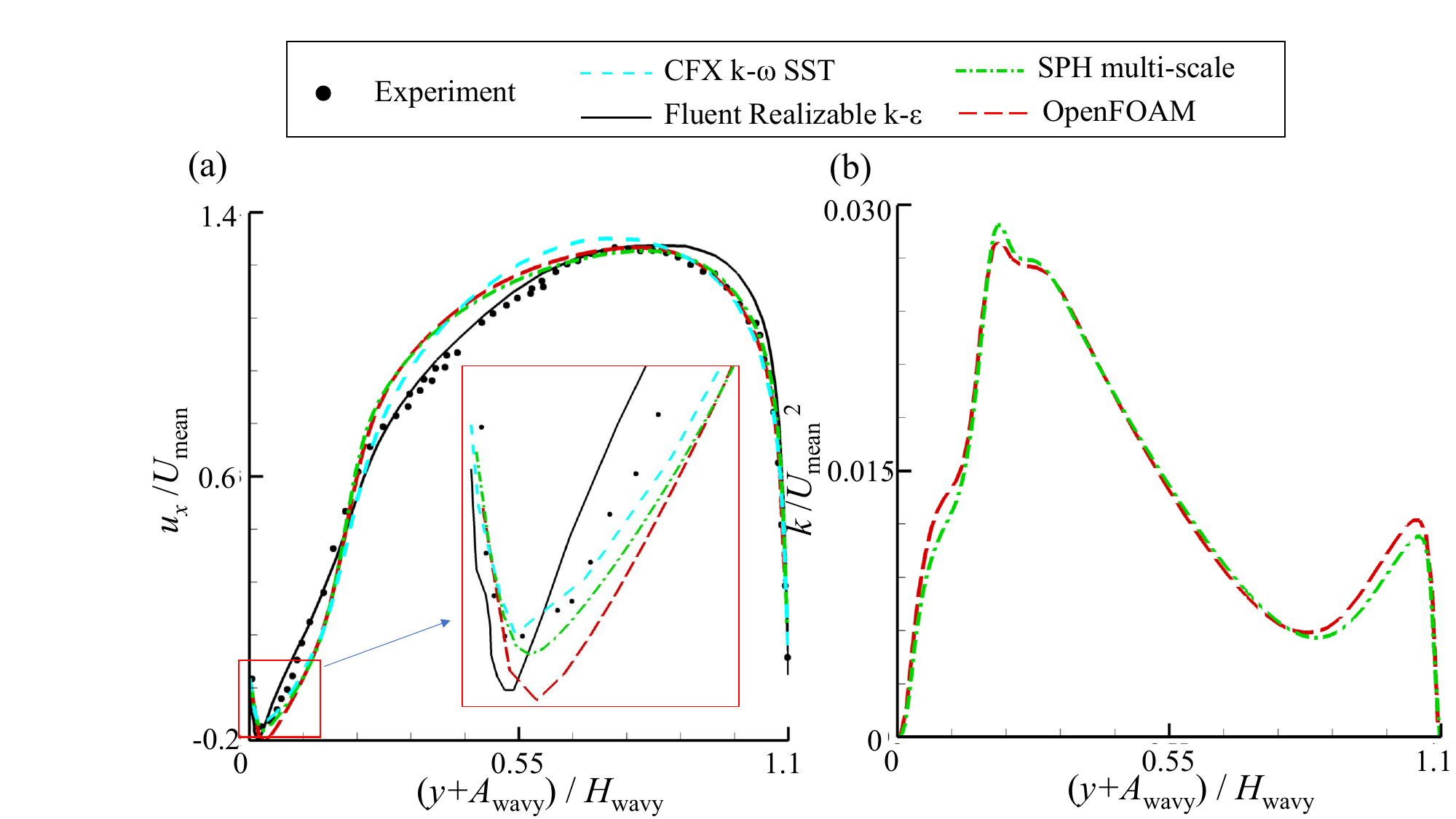}
	\caption{
		Wavy channel: comparison of profiles at the trough section: (a) streamwise velocity profile, including a local enlargement of the near-wall reversed-flow region, compared with experimental data~\cite{kuzan1986velocity}, ANSYS Fluent realizable $k$--$\varepsilon$ result~\cite{ansys_vmfl012_wavy_channel}, ANSYS CFX $k$--$\omega$ SST result~\cite{ansys_vmfl012_wavy_channel}, OpenFOAM $k$--$\omega$ 2006 result, and the SPH result with the proposed multi-scale approach; and (b) turbulent kinetic energy profile compared between OpenFOAM and the SPH result with the proposed multi-scale approach.
	}
	\label{fig-wavy-OF-SPH-trough-vel-k-comp}
\end{figure}

The comparison is then extended to the trough section, as shown in
Fig. \ref{fig-wavy-OF-SPH-trough-vel-k-comp}.
The proposed SPH approach and OpenFOAM show very close agreement for both
the streamwise velocity and turbulent kinetic energy profiles.
For the streamwise velocity profile, the $k$--$\omega$-based results,
including SPH, OpenFOAM, and the CFX $k$--$\omega$ SST reference result,
agree better with the experimental data near the upper wall.
However, in the lower-wall shear layer, where the velocity rapidly recovers
from the reversed-flow region, the Fluent realizable $k$--$\varepsilon$ result
is closer to the experimental measurements.
By contrast, the $k$--$\omega$-based results tend to predict a steeper velocity
variation in this region.
In the reversed-flow region shown in the local enlargement, the SPH and CFX
$k$--$\omega$ SST results lie on opposite sides of the experimental data,
with the CFX result above and the SPH result below the measurements, while
both remain relatively close to the experimental data.

Finally, Table~\ref{tab-wavy-cpu-time} summarizes the computational cost of the
wavy-channel simulations with and without the proposed sublayer model.
The higher relative cost at $N_f=160$ may be attributed to the increased
number of wall-adjacent particles, which requires more local sublayer solves.
Additionally, as shown in Fig.~\ref{fig-wavy-sph-converge-vel-tke}, the recirculation region becomes better resolved at $N_f=160$, producing more complex near-wall flow conditions and potentially more demanding local sublayer solves.

However, this increase in relative computational cost does not persist at
$N_f=320$. Although the number of local sublayer solves increases with the
number of wall-adjacent particles, the total number of fluid particles grows
much more rapidly, substantially reducing the proportion of wall-adjacent
particles. Consequently, full-domain SPH operations, particularly particle
neighbor searches and configuration updates, dominate the overall
computational cost, reducing the relative overhead of the sublayer model
to approximately $1.8\%$.
Since wall-adjacent particles typically account for only a small fraction of
the total fluid particles in engineering-scale simulations, the proposed
method is expected to introduce only limited additional computational
overhead.

\begin{table}[htbp]
	\centering
	\small
	\setlength{\tabcolsep}{5pt}
	\caption{
		Wall-clock time for simulating the turbulent flow in the wavy channel
		at three spatial resolutions with and without the proposed approach.
		All simulations are advanced to 300 seconds.
		All simulations were performed on the same computer equipped with an
		Intel Core Ultra 9 275HX processor.
	}
	\label{tab-wavy-cpu-time}
	\begin{tabular}{@{}lccc@{}}
		\toprule
		Resolution $N_f$
		 & 80     & 160     & 320      \\
		\midrule
		Total fluid particles
		 & 6400   & 25600   & 102400   \\
		Wall-adjacent fluid particles
		 & 164    & 328     & 663      \\
		Wall-adjacent particle fraction (\%)
		 & 2.56   & 1.28    & 0.65     \\
		\midrule
		Without sublayer model (s)
		 & 231.87 & 1303.09 & 22541.97 \\
		With sublayer model (s)
		 & 246.64 & 1466.80 & 22938.16 \\
		Relative computational cost (With/Without)
		 & 1.064  & 1.126   & 1.018    \\
		\bottomrule
	\end{tabular}
\end{table}


\section{Conclusion}
\label{section-conclusion}

In this work, a multi-scale near-wall approach is developed for wall-bounded WCSPH--RANS simulations.
A local 1D sublayer solver is coupled with each wall-adjacent fluid particle to solve the simplified steady $k$--$\omega$ equations.
The formulation builds on the established WCSPH--RANS framework,
with supporting numerical techniques including the de-noised TVF
technique, the enhanced ARD scheme, boundary-offset technique,
and dual-criteria time-stepping scheme to improve stability
and efficiency.
This approach enhances the near-wall resolving capability without
refining the SPH particle distribution or introducing the additional
time-step restriction associated with near-wall particle refinement.

First, for the 1D sublayer solver, an additional constraint on
local flow rate is introduced to close the under-determined system.
A friction-velocity-based fixed-point iteration scheme is further
developed to satisfy this constraint while adaptively determining
the local pressure gradient. This scheme improves numerical
stability and avoids the additional discretization errors associated
with back-calculating the friction velocity from the pressure gradient.

Second, for the coupling between the two scales, corrections to
both the wall and inner shear stresses in the SPH formulation
are developed to incorporate the sublayer solution into the
SPH-scale momentum balance.
The associated local feedback mechanism improves coupling
consistency and partially compensates for errors in the estimated
local flow rate.

The proposed approach is evaluated using four cases:
turbulent straight-channel flows at three representative and
challenging Reynolds numbers, and a turbulent wavy-channel flow
involving separation and recirculation.

For the straight channel at $Re = 5714$, the friction coefficient
converges at $N_f = 40$, whereas the original WCSPH--RANS method
still exhibits resolution dependence at $N_f = 160$.
Comparable velocity predictions are therefore obtained with
a fourfold reduction in wall-normal particle resolution.

At the practical Reynolds number of $Re = 40{,}000$, the proposed
approach achieves monotonic convergence even when wall-adjacent
SPH particles lie within the buffer layer, where the tested
conventional wall treatments exhibit non-monotonic convergence.
The converged profiles agree well with the reference solution.
At an intermediate resolution, the sharper near-wall turbulent
kinetic energy profile also suggests that the Lagrangian
formulation may alleviate the excessive smoothing observed
in the one-dimensional $k$--$\omega$ solution.

At $Re = 8.0 \times 10^{7}$, the proposed approach yields
convergent near-wall predictions, overcoming the persistent
near-wall resolution sensitivity of the original WCSPH--RANS
method. The friction coefficient differs from the NASA TMR
reference by only $0.53\%$ at $N_f = 320$, without near-wall
particle refinement. By comparison, the NASA reference grid
uses a maximum-to-minimum wall-normal cell-size ratio of
approximately $2.56 \times 10^{5}$.

For the wavy channel, the proposed approach captures a coherent
recirculation structure already at $N_f = 40$ and achieves
satisfactory convergence upon refinement.
The converged profiles agree closely with the OpenFOAM results
and reproduce the main experimental velocity trends.
Convergence comparable to that of the locally refined FVM
simulation is obtained with an approximately uniform particle
distribution, reducing the need for case-specific near-wall
discretization design.

Additionally, the sublayer model introduces little computational
overhead in the baseline straight-channel case (approximately
$1\%$ or less) and in the wavy-channel case at $N_f = 320$
(approximately $1.8\%$).
As the proportion of wall-adjacent particles decreases with
increasing resolution, the relative overhead is expected to
become less significant, supporting the applicability of the
approach to engineering-scale simulations.

Although the present validation is limited to two-dimensional
flows, the local and independent formulation of each sublayer
solver facilitates application to complex geometries and
extension to three dimensions.
Future work will investigate three-dimensional turbulent flows
and fluid--structure interactions.

\clearpage
\appendix
\section{Turbulence-model coefficients}
\label{appendix}

\begin{table}[htbp]
	\scriptsize
	\centering
	\caption{Coefficients for the $k-\omega$ RANS model, in which \( \sigma_d = \sigma_{d_0} \) if \( \nabla k \cdot \nabla \omega > 0 \), and \( \sigma_d = 0 \) otherwise.}
	\begin{tabularx}{8.5cm}{@{\extracolsep{\fill}}lc}
		\hline
		Name of the coefficients & Value    \\
		\hline
		$\beta^*$                & $0.09$   \\
		\hline
		$\sigma^*$               & $0.6$    \\
		\hline
		$\alpha$                 & $0.52$   \\
		\hline
		$\beta$                  & $0.0708$ \\
		\hline
		$\sigma$                 & $0.5$    \\
		\hline
		$\sigma_{d_0}$           & $0.125$  \\
		\hline
		$C_{\mathrm{lim}}$       & $0.875$  \\
		\hline
	\end{tabularx}
	\label{tab-coeff-kw}
\end{table}

\bibliographystyle{elsarticle-num}
\bibliography{p_refinement}

\begin{thebibliography}{10}
\expandafter\ifx\csname url\endcsname\relax
  \def\url#1{\texttt{#1}}\fi
\expandafter\ifx\csname urlprefix\endcsname\relax\def\urlprefix{URL }\fi
\expandafter\ifx\csname href\endcsname\relax
  \def\href#1#2{#2} \def\path#1{#1}\fi

\bibitem{nguyen2020dual}
P.~T. Nguyen, J.~C. Uribe, I.~Afgan, D.~R. Laurence, A dual-grid hybrid
  {RANS}/{LES} model for under-resolved near-wall regions and its application
  to heated and separating flows, Flow, turbulence and combustion 104~(4)
  (2020) 835--859.
\newblock \href {https://doi.org/10.1007/s10494-019-00070-8}
  {\path{doi:10.1007/s10494-019-00070-8}}.

\bibitem{ecca2018viscous}
L.~E{\c{c}}a, F.~Pereira, G.~Vaz, Viscous flow simulations at high {Reynolds}
  numbers without wall functions: Is y+ $\approx$ 1 enough for the near-wall
  cells?, Computers \& Fluids 170 (2018) 157--175.
\newblock \href {https://doi.org/10.1016/j.compfluid.2018.04.035}
  {\path{doi:10.1016/j.compfluid.2018.04.035}}.

\bibitem{wang6271943unified}
F.~Wang, X.~Hu, A unified {Riemann}-based k-omega {WCSPH} framework for
  simulating turbulent flow with separation, Available at SSRN 6271943,
  doi:10.2139/ssrn.6271943.

\bibitem{gingold1977smoothed}
R.~A. Gingold, J.~J. Monaghan, Smoothed particle hydrodynamics: theory and
  application to non-spherical stars, Monthly notices of the royal astronomical
  society 181~(3) (1977) 375--389.
\newblock \href {https://doi.org/10.1093/mnras/181.3.375}
  {\path{doi:10.1093/mnras/181.3.375}}.

\bibitem{lucy1977numerical}
L.~B. {Lucy}, A numerical approach to the testing of the fission hypothesis,
  The astronomical journal 82 (1977) 1013--1024.
\newblock \href {https://doi.org/10.1086/112164} {\path{doi:10.1086/112164}}.

\bibitem{koshizuka1996moving}
S.~Koshizuka, Y.~Oka, Moving-particle semi-implicit method for fragmentation of
  incompressible fluid, Nuclear science and engineering 123~(3) (1996)
  421--434.
\newblock \href {https://doi.org/10.13182/nse96-a24205}
  {\path{doi:10.13182/nse96-a24205}}.

\bibitem{khayyer2019multi}
A.~Khayyer, N.~Tsuruta, Y.~Shimizu, H.~Gotoh, Multi-resolution {MPS} for
  incompressible fluid-elastic structure interactions in ocean engineering,
  Applied Ocean Research 82 (2019) 397--414.
\newblock \href {https://doi.org/10.1016/j.apor.2018.10.020}
  {\path{doi:10.1016/j.apor.2018.10.020}}.

\bibitem{tang2016numerical}
Z.~Tang, D.~Wan, G.~Chen, Q.~Xiao, Numerical simulation of {3D} violent
  free-surface flows by multi-resolution {MPS} method, Journal of ocean
  engineering and marine energy 2~(3) (2016) 355--364.
\newblock \href {https://doi.org/10.1007/s40722-016-0062-6}
  {\path{doi:10.1007/s40722-016-0062-6}}.

\bibitem{tanaka2018multi}
M.~Tanaka, R.~Cardoso, H.~Bahai, Multi-resolution {MPS} method, Journal of
  computational physics 359 (2018) 106--136.
\newblock \href {https://doi.org/10.1016/j.jcp.2017.12.042}
  {\path{doi:10.1016/j.jcp.2017.12.042}}.

\bibitem{shibata2017overlapping}
K.~Shibata, S.~Koshizuka, T.~Matsunaga, I.~Masaie, The overlapping particle
  technique for multi-resolution simulation of particle methods, Computer
  Methods in Applied Mechanics and Engineering 325 (2017) 434--462.
\newblock \href {https://doi.org/10.1016/j.cma.2017.06.030}
  {\path{doi:10.1016/j.cma.2017.06.030}}.

\bibitem{khorasanizade2016dynamic}
S.~Khorasanizade, J.~Sousa, Dynamic flow-based particle splitting in smoothed
  particle hydrodynamics, International Journal for Numerical Methods in
  Engineering 106~(5) (2016) 397--410.
\newblock \href {https://doi.org/10.1002/nme.5128}
  {\path{doi:10.1002/nme.5128}}.

\bibitem{rahmani2025anisotropic}
S.~Rahmani, J.~Baiges, J.~Principe, Anisotropic variational mesh adaptation for
  embedded finite element methods, Computer Methods in Applied Mechanics and
  Engineering 433 (2025) 117504.
\newblock \href {https://doi.org/10.1016/j.cma.2024.117504}
  {\path{doi:10.1016/j.cma.2024.117504}}.

\bibitem{chiron2018analysis}
L.~Chiron, G.~Oger, M.~De~Leffe, D.~Le~Touz{\'e}, Analysis and improvements of
  adaptive particle refinement ({APR}) through {CPU} time, accuracy and
  robustness considerations, Journal of Computational Physics 354 (2018)
  552--575.
\newblock \href {https://doi.org/10.1016/j.jcp.2017.10.041}
  {\path{doi:10.1016/j.jcp.2017.10.041}}.

\bibitem{barcarolo2014adaptive}
D.~A. Barcarolo, D.~Le~Touz{\'e}, G.~Oger, F.~De~Vuyst, Adaptive particle
  refinement and derefinement applied to the smoothed particle hydrodynamics
  method, Journal of Computational Physics 273 (2014) 640--657.
\newblock \href {https://doi.org/10.1016/j.jcp.2014.05.040}
  {\path{doi:10.1016/j.jcp.2014.05.040}}.

\bibitem{vacondio2012accurate}
R.~Vacondio, B.~Rogers, P.~K. Stansby, Accurate particle splitting for smoothed
  particle hydrodynamics in shallow water with shock capturing, International
  Journal for Numerical Methods in Fluids 69~(8) (2012) 1377--1410.
\newblock \href {https://doi.org/10.1002/fld.2646}
  {\path{doi:10.1002/fld.2646}}.

\bibitem{reyes2013dynamic}
Y.~Reyes~L{\'o}pez, D.~Roose, C.~Recarey~Morfa, Dynamic particle refinement in
  {SPH}: application to free surface flow and non-cohesive soil simulations,
  Computational Mechanics 51~(5) (2013) 731--741.
\newblock \href {https://doi.org/10.1007/s00466-012-0748-0}
  {\path{doi:10.1007/s00466-012-0748-0}}.

\bibitem{omidvar2012wave}
P.~Omidvar, P.~K. Stansby, B.~D. Rogers, Wave body interaction in {2D} using
  smoothed particle hydrodynamics ({SPH}) with variable particle mass,
  International journal for numerical methods in fluids 68~(6) (2012) 686--705.
\newblock \href {https://doi.org/10.1002/fld.2528}
  {\path{doi:10.1002/fld.2528}}.

\bibitem{bonet2005hamiltonian}
J.~Bonet, M.~X. Rodr{\'\i}guez-Paz, {Hamiltonian} formulation of the variable-h
  {SPH} equations, Journal of Computational Physics 209~(2) (2005) 541--558.
\newblock \href {https://doi.org/10.1016/j.jcp.2005.03.030}
  {\path{doi:10.1016/j.jcp.2005.03.030}}.

\bibitem{zhang2025multi}
Y.~Zhang, J.~Pan, M.~Song, H.~Jiang, F.~He, C.~Huang, A.~Shakibaeinia, A
  multi-phase {SPH} model for simulating the floating {OWC}-breakwater
  integrated systems, Coastal Engineering 197 (2025) 104658.
\newblock \href {https://doi.org/10.1016/j.coastaleng.2024.104658}
  {\path{doi:10.1016/j.coastaleng.2024.104658}}.

\bibitem{yang2021smoothed}
X.~Yang, S.-C. Kong, M.~Liu, Q.~Liu, Smoothed particle hydrodynamics with
  adaptive spatial resolution ({SPH}-{ASR}) for free surface flows, Journal of
  Computational Physics 443 (2021) 110539.
\newblock \href {https://doi.org/10.1016/j.jcp.2021.110539}
  {\path{doi:10.1016/j.jcp.2021.110539}}.

\bibitem{zhang2024numerical}
G.~Zhang, G.~Liang, X.~Yang, Z.~Zhang, Numerical investigations on water entry
  and/or exit problems using a multi-resolution delta-plus-{SPH} model with
  {TIC}, Ocean Engineering 292 (2024) 116560.
\newblock \href {https://doi.org/10.1016/j.oceaneng.2023.116560}
  {\path{doi:10.1016/j.oceaneng.2023.116560}}.

\bibitem{yang2019adaptive}
X.~Yang, S.-C. Kong, Adaptive resolution for multiphase smoothed particle
  hydrodynamics, Computer Physics Communications 239 (2019) 112--125.
\newblock \href {https://doi.org/10.1016/j.cpc.2019.01.002}
  {\path{doi:10.1016/j.cpc.2019.01.002}}.

\bibitem{sun2019study}
P.~Sun, D.~Le~Touz{\'e}, A.-M. Zhang, Study of a complex fluid-structure
  dam-breaking benchmark problem using a multi-phase {SPH} method with {APR},
  Engineering Analysis with Boundary Elements 104 (2019) 240--258.
\newblock \href {https://doi.org/10.1016/j.enganabound.2019.03.033}
  {\path{doi:10.1016/j.enganabound.2019.03.033}}.

\bibitem{ju2023study}
X.-Y. Ju, Y.-M. Shen, W.-K. Shi, P.-N. Sun, H.~Tang, Study on the effect of
  cavity oscillation on wedge water entry with a multiphase smoothed particle
  hydrodynamics model, Physics of Fluids 35~(11) (2023) 113312.
\newblock \href {https://doi.org/10.1063/5.0174222}
  {\path{doi:10.1063/5.0174222}}.

\bibitem{sun2018multi}
P.~Sun, A.~Colagrossi, S.~Marrone, M.~Antuono, A.~Zhang, Multi-resolution
  delta-plus-{SPH} with tensile instability control: Towards high {Reynolds}
  number flows, Computer Physics Communications 224 (2018) 63--80.
\newblock \href {https://doi.org/10.1016/j.cpc.2017.11.016}
  {\path{doi:10.1016/j.cpc.2017.11.016}}.

\bibitem{vacondio2016variable}
R.~Vacondio, B.~D. Rogers, P.~K. Stansby, P.~Mignosa, Variable resolution for
  {SPH} in three dimensions: Towards optimal splitting and coalescing for
  dynamic adaptivity, Computer Methods in Applied Mechanics and Engineering 300
  (2016) 442--460.
\newblock \href {https://doi.org/10.1016/j.cma.2015.11.021}
  {\path{doi:10.1016/j.cma.2015.11.021}}.

\bibitem{vacondio2013variable}
R.~Vacondio, B.~D. Rogers, P.~K. Stansby, P.~Mignosa, J.~Feldman, Variable
  resolution for {SPH}: a dynamic particle coalescing and splitting scheme,
  Computer Methods in Applied Mechanics and Engineering 256 (2013) 132--148.
\newblock \href {https://doi.org/10.1016/j.cma.2012.12.014}
  {\path{doi:10.1016/j.cma.2012.12.014}}.

\bibitem{feldman2007dynamic}
J.~Feldman, J.~Bonet, Dynamic refinement and boundary contact forces in {SPH}
  with applications in fluid flow problems, International journal for numerical
  methods in engineering 72~(3) (2007) 295--324.
\newblock \href {https://doi.org/10.1002/nme.2010}
  {\path{doi:10.1002/nme.2010}}.

\bibitem{xiong2013gpu}
Q.~Xiong, B.~Li, J.~Xu, {GPU}-accelerated adaptive particle splitting and
  merging in {SPH}, Computer Physics Communications 184~(7) (2013) 1701--1707.
\newblock \href {https://doi.org/10.1016/j.cpc.2013.02.021}
  {\path{doi:10.1016/j.cpc.2013.02.021}}.

\bibitem{hu2017consistent}
W.~Hu, W.~Pan, M.~Rakhsha, Q.~Tian, H.~Hu, D.~Negrut, A consistent
  multi-resolution smoothed particle hydrodynamics method, Computer Methods in
  Applied Mechanics and Engineering 324 (2017) 278--299.
\newblock \href {https://doi.org/10.1016/j.cma.2017.06.010}
  {\path{doi:10.1016/j.cma.2017.06.010}}.

\bibitem{bian2015multi}
X.~Bian, Z.~Li, G.~E. Karniadakis, Multi-resolution flow simulations by
  smoothed particle hydrodynamics via domain decomposition, Journal of
  Computational Physics 297 (2015) 132--155.
\newblock \href {https://doi.org/10.1016/j.jcp.2015.04.044}
  {\path{doi:10.1016/j.jcp.2015.04.044}}.

\bibitem{muta2022efficient}
A.~Muta, P.~Ramachandran, Efficient and accurate adaptive resolution for
  weakly-compressible {SPH}, Computer Methods in Applied Mechanics and
  Engineering 395 (2022) 115019.
\newblock \href {https://doi.org/10.1016/j.cma.2022.115019}
  {\path{doi:10.1016/j.cma.2022.115019}}.

\bibitem{gao2023multi}
T.~Gao, H.~Qiu, L.~Fu, Multi-level adaptive particle refinement method with
  large refinement scale ratio and new free-surface detection algorithm for
  complex fluid-structure interaction problems, Journal of Computational
  Physics 473 (2023) 111762.
\newblock \href {https://doi.org/10.1016/j.jcp.2022.111762}
  {\path{doi:10.1016/j.jcp.2022.111762}}.

\bibitem{gao2022block}
T.~Gao, H.~Qiu, L.~Fu, A block-based adaptive particle refinement {SPH} method
  for fluid--structure interaction problems, Computer Methods in Applied
  Mechanics and Engineering 399 (2022) 115356.
\newblock \href {https://doi.org/10.1016/j.cma.2022.115356}
  {\path{doi:10.1016/j.cma.2022.115356}}.

\bibitem{kitsionas2002smoothed}
S.~Kitsionas, A.~Whitworth, Smoothed particle hydrodynamics with particle
  splitting, applied to self-gravitating collapse, Monthly Notices of the Royal
  Astronomical Society 330~(1) (2002) 129--136.
\newblock \href {https://doi.org/10.1046/j.1365-8711.2002.05115.x}
  {\path{doi:10.1046/j.1365-8711.2002.05115.x}}.

\bibitem{bate1995modelling}
M.~R. Bate, I.~A. Bonnell, N.~M. Price, Modelling accretion in protobinary
  systems, Monthly Notices of the Royal Astronomical Society 277~(2) (1995)
  362--376.
\newblock \href {https://doi.org/10.1093/mnras/277.2.362}
  {\path{doi:10.1093/mnras/277.2.362}}.

\bibitem{meglicki19933d}
Z.~Meglicki, D.~Wickramasinghe, G.~V. Bicknell, {3D} structure of truncated
  accretion discs in close binaries, Monthly Notices of the Royal Astronomical
  Society 264~(3) (1993) 691--704.
\newblock \href {https://doi.org/10.1093/mnras/264.3.691}
  {\path{doi:10.1093/mnras/264.3.691}}.

\bibitem{monaghan1988dynamics}
J.~Monaghan, S.~Varnas, The dynamics of interstellar cloud complexes, Monthly
  Notices of the Royal Astronomical Society 231~(3) (1988) 515--534.
\newblock \href {https://doi.org/10.1093/mnras/231.3.515}
  {\path{doi:10.1093/mnras/231.3.515}}.

\bibitem{alimi2003smooth}
J.-M. Alimi, A.~Serna, C.~Pastor, G.~Bernabeu, Smooth particle hydrodynamics:
  importance of correction terms in adaptive resolution algorithms, Journal of
  Computational Physics 192~(1) (2003) 157--174.
\newblock \href {https://doi.org/10.1016/s0021-9991(03)00351-6}
  {\path{doi:10.1016/s0021-9991(03)00351-6}}.

\bibitem{owen1998adaptive}
J.~M. Owen, J.~V. Villumsen, P.~R. Shapiro, H.~Martel, Adaptive smoothed
  particle hydrodynamics: Methodology. {II}., The Astrophysical Journal
  Supplement Series 116~(2) (1998) 155--209.
\newblock \href {https://doi.org/10.1086/313100} {\path{doi:10.1086/313100}}.

\bibitem{lastiwka2005adaptive}
M.~Lastiwka, N.~Quinlan, M.~Basa, Adaptive particle distribution for smoothed
  particle hydrodynamics, International Journal for Numerical Methods in Fluids
  47~(10-11) (2005) 1403--1409.
\newblock \href {https://doi.org/10.1002/fld.891} {\path{doi:10.1002/fld.891}}.

\bibitem{shapiro1996adaptive}
P.~R. Shapiro, H.~Martel, J.~V. Villumsen, J.~M. Owen, Adaptive smoothed
  particle hydrodynamics, with application to cosmology: methodology,
  Astrophysical Journal Supplement v. 103, p. 269 103 (1996) 269.
\newblock \href {https://doi.org/10.1086/192279} {\path{doi:10.1086/192279}}.

\bibitem{nelson1994variable}
R.~P. Nelson, J.~C. Papaloizou, Variable smoothing lengths and energy
  conservation in smoothed particle hydrodynamics, Monthly Notices of the Royal
  Astronomical Society 270~(1) (1994) 1--20.
\newblock \href {https://doi.org/10.1093/mnras/270.1.1}
  {\path{doi:10.1093/mnras/270.1.1}}.

\bibitem{borve2005regularized}
S.~B{\o}rve, M.~Omang, J.~Trulsen, Regularized smoothed particle hydrodynamics
  with improved multi-resolution handling, Journal of Computational Physics
  208~(1) (2005) 345--367.
\newblock \href {https://doi.org/10.1016/j.jcp.2005.02.018}
  {\path{doi:10.1016/j.jcp.2005.02.018}}.

\bibitem{liu2006adaptive}
M.~Liu, G.~Liu, K.~Lam, Adaptive smoothed particle hydrodynamics for high
  strain hydrodynamics with material strength, Shock Waves 15~(1) (2006)
  21--29.
\newblock \href {https://doi.org/10.1007/s00193-005-0002-1}
  {\path{doi:10.1007/s00193-005-0002-1}}.

\bibitem{spreng2014local}
F.~Spreng, D.~Schnabel, A.~Mueller, P.~Eberhard, A local adaptive
  discretization algorithm for smoothed particle hydrodynamics: For the
  inaugural issue, Computational Particle Mechanics 1~(2) (2014) 131--145.
\newblock \href {https://doi.org/10.1007/s40571-014-0015-6}
  {\path{doi:10.1007/s40571-014-0015-6}}.

\bibitem{vacondio2013shallow}
R.~Vacondio, B.~Rogers, P.~Stansby, P.~Mignosa, Shallow water {SPH} for
  flooding with dynamic particle coalescing and splitting, Advances in Water
  Resources 58 (2013) 10--23.
\newblock \href {https://doi.org/10.1016/j.advwatres.2013.04.007}
  {\path{doi:10.1016/j.advwatres.2013.04.007}}.

\bibitem{oger2006two}
G.~Oger, M.~Doring, B.~Alessandrini, P.~Ferrant, Two-dimensional {SPH}
  simulations of wedge water entries, Journal of computational physics 213~(2)
  (2006) 803--822.
\newblock \href {https://doi.org/10.1016/j.jcp.2005.09.004}
  {\path{doi:10.1016/j.jcp.2005.09.004}}.

\bibitem{lyu20223d}
H.-G. Lyu, P.-N. Sun, J.-M. Miao, A.-M. Zhang, {3D} multi-resolution {SPH}
  modeling of the water entry dynamics of free-fall lifeboats, Ocean
  Engineering 257 (2022) 111648.
\newblock \href {https://doi.org/10.1016/j.oceaneng.2022.111648}
  {\path{doi:10.1016/j.oceaneng.2022.111648}}.

\bibitem{liang2023pressure}
C.~Liang, W.~Huang, D.~Chen, A pressure-dependent adaptive resolution scheme
  for smoothed particle hydrodynamics simulation of underwater explosion, Ocean
  Engineering 270 (2023) 113695.
\newblock \href {https://doi.org/10.1016/j.oceaneng.2023.113695}
  {\path{doi:10.1016/j.oceaneng.2023.113695}}.

\bibitem{sun2021accurate}
P.-N. Sun, D.~Le~Touz{\'e}, G.~Oger, A.-M. Zhang, An accurate {SPH} volume
  adaptive scheme for modeling strongly-compressible multiphase flows. part 1:
  Numerical scheme and validations with basic {1D} and {2D} benchmarks, Journal
  of Computational Physics 426 (2021) 109937.
\newblock \href {https://doi.org/10.1016/j.jcp.2020.109937}
  {\path{doi:10.1016/j.jcp.2020.109937}}.

\bibitem{sun2021accurateb}
P.-N. Sun, D.~Le~Touz{\'e}, G.~Oger, A.-M. Zhang, An accurate {SPH} volume
  adaptive scheme for modeling strongly-compressible multiphase flows. part 2:
  Extension of the scheme to cylindrical coordinates and simulations of {3D}
  axisymmetric problems with experimental validations, Journal of Computational
  Physics 426 (2021) 109936.
\newblock \href {https://doi.org/10.1016/j.jcp.2020.109936}
  {\path{doi:10.1016/j.jcp.2020.109936}}.

\bibitem{zhuang2020smoothed}
T.~Zhuang, X.~Dong, Smoothed particle hydrodynamics simulation of underwater
  explosions with dynamic particle refinement, AIP Advances 10~(11) (2020)
  115314.
\newblock \href {https://doi.org/10.1063/5.0029472}
  {\path{doi:10.1063/5.0029472}}.

\bibitem{yi2026novel}
C.~Yi, J.~Chen, D.~Feng, A novel {GPU}-accelerated debris flow-turbidity
  currents transition model for simulating underwater sliding impacts on
  deformable pipelines using total and updated {Lagrangian} {WCSPH} method,
  Computers and Geotechnics 192 (2026) 107923.
\newblock \href {https://doi.org/10.1016/j.compgeo.2026.107923}
  {\path{doi:10.1016/j.compgeo.2026.107923}}.

\bibitem{zienkiewicz1983hierarchical}
O.~C. Zienkiewicz, J.~D.~S. Gago, D.~W. Kelly, The hierarchical concept in
  finite element analysis, Computers \& Structures 16~(1-4) (1983) 53--65.
\newblock \href {https://doi.org/10.1016/0045-7949(83)90147-5}
  {\path{doi:10.1016/0045-7949(83)90147-5}}.

\bibitem{hillman2021consistent}
M.~Hillman, K.-C. Lin, Consistent weak forms for meshfree methods: Full
  realization of h-refinement, p-refinement, and a-refinement in strong-type
  essential boundary condition enforcement, Computer Methods in Applied
  Mechanics and Engineering 373 (2021) 113448.
\newblock \href {https://doi.org/10.1016/j.cma.2020.113448}
  {\path{doi:10.1016/j.cma.2020.113448}}.

\bibitem{perazzo2008adaptive}
F.~Perazzo, R.~L{\"o}hner, L.~Perez-Pozo, Adaptive methodology for meshless
  finite point method, Advances in Engineering Software 39~(3) (2008) 156--166.
\newblock \href {https://doi.org/10.1016/j.advengsoft.2007.02.007}
  {\path{doi:10.1016/j.advengsoft.2007.02.007}}.

\bibitem{zhang2019weakly}
C.~Zhang, G.~Xiang, B.~Wang, X.~Hu, N.~A. Adams, A weakly compressible {SPH}
  method with {WENO} reconstruction, Journal of computational physics 392
  (2019) 1--18.
\newblock \href {https://doi.org/10.1016/j.jcp.2019.04.038}
  {\path{doi:10.1016/j.jcp.2019.04.038}}.

\bibitem{renaut2015high}
G.~Renaut, S.~Aubert, J.~Marongiu,
  \href{https://www.euroturbo.eu/paper/ETC2015-077.pdf}{High order {SPH}-{ALE}
  method for hydraulic turbine simulations}, in: 11 th European Conference on
  Turbomachinery Fluid dynamics \& Thermodynamics, EUROPEAN TURBOMACHINERY
  SOCIETY, 2015, pp. 1--11, paper ETC2015-077.
\newline\urlprefix\url{https://www.euroturbo.eu/paper/ETC2015-077.pdf}

\bibitem{kumar2009partition}
M.~Kumar, S.~Chakravorty, P.~Singla, J.~L. Junkins, The partition of unity
  finite element approach with hp-refinement for the stationary
  {Fokker}--{Planck} equation, Journal of Sound and Vibration 327~(1-2) (2009)
  144--162.
\newblock \href {https://doi.org/10.1016/j.jsv.2009.05.033}
  {\path{doi:10.1016/j.jsv.2009.05.033}}.

\bibitem{schweitzer2009adaptive}
M.~A. Schweitzer, An adaptive hp-version of the multilevel particle--partition
  of unity method, Computer methods in applied mechanics and engineering
  198~(13-14) (2009) 1260--1272.
\newblock \href {https://doi.org/10.1016/j.cma.2008.01.009}
  {\path{doi:10.1016/j.cma.2008.01.009}}.

\bibitem{wu2014adaptive}
C.~Wu, D.~Young, H.~Hong, Adaptive meshless local maximum-entropy finite
  element method for convection--diffusion problems, Computational Mechanics
  53~(1) (2014) 189--200.
\newblock \href {https://doi.org/10.1007/s00466-013-0901-4}
  {\path{doi:10.1007/s00466-013-0901-4}}.

\bibitem{wang2025weakly}
F.~Wang, Z.~Sun, X.~Hu, A weakly compressible {SPH} method for {RANS}
  simulation of wall-bounded turbulent flows, Journal of Computational Physics
  547 (2026) 114532.
\newblock \href {https://doi.org/10.1016/j.jcp.2025.114532}
  {\path{doi:10.1016/j.jcp.2025.114532}}.

\bibitem{wang2026effective}
F.~Wang, X.~Hu, An effective implementation of the bidirectional buffer:
  Towards laminar and turbulent open-boundary flows, Computer Physics
  Communications 323 (2026) 110116.
\newblock \href {https://doi.org/10.1016/j.cpc.2026.110116}
  {\path{doi:10.1016/j.cpc.2026.110116}}.

\bibitem{wang2025zero}
F.~Wang, X.~Hu, On zero-order consistency residue and background pressure for
  the conservative {SPH} fluid dynamics, arXiv preprint arXiv:2507.18210
  (2025).
\newblock \href {https://doi.org/10.48550/arXiv.2507.18210}
  {\path{doi:10.48550/arXiv.2507.18210}}.

\bibitem{yu2023level}
Y.~Yu, Y.~Zhu, C.~Zhang, O.~J. Haidn, X.~Hu, Level-set based pre-processing
  techniques for particle methods, Computer Physics Communications 289 (2023)
  108744.
\newblock \href {https://doi.org/10.1016/j.cpc.2023.108744}
  {\path{doi:10.1016/j.cpc.2023.108744}}.

\bibitem{zhu2021cad}
Y.~Zhu, C.~Zhang, Y.~Yu, X.~Hu, A {CAD}-compatible body-fitted particle
  generator for arbitrarily complex geometry and its application to
  wave-structure interaction, Journal of Hydrodynamics 33~(2) (2021) 195--206.
\newblock \href {https://doi.org/10.1007/s42241-021-0031-y}
  {\path{doi:10.1007/s42241-021-0031-y}}.

\bibitem{zhang2025towards}
B.~Zhang, N.~Adams, X.~Hu, Towards high-order consistency and convergence of
  conservative {SPH} approximations, Computer Methods in Applied Mechanics and
  Engineering 433 (2025) 117484.
\newblock \href {https://doi.org/10.1016/j.cma.2024.117484}
  {\path{doi:10.1016/j.cma.2024.117484}}.

\bibitem{ren2023efficient}
Y.~Ren, P.~Lin, C.~Zhang, X.~Hu, An efficient correction method in {Riemann}
  {SPH} for the simulation of general free surface flows, Computer Methods in
  Applied Mechanics and Engineering 417 (2023) 116460.
\newblock \href {https://doi.org/10.1016/j.cma.2023.116460}
  {\path{doi:10.1016/j.cma.2023.116460}}.

\bibitem{wilcox2008formulation}
D.~C. Wilcox, Formulation of the {$k$--$\omega$} turbulence model revisited,
  AIAA Journal 46~(11) (2008) 2823--2838.
\newblock \href {https://doi.org/10.2514/1.36541} {\path{doi:10.2514/1.36541}}.

\bibitem{wilcox1998turbulence}
D.~C. Wilcox, et~al., Turbulence modeling for {CFD}, Vol.~2, DCW industries La
  Canada, CA, 1998.

\bibitem{patankar2018numerical}
S.~Patankar, Numerical heat transfer and fluid flow, CRC press, 2018.
\newblock \href {https://doi.org/10.1201/9781482234213}
  {\path{doi:10.1201/9781482234213}}.

\bibitem{adami2013transport}
S.~Adami, X.~Hu, N.~A. Adams, A transport-velocity formulation for smoothed
  particle hydrodynamics, Journal of Computational Physics 241 (2013) 292--307.
\newblock \href {https://doi.org/10.1016/j.jcp.2013.01.043}
  {\path{doi:10.1016/j.jcp.2013.01.043}}.

\bibitem{zhang2020dual}
C.~Zhang, M.~Rezavand, X.~Hu, Dual-criteria time stepping for weakly
  compressible smoothed particle hydrodynamics, Journal of Computational
  Physics 404 (2020) 109135.
\newblock \href {https://doi.org/10.1016/j.jcp.2019.109135}
  {\path{doi:10.1016/j.jcp.2019.109135}}.

\bibitem{quinlan2006truncation}
N.~J. Quinlan, M.~Basa, M.~Lastiwka, Truncation error in mesh-free particle
  methods, International Journal for Numerical Methods in Engineering 66~(13)
  (2006) 2064--2085.
\newblock \href {https://doi.org/10.1002/nme.1617}
  {\path{doi:10.1002/nme.1617}}.

\bibitem{lee2015direct}
M.~Lee, R.~D. Moser, Direct numerical simulation of turbulent channel flow up
  to {$Re_{\tau} \approx 5200$}, Journal of Fluid Mechanics 774 (2015)
  395--415.
\newblock \href {https://doi.org/10.1017/jfm.2015.268}
  {\path{doi:10.1017/jfm.2015.268}}.

\bibitem{zhang2025dynamical}
S.~Zhang, Y.~Fan, D.~Wu, C.~Zhang, X.~Hu, Dynamical pressure boundary condition
  for weakly compressible smoothed particle hydrodynamics, Physics of Fluids
  37~(2) (2025) 027193.
\newblock \href {https://doi.org/10.1063/5.0254575}
  {\path{doi:10.1063/5.0254575}}.

\bibitem{NASA_TMR_channel}
{NASA}, \href{https://tmbwg.github.io/turbmodels/channelflow_val_w06.html}{{2D
  Fully-Developed Channel Flow at High Reynolds Number: Wilcox2006-klim-m Model
  Results}}, Turbulence Modeling Resource, accessed September 8, 2026.
\newline\urlprefix\url{https://tmbwg.github.io/turbmodels/channelflow_val_w06.html}

\bibitem{ansys_vmfl012_wavy_channel}
ANSYS, Inc.,
  \href{https://ansyshelp.ansys.com/public/Views/Secured/corp/v242/en/fbu_vm/Hlp_VMFL012.html}{{VMFL012}:
  Turbulent Flow in a Wavy Channel}, release 2024 R2 (2024).
\newline\urlprefix\url{https://ansyshelp.ansys.com/public/Views/Secured/corp/v242/en/fbu_vm/Hlp_VMFL012.html}

\bibitem{wang2022isph}
D.~Wang, P.~L.-F. Liu, An {ISPH} with modified k--$\varepsilon$ closure for
  simulating breaking periodic waves, Coastal Engineering 178 (2022) 104191.
\newblock \href {https://doi.org/10.1016/j.coastaleng.2022.104191}
  {\path{doi:10.1016/j.coastaleng.2022.104191}}.

\bibitem{bao2023pof}
T.~Bao, J.~Hu, C.~Huang, Y.~Yu, Smoothed particle hydrodynamics with
  k-$\varepsilon$ closure for simulating wall-bounded turbulent flows at medium
  and high {Reynolds} numbers, Physics of Fluids 35~(8) (2023) 085114.
\newblock \href {https://doi.org/10.1063/5.0158301}
  {\path{doi:10.1063/5.0158301}}.

\bibitem{kuzan1986velocity}
J.~D. Kuzan, Velocity measurements for turbulent separated and near-separated
  flows over solid waves, Ph.d. thesis, University of Illinois at
  Urbana-Champaign (1986).

\end{thebibliography}
%
%
\end{document}